\documentclass{article}
\usepackage[utf8]{inputenc}
\usepackage{authblk}
\usepackage{setspace}
\usepackage[margin=1.25in]{geometry}
\usepackage{graphicx}
\graphicspath{ {./figures/} }
\usepackage{subcaption}
\usepackage{amsmath}
\usepackage{lineno}
\usepackage{float}
\usepackage{hyperref}
\usepackage{engord}
\usepackage{booktabs}
\usepackage{soul}
\usepackage{fancyhdr}
\usepackage{multirow}

\usepackage[table]{xcolor}
\usepackage{pgfplots}
\pgfplotsset{compat=1.17}

\newcommand{\colcell}[1]{%
    \pgfmathparse{#1*100}%
    \edef\temp{\noexpand\cellcolor{green!\pgfmathresult!yellow}}%
    \temp #1%
}

\engordraisetrue

\usepackage[bottom]{footmisc}

\usepackage{tikz}
\usetikzlibrary{arrows.meta,positioning,fit,backgrounds,calc}

\usepackage[style=nejm, 
citestyle=numeric-comp,
sorting=none, eprint=true]{biblatex}
\bibliography{cites_sources}

\title{Benchmarking Quantum Computers via Protocols\\Comparing IBM's Heron vs IBM's Eagle}
\author{$^*$Nitay Mayo, $^{*\dagger}$Tal Mor and $^*$Yossi Weinstein}
\affil{$^*$Department of Computer Science, Technion University, Haifa, Israel.}
\affil{$^\dagger$The Helen Diller Quantum Center}
\date{June 12, 2026}

\begin{document}

\maketitle

\begin{abstract}
As quantum computing hardware rapidly advances, objectively evaluating the capabilities and error rates of new processors remains a critical challenge for the field. A clear and realistic understanding of current quantum performance is essential for guiding research priorities and driving meaningful progress. In this work, we apply and extend a protocol-based benchmarking methodology (Meirom, Mor, Weinstein Arxiv 2505.12441) that utilizes well-defined \underline{quantumness} thresholds. By evaluating performance at protocol level rather than the gate level, this approach provides a transparent and intuitive assessment of whether specific quantum processors, or isolated sub-chips within them, can demonstrate a practical quantum advantage. To illustrate the utility of this method, we compare two generations of IBM quantum computers: the older Eagle architecture and the newer Heron architecture. Our findings reveal the genuine operational strengths and limitations of these devices, demonstrating substantial performance improvements in the newer Heron generation. 
This work was made possible by IBM Quantum policies that enable independent and objective assessment of its quantum computers and sub-chips. We strongly encourage other companies to emulate the independent qubit availability and the fair pricing that allow researchers to perform such assessments.

\end{abstract}
\newpage
\tableofcontents

\section{Introduction}
In this work, we present a benchmarking study conducted on two IBM's quantum computers: Heron-r2 chip and Eagle-r3 chip. These systems were specifically selected due to IBM's cost-effective pricing models and their provision of granular, qubit-specific control—distinct advantages over many other hardware vendors that are essential for executing our targeted benchmarking methodology. The Heron-r2 represents IBM's newer generation of superconducting quantum computers, designed with an expanded qubit count intended to improve performance and scalability. In contrast, the Eagle-r3 is from the older generation, with a smaller design that has fewer qubits. By attempting to benchmark both “new” and “old” generations side by side, this work provides a clearer picture of IBM’s hardware by assessing the quantumness of its machines. The necessity for robust hardware evaluation has been extensively discussed in recent literature, with various methodologies proposed to characterize quantum performance~\cite{Cross_2019, Emerson_2005, Erhard_2019, helsen2019newclassefficientrandomized}. Building upon this broader context, our work focuses specifically on protocol-level evaluation to provide a clearer picture of the quantum hardware.

During the second and third quarters of 2025, IBM has improved its quantum computers, noting a significant difference between the old and the new versions of some of its chips. In this work we present a comparison between the older and the newer results obtained on Brisbane as can be seen in section \ref{sec:motivation_to_transmit}.
For the sake of consistency in terminology we use the term Old Brisbane for results obtained before the beginning of August 2025 and Modified Brisbane for results obtained after that date. Notably, IBM did not officially declare any improvement in Brisbane on that date, our analysis indicated a significant performance improvement. Hence we decided to make a temporal comparison to illustrate the change.

Originally we tested a third quantum computer from IBM - Sherbrooke from the Eagle-r3 series. Unfortunately it was retired by IBM before we completed its assessment, its partial assessment is presented in the appendix~\ref{sec:comparing_brisbane_sherbrooke} side by side with the older version of Brisbane. 

While assessing the different quantum chips it was apparent that some of them couldn't handle even the simplest protocols of our benchmarking procedure, presented in~\cite{Bench1_arxiv}, so we defined a new protocol - transmit. The motivation for defining this protocol is discussed in section \ref{sec:motivation_to_transmit}.

The rest of the paper is structured as follows:
\begin{itemize}
\setlength{\itemsep}{0pt}
    \item Section~\ref{sec:the_protocols} reviews the five benchmarking protocols defined in the earlier publication
    \item In section~\ref{sec:optimal_lookup_workflow} we define the optimal lookup workflow, explain how it is used to evaluate the different hardware and how it eventually constructs the protocol vector
    \item Section~\ref{sec:motivation_to_transmit} presents results of Brisbane quantum computer taken four months apart, explains the motivation for a new protocol and formally defines it
    \item Sections~\ref{sec:Brisbane_results_section} and~\ref{sec:Kingston_results_section} provides the full benchmarking results of Eagle r3 - Brisbane and Heron r2 - Kingston, respectively
    \item Section~\ref{sec:comparison_section} compares the two architectures at both protocol and sub-chip level
    \item Section~\ref{sec:limitations_and_malfunctiuons} explains the limitations and malfunctions we encountered during this research
    \item Section~\ref{sec:conclusions} concludes the study
    \item The appendices include a comparison between Brisbane and another Eagle r3 quantum computer - Sherbrooke, as well as extended comparison results, consistency checks, and supporting figures not shown in the main text
\end{itemize}

\section{The Protocols}\label{sec:the_protocols}
\begin{figure}[H]
    \centering

    \begin{subfigure}[t]{0.48\linewidth}
        \centering
        \includegraphics[width=\linewidth]{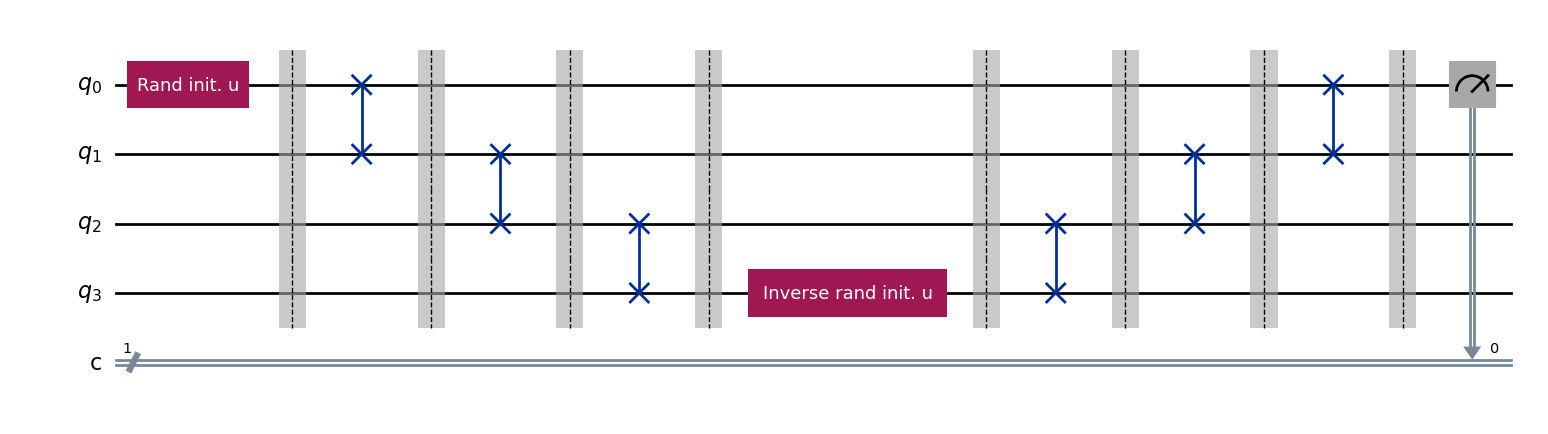}
        \caption{Do-nothing}
    \end{subfigure}
    \hfill
    \begin{subfigure}[t]{0.48\linewidth}
        \centering
        \includegraphics[width=\linewidth]{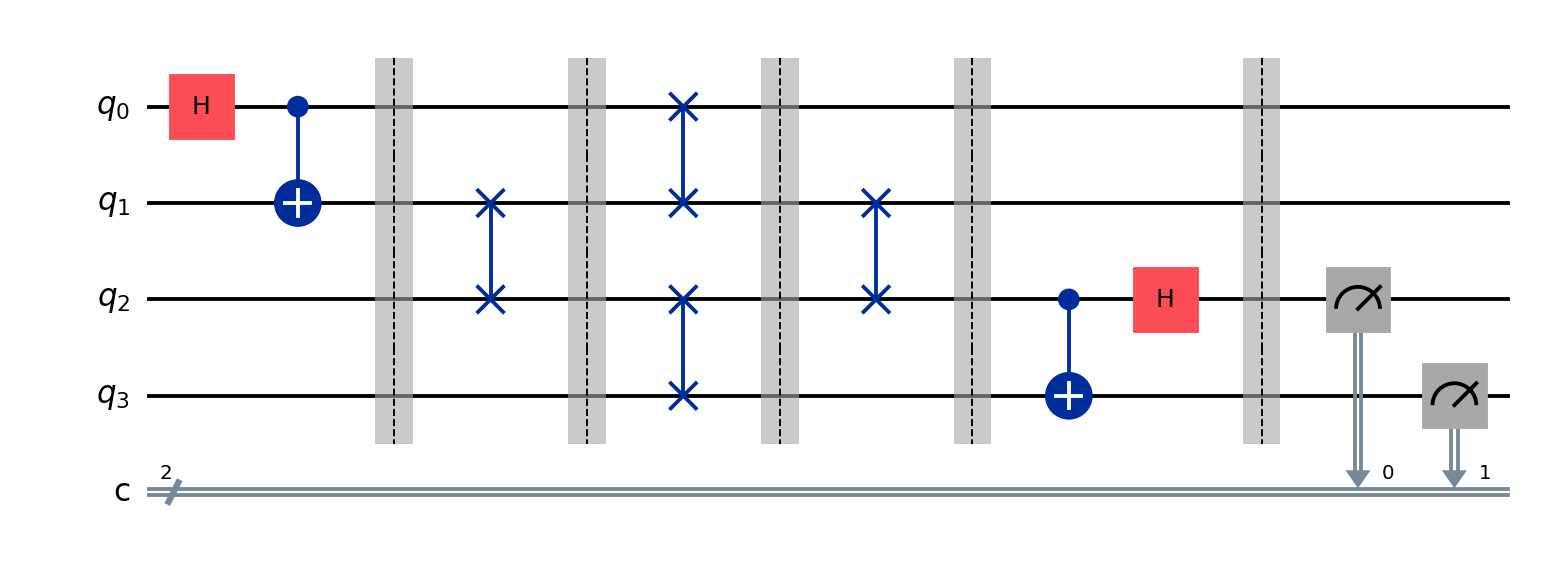}
        \caption{Bell-state transfer}
    \end{subfigure}

    \begin{subfigure}[t]{0.48\linewidth}
        \centering
        \includegraphics[width=\linewidth]{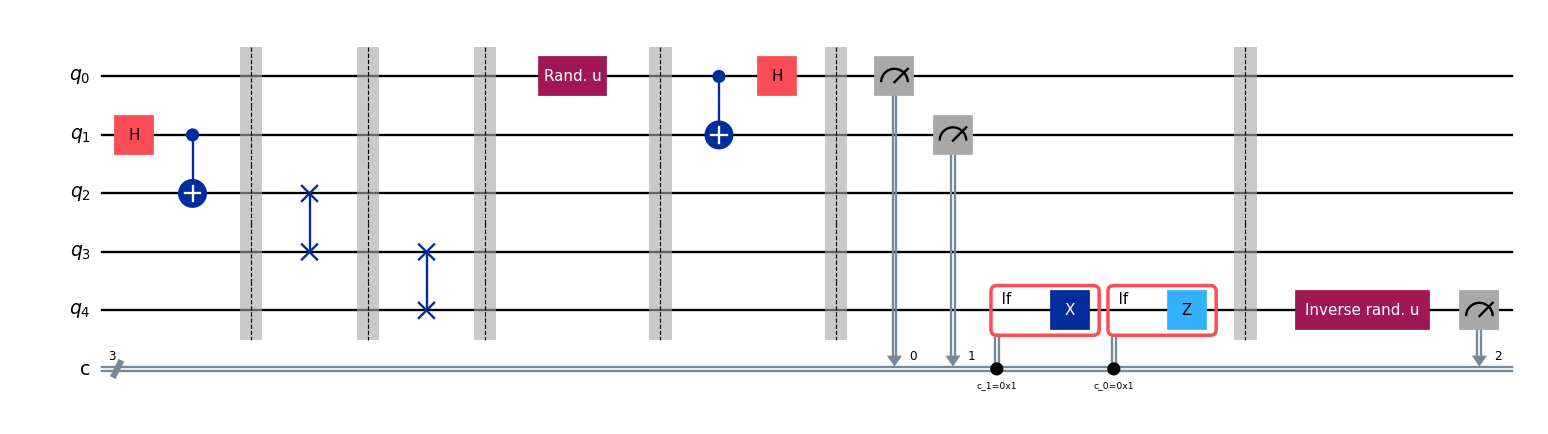}
        \caption{Teleportation}
    \end{subfigure}
    \hfill
    \begin{subfigure}[t]{0.48\linewidth}
        \centering
        \includegraphics[width=\linewidth]{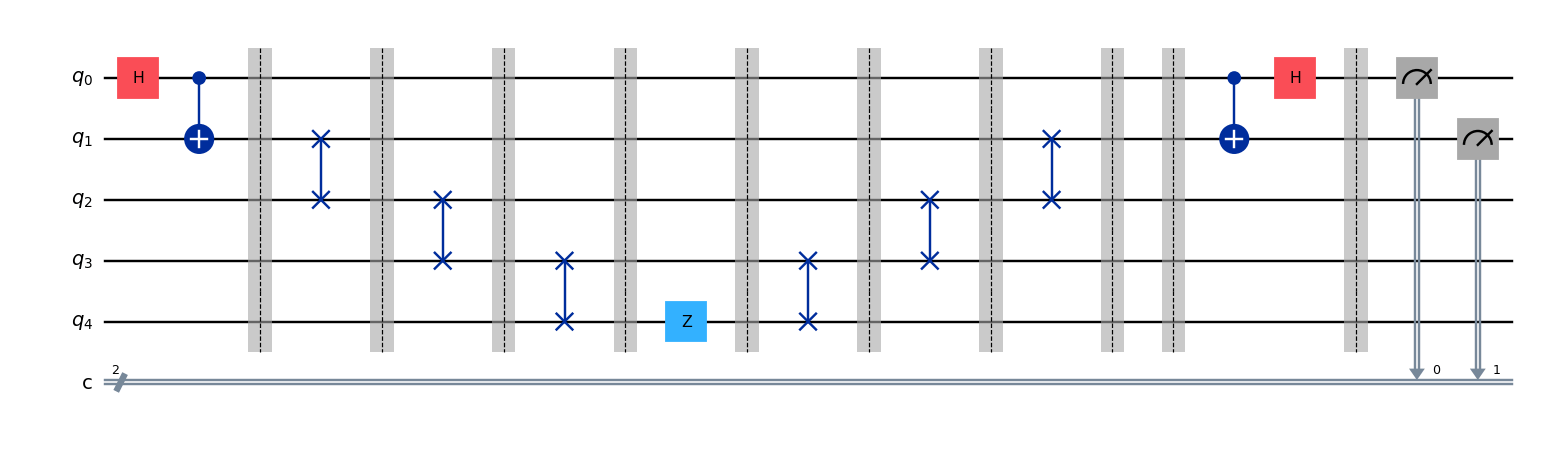}
        \caption{Super-dense coding}
    \end{subfigure}
    
    \begin{subfigure}[t]{0.48\linewidth}
        \centering
        \includegraphics[width=\linewidth]{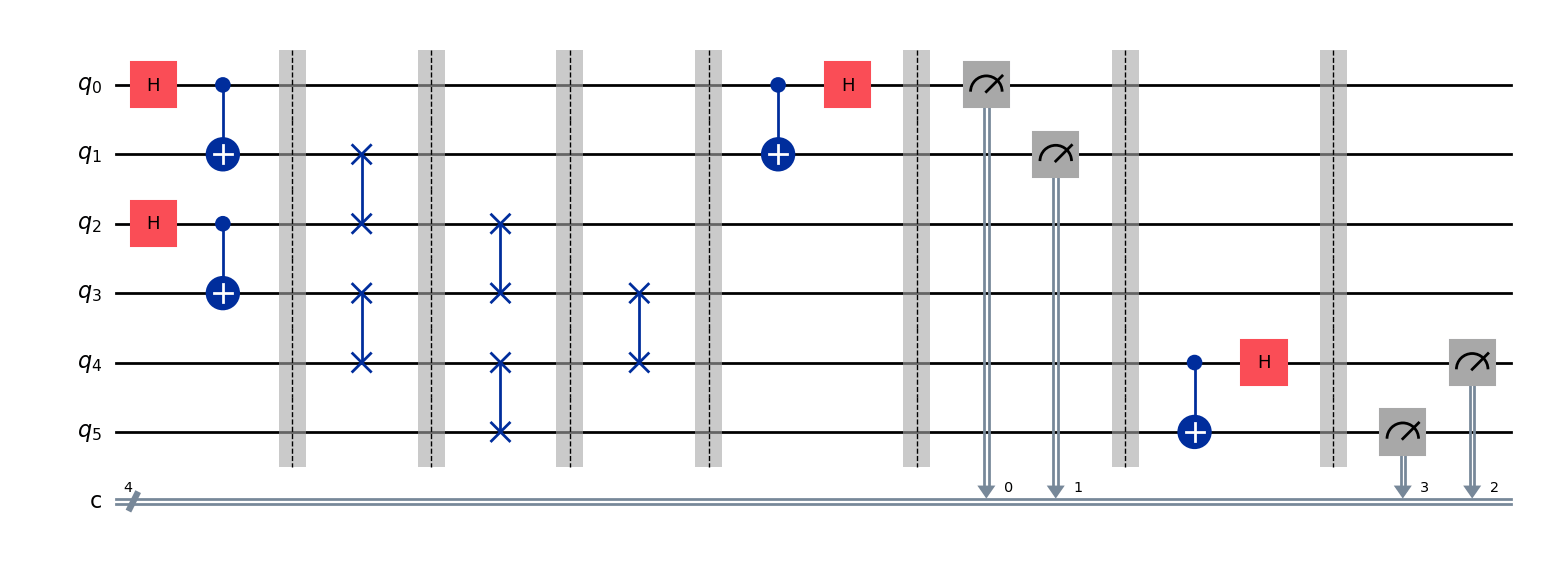}
        \caption{Entanglement swapping}
    \end{subfigure}
    
    \caption{The five protocols applied on the quantum computers. These protocols were defined in previous work done in this subject}
    \label{fig:the-six-protocols}
\end{figure}

In this paper we rely on previous conceptual work~\cite{Bench1_arxiv} which defined 5 basic protocols (see Figure~\ref{fig:the-six-protocols}). This section discusses them briefly and points out the key attributes of the first two protocols - \textit{do-nothing} and \textit{Bell-state transfer}.
\subsection{Basic Protocols}

\subsubsection{Do-nothing}
This protocol is the baseline of our assessment, each benchmarking of a quantum computer starts with a series of do nothing experiments. We began this work relying on \textit{do-nothing} as the baseline protocol, but as the results came in we noticed that some quantum computers failed to demonstrate quantum advantages even in this basic task. This led us to define a new, simpler protocol which will be discussed and defined in section~\ref{sec:TRANSMIT_definition}

\subsubsection{Bell-state transfer}\label{sec:bell_swap_explanation}
As this protocol was defined in earlier work~\cite{Bench1_arxiv} we briefly explain it here as its results shed light on the differences between the two tested quantum computers. 
We start this protocol by initializing a maximally entangled pair of qubits (a bell state) on Alice's side, then she transmits the bell-state to Bob's camp on the other side of the physical circuit. Bob measures in the bell basis, the threshold for quantumness on this protocol is 0.5. This threshold for entangled pair state transfer was inspired by the threshold for entanglement stated by the \textit{Werner States}~\cite{PhysRevA.40.4277} and is used in many cases to prove the presence of entanglement~\cite{Peters_2004, kwiat_2004, zeilinger_2012}.

\subsubsection{Teleportation, Super-dense coding and Entanglement swapping}
The rest of the protocols are well-known and thoroughly explained in the previous paper. In this work we have used all of them, while only the newer generation of IBM's quantum computers, Heron, managed to demonstrate quantum advantage when performing them as we'll present in sections \ref{sec:mod_brisbane_results_single_and_pairs} and \ref{sec:kinsgton_results}.

\section{The Optimal Lookup Workflow}\label{sec:optimal_lookup_workflow}
In order to compare different architectures and technologies we developed an `Optimal lookup' workflow. This approach produces a \textbf{protocol vector} - a compact yet comprehensive way to view the chip's performance across multiple aspects of quantum capabilities. The workflow is designed to be budget efficient while still providing a complete benchmarking of the subjected chip. 

\begin{figure}
\centering
\begin{tikzpicture}[
  >=Latex,
  node distance=10mm and 10mm,
  block/.style={draw, rounded corners, minimum width=42mm, minimum height=10mm, align=center},
  wideblock/.style={draw, rounded corners, minimum width=60mm, minimum height=10mm, align=center},
  smallblock/.style={draw, rounded corners, minimum width=38mm, minimum height=10mm, align=center},
  groupbox/.style={draw, rounded corners=6pt, line width=2pt, inner sep=6mm},
  titlelabel/.style={font=\bfseries\large, align=center}, 
  note/.style={align=left}
]

\coordinate (L0) at (0,0);
\coordinate (L1) at ($(L0)+(52mm,0)$);

\node (donothing) [block] at (L1) {Do-nothing\\\small c2c};

\node (eachrect)  [wideblock, below=5mm of donothing] {Full assessment of\\telep., bell-state transfer,\\ent. swapping, s-d coding \\and the rest of do-nothing \\assessment};
\draw[->] (donothing) -- (eachrect);

\node (eachpair)  [wideblock, below=20mm of eachrect] {For each protocol and\\neighbors that passed};
\node (fa_pair)   [block, below=8mm of eachpair] {Full assessment};
\draw[->]  (eachrect) -- (eachpair);
\draw[->] (eachpair) -- (fa_pair);

\node (output) [wideblock, below=14mm of fa_pair,
  minimum width=78mm, text width=78mm, align=center, inner sep=3mm]
{%
  Output\\[-0.2ex]
  \rule{\linewidth}{0.5pt}\\
  Max capabilities vector of each single/pair
};

\draw[->] (fa_pair) -- (output);


\coordinate (old_start) at ($(donothing.north)+(0,12mm)$);
\node at ([yshift=4mm]old_start) {Start};
\draw[->] (old_start) -- (donothing.north);

\node[groupbox, draw=cyan!60!black, fit=(donothing) (eachrect)] (g_single) {};
\node[titlelabel, text=cyan!60!black, anchor=south west]
  at ([xshift=2mm,yshift=1mm] g_single.south west) {Single rectangle};

\node[groupbox, draw=green!50!black, fit=(eachpair) (fa_pair)] (g_pair) {};
\node[titlelabel, text=green!50!black, anchor=south west]
  at ([xshift=2mm,yshift=0.5mm] g_pair.south west) {Rectangles pair};

\coordinate (Rmid) at ($(donothing.east)+(45mm,2mm)$);

\node (c2c) [smallblock] at (Rmid) {c2c};
\node (ml)  [smallblock, below=8mm of c2c] {M-L};
\node (al)  [smallblock, below=8mm of ml] {A-L};
\draw[->] (c2c) -- (ml);
\draw[->] (ml) -- (al);

\def\FApadRight{24mm} 
\coordinate (fa_rightpad) at ($(al.east)+(\FApadRight,0)$);

\node (fa_box) [groupbox, fit=(c2c) (ml) (al) (fa_rightpad), inner sep=4mm, rounded corners=10pt] {};

\node[font=\bfseries\large, anchor=north east, align=center]
  at ([xshift=-1.5mm, yshift=-1.5mm] fa_box.north east) {Full\\assessment};

\coordinate (fa_start) at ($(c2c.north)+(0,12mm)$);
\node at ([yshift=4mm]fa_start) {};
\draw[->] (fa_start) -- (c2c.north);


\end{tikzpicture}%

\caption{Flowchart of the optimal lookup workflow. Each arrow in this flowchart has the same meaning - proceed to next stage only with sub-chips that passed the threshold in the previous stage}
\label{fig:optimal_workflow}
\end{figure}
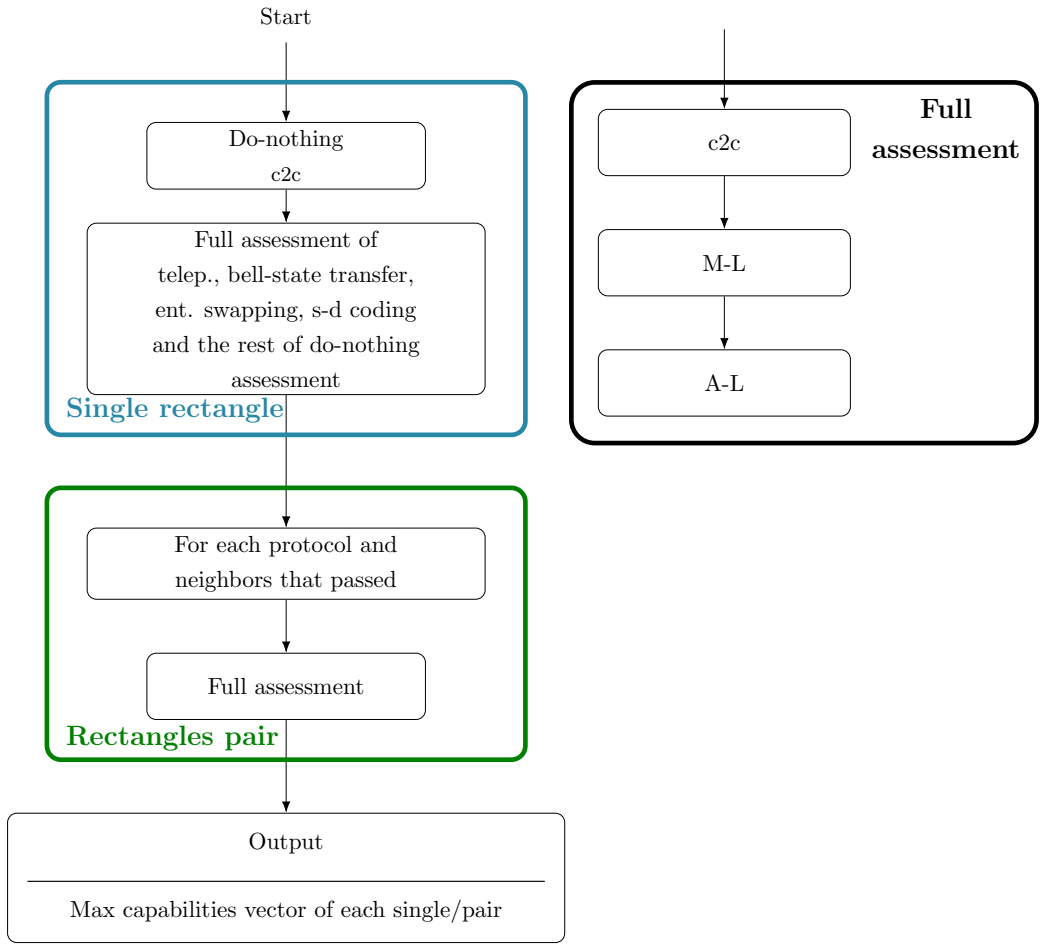

As the chips of the Eagle and Heron series are organized in a repeating rectangles pattern. The workflow as seen in figure~\ref{fig:optimal_workflow} is composed of 2 stages: assessing single rectangles and pairs of rectangles. We use the term sub-chip for any sub-set of qubits selected from the whole chip, in this work we'll use this term for single and pairs of rectangles.
Each stage in the workflow is composed of a ``Full Assessment" applied to the tested sub-chips. 

\begin{figure}[H]
    \centering
    \includegraphics[width=0.5\linewidth]{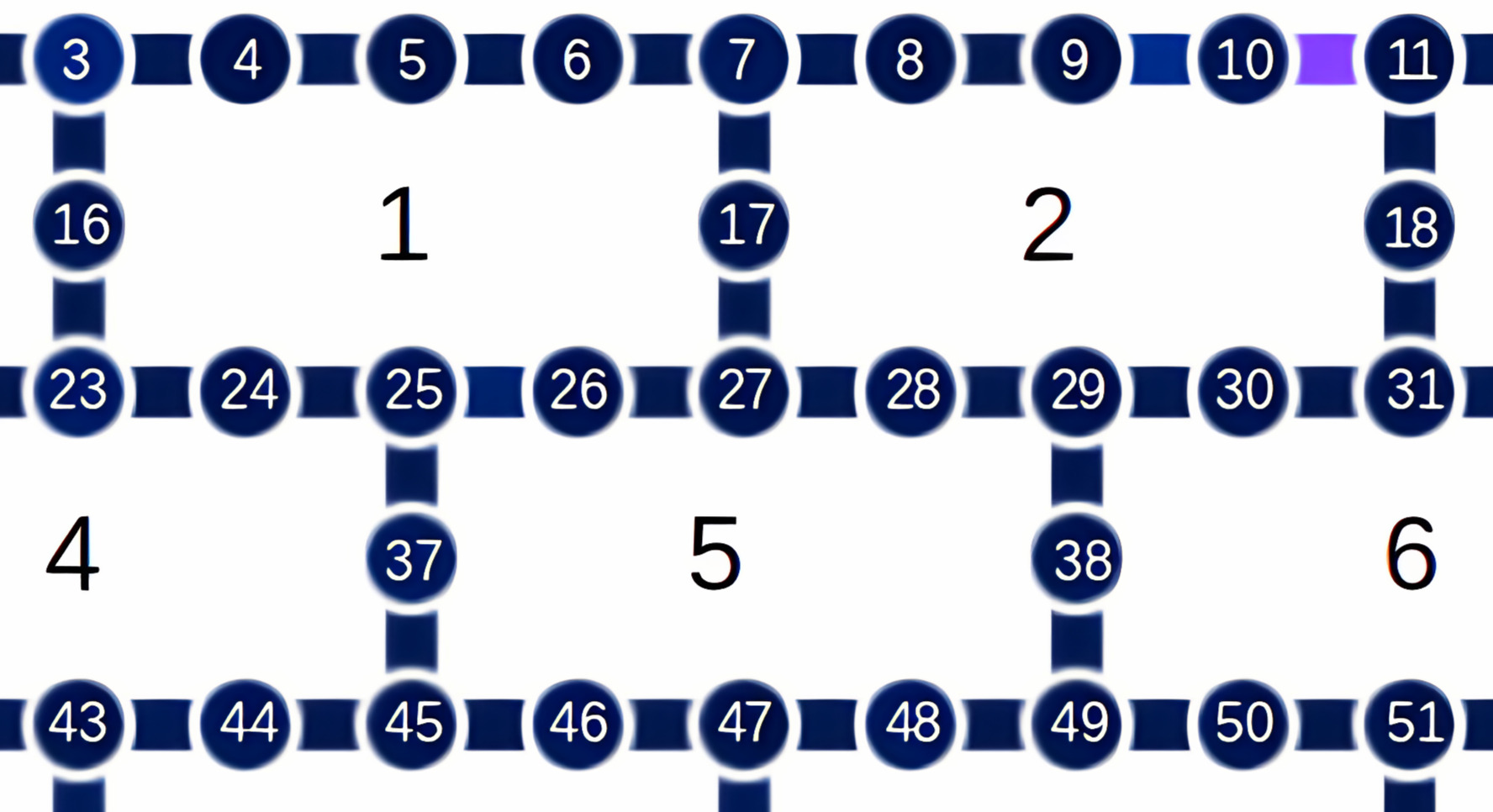}
    \caption{Snapshot from the Heron r2 chip architecture. Qubits are arranged in repeating rectangular layout, we numbered each rectangle for consistent reference}
    \label{fig:rectangle_sample}
\end{figure}
The three stages of full assessment of a sub-chip are:

\begin{enumerate}
    \item c2c - shorthand for ``corner to corner". In this test we obtain a first impression of the tested sub-chips (rectangles). In this test we run the currently applied protocol over every path (sequence of consecutive qubits) that starts at one corner of a rectangle and ends at the furthest corner.
 For example, consider rectangle 1 in figure \ref{fig:rectangle_sample}, applying c2c will include both routes between qubit 3 and 27 and both routes from 7 to 23. each route is tested on both directions so the total number of circuits per rectangle is eight. If the tested sub-chip is a pair of rectangles, we separate the case into two: if the rectangles are side by side like rectangles 1 and 2 - we treat them like one long rectangle. We run the protocol over the paths connecting opposing corners of this long rectangle. If the two rectangles are connected diagonally, then we test only the paths that go between the two most distant corners. For example in \ref{fig:rectangle_sample} suppose we check rectangles 1 and 5 as a pair, we'll only test the three paths from qubit 3 to 49, which are $3\rightarrow23\rightarrow25\rightarrow45\rightarrow49, \ 3\rightarrow23\rightarrow29\rightarrow49,\ 3\rightarrow7\rightarrow27\rightarrow29\rightarrow49$, and the paths back - six paths in total.
    \item M-L - ``Maximal Lengths", The objective of this test is to employ a budget-efficient approach, similar to the c2c method, ensuring that every qubit serves as the measured target. The maximal lengths test go's over all paths of the maximal length in a rectangle, where the closed loop shape of a single rectangle allows each qubit to be the start and end of an inner route. For example in fig \ref{fig:rectangle_sample}, looking at rectangle 1, in addition to the eight paths we apply the protocol to in the c2c stage, we now run both paths between qubit 4 and qubit 26, and both paths between qubit 5 and qubit 25, and so on. Totaling at 24 paths per rectangle. Because treating a pair as a sub-chip breaks the cyclical form of the sub-chip the maximal lengths test gets degenerated to only a few paths and is missing the point of measuring all qubits in the sub-chip. In the example of figure~\ref{fig:rectangle_sample} for rectangles 1 and 5 as pair there are only sixteen maximal length paths: $3 \times(3\rightarrow49), \ 2\times (4\rightarrow48), \ 3 \times (5 \rightarrow 47)$ and back, thus sixteen paths in total. A pair of rectangles contains twenty-one qubits, and as shown in the M-L stage only six of them act as measured qubits. Because of this we decided not to run this stage on pairs of rectangles, in this case we jump from c2c straight to the following stage.
    \item A-L - ``All Lengths", this is the final stage in the full assessment method. In A-L we run all the inner routes a single rectangle has. Note that this test has a different number of paths for each protocol, for example some protocols can be done on a path of length two and some can't. This test gives us a full view on the rectangle performance, and a rectangle that passed A-L on a specific protocol is consider to have true quantum advantage on the aspect of that protocol, e.g. if a rectangle passed A-L for do-nothing then we say that this rectangle is a do-nothing-capable sub-chip.
\end{enumerate}

It is important to note that in each test we proceed only with rectangles that passed the previous stage, e.g. only rectangles who passed c2c move on to M-L, only those who passed M-L move to A-L, only two neighboring rectangles that passed A-L as single rectangles proceed to be tested with c2c as rectangle pairs and only the pairs that passed c2c are tested in the A-L stage as pair. 

\section{Brisbane timeline and Motivation for Transmit} \label{sec:motivation_to_transmit}
Assessment of an older version of Brisbane using the workflow yielded poor results. Specifically, the optimal lookup workflow process had to be stopped at an early stage because almost no sub-chip managed to pass the first stages of the assessments. In this section we present a timeline of the results from the do-nothing protocol on the Brisbane chip at different times, four months apart, before and after the modification. While the old results motivated us to define a new protocol - ``Transmit", the performance of the modified chip suggests this protocol is not strictly necessary for the optimal lookup workflow. However, we define and present the Transmit protocol results here to present the complete results obtained on each quantum computer and to document the development of the workflow.

\subsection{Defining Transmit}\label{sec:TRANSMIT_definition}
While working and assessing the different hardware we encountered a problem where the protocols were too complicated and only a few chips managed to pass them. This led us to define transmit, which is the simplest protocol that can be thought of. This protocol is simpler from all the rest, being merely a component of do nothing. Our intentions for defining this protocol is to refine the result's resolution on badly performing quantum computers. Although the simplicity of this protocol, we still found many qubits-clusters within each of the proclaimed 100+ qubits quantum computers that were not successful in performing the transmit protocol.

\begin{figure}[H]
    \centering
    \includegraphics[width=0.8\linewidth]{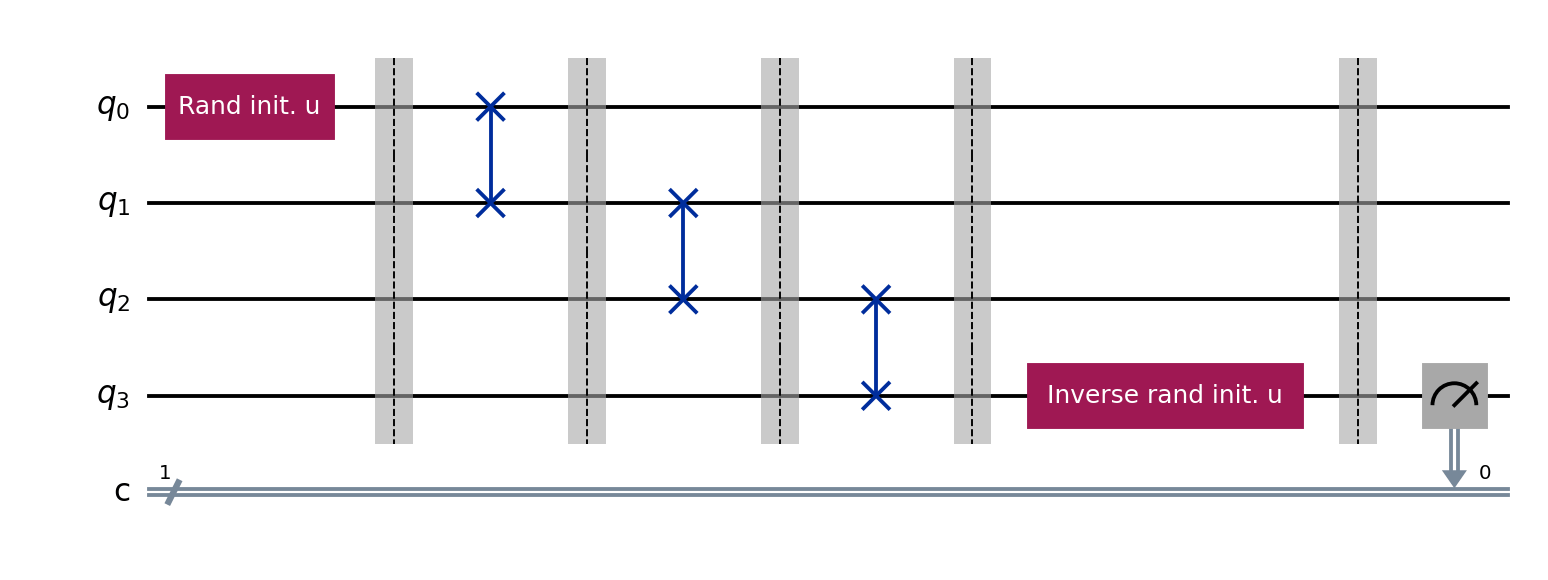}
    \caption{Transmit protocol}
    \label{fig:transmit_protocol_image}
\end{figure}

The transmit protocol for a circuit of size n is defined as follows:
\begin{enumerate}
    \item Reset the single work qubit $q_0$ and the $n-1$ ancilla qubits to zero
    \item Alice applies a random 1-qubit gate $U_{work}$  to $q_0$, the work qubit.
    \item Apply swap gates to move $q_0$'s state to Bob's camp, i.e. the n-th qubit. In the example shown in fig \ref{fig:transmit_protocol_image} we set $n=4$ thus 3 swap gates are needed to move the state of the work qubit we say that the swap distance between the camps is 3, or more generally $n-1$
    \item Bob applies the inverse gate $U^\dagger_{work}$ and measures his qubit in the computation basis
\end{enumerate}
The required threshold for quantumness is a fidelity of at least $\frac{2}{3}$ for the work qubit. This threshold was originally derived by Massar and Popescu~\cite{PhysRevLett.74.1259_fid_thresh}, and subsequently inspired Meirom, Mor and Weinstein~\cite{Bench1_arxiv} to define this threshold as the quantumness boundary for degraded state transfer. There are more noted cases where this threshold is used to prove the presence of quantum phenomena~\cite{Pfaff_2014} 

We note that IBM managed to improve the chip significantly, hence we present the comparison between Heron and Eagle over the old workflow, which did not include transmit in the selection process. Still, the results of transmit are presented as a refined baseline for the quantum computers assessments.

\subsection{Old vs Modified Brisbane - do-nothing protocol - c2c}
\begin{figure}[H]
    \centering

    \begin{subfigure}[t]{0.48\linewidth}
        \centering
        \includegraphics[width=\linewidth]{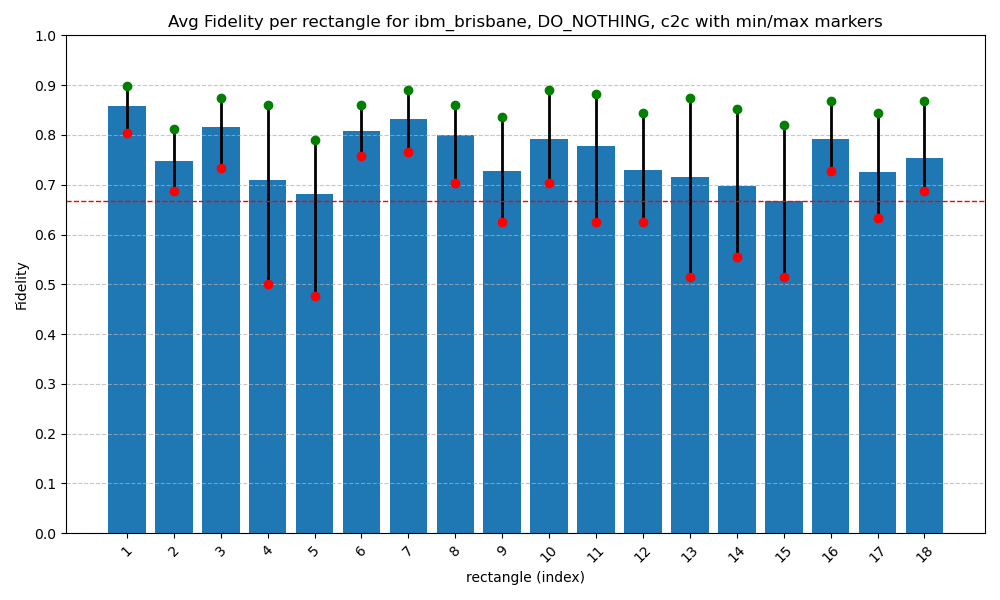}
        \caption{Old Brisbane results from 29 May 2025}
        \label{fig:old_brisbane_new_v_old_do_nothing_c2c}
    \end{subfigure}
    \hfill
    \begin{subfigure}[t]{0.48\linewidth}
        \centering
        \includegraphics[width=\linewidth]{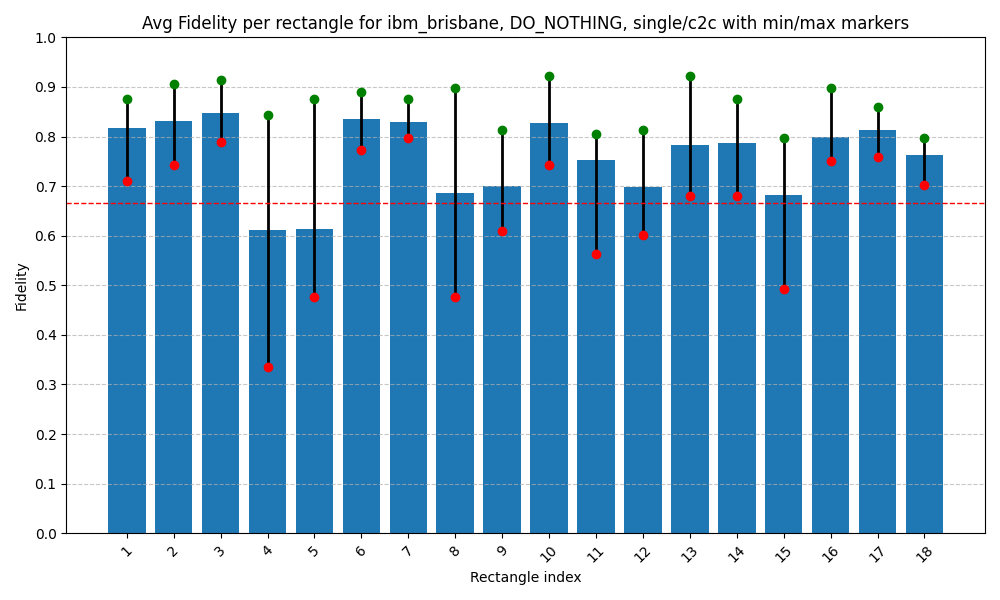}
        \caption{Modified Brisbane results from 28 Sep 2025}
        \label{fig:new_brisbane_new_v_old_do_nothing_c2c}
    \end{subfigure}
    
    \caption{Results of Brisbane c2c do-nothing protocol, taken four months apart. In this test all the rectangles in the quantum computer are tested, 18 rectangles for each test. The blue bars present the mean fidelity while the green and the red dots are the max and min respectively. The red dotted line marks the fidelity threshold for this protocol, in do-nothing protocol the threshold is $\frac{2}{3}$}
    \label{fig:brisbane_old_v_new_do_nothing_c2c}
\end{figure}

We ran do-nothing c2c on both old and new Brisbane four months apart, the results are presented in figure~\ref{fig:brisbane_old_v_new_do_nothing_c2c}. This initial experiment does not show a significant difference between old and modified Brisbane results.
In the modified version we have 11 rectangles that passed the $\frac{2}{3}$ fidelity threshold for quantumness and in the old version only 9 rectangles passed. We see a consistent behavior across runs in the Brisbane chip with eight rectangles that pass in both experiments. In the old Brisbane the failed rectangles are \textbf{4}, \textbf{5}, \textbf{9}, \textbf{11}, \textbf{12}, 13, 14, \textbf{15}, 17. In the modified Brisbane the failed ones are \textbf{4}, \textbf{5}, 8, \textbf{9}, \textbf{11}, \textbf{12}, \textbf{15}. Marked in \textbf{bold} are the rectangles that failed both experiments. 

\subsection{Old vs Modified Brisbane - do-nothing protocol - M-L}
\begin{figure}[H]
    \centering

    \begin{subfigure}[t]{0.48\linewidth}
        \centering
        \includegraphics[width=\linewidth]{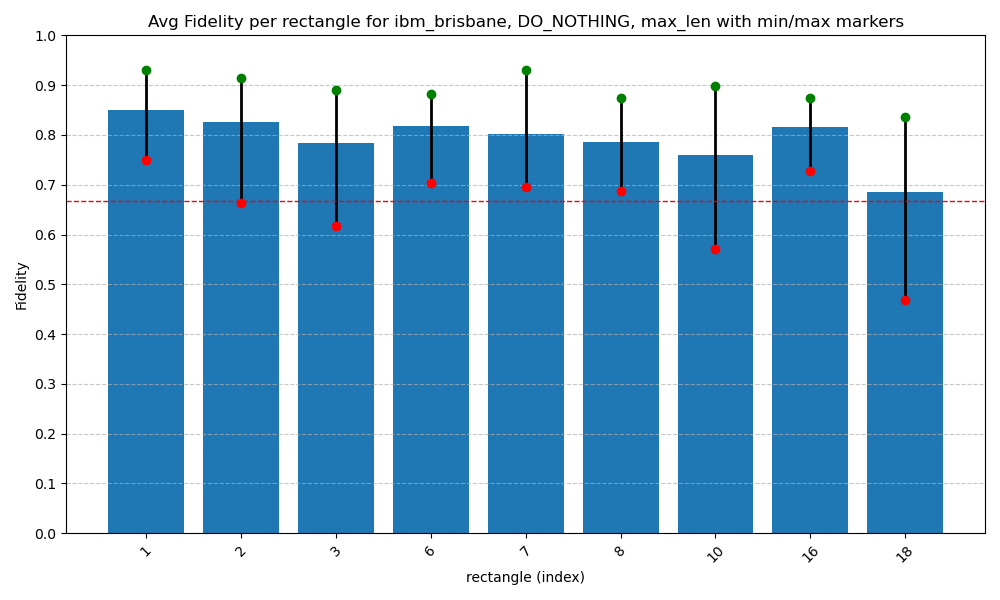}
        \caption{Old Brisbane results from 29 May 2025}
        \label{fig:old_brisbane_new_v_old_do_nothing_ML}
    \end{subfigure}
    \hfill
    \begin{subfigure}[t]{0.48\linewidth}
        \centering
        \includegraphics[width=\linewidth]{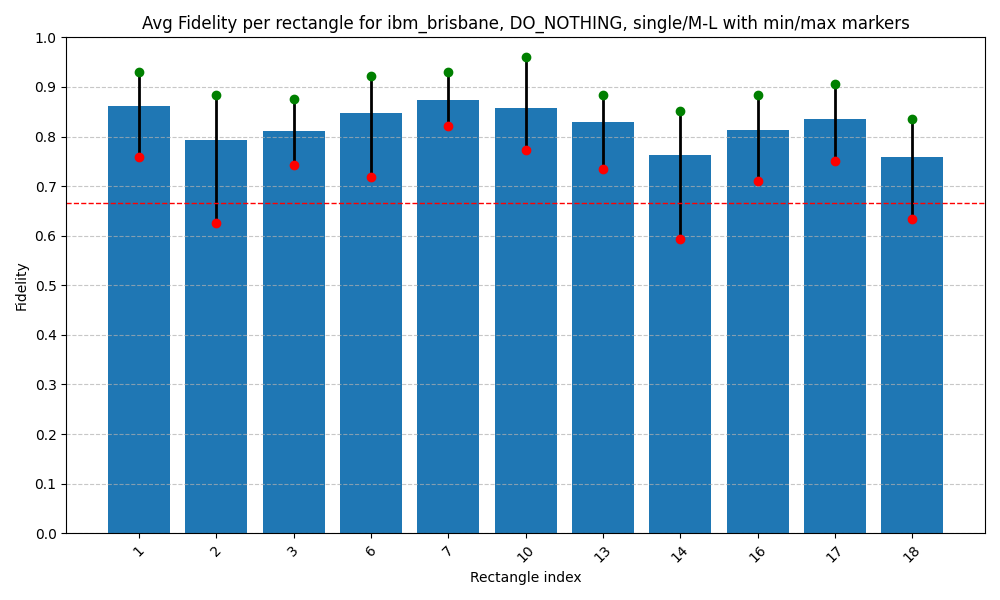}
        \caption{Modified Brisbane results from 28 Sep 2025}
        \label{fig:new_brisbane_new_v_old_do_nothing_ML}
    \end{subfigure}
    
    \caption{Results of Brisbane M-L do nothing protocol on rectangles that passed the c2c stage in figure \ref{fig:brisbane_old_v_new_do_nothing_c2c}. In this stage there are 24 different paths for every tested rectangle, and each qubit of the rectangle gets to be the measured qubit}
    \label{fig:brisbane_old_v_new_do_nothing_ML}
\end{figure}
In the M-L experiment we see similar results for both chips. Here it is important to note that while do-nothing is a very simple protocol, only five rectangles from Old Brisbane's 18 rectangles managed to successfully show quantumness capabilities in this simple task. The rectangles that failed on old Brisbane are \textbf{2}, 3, 10, \textbf{18} and the ones who failed on the modified Brisbane are \textbf{2}, 14, \textbf{18}, marked in \textbf{bold} are the rectangles that failed on both quantum computers on this stage. Rectangle 14 which failed M-L on the modified Brisbane failed on the old Brisbane in c2c stage (see figure~\ref{fig:old_brisbane_new_v_old_do_nothing_c2c}). We see that in the modified Brisbane eight rectangles passed the M-L stage, in contrast to old Brisbane where only five rectangles passed.
Following the initial experiments on the older iteration of Brisbane, we concluded that the baseline fidelity of the available hardware was insufficient to reliably execute the Do-nothing protocol. Consequently, we developed the even more fundamental Transmit protocol to establish a lower baseline.

\subsection{Old vs Modified Brisbane - do-nothing protocol - A-L}
\begin{figure}[H]
    \centering

    \begin{subfigure}[t]{0.48\linewidth}
        \centering
        \includegraphics[width=\linewidth]{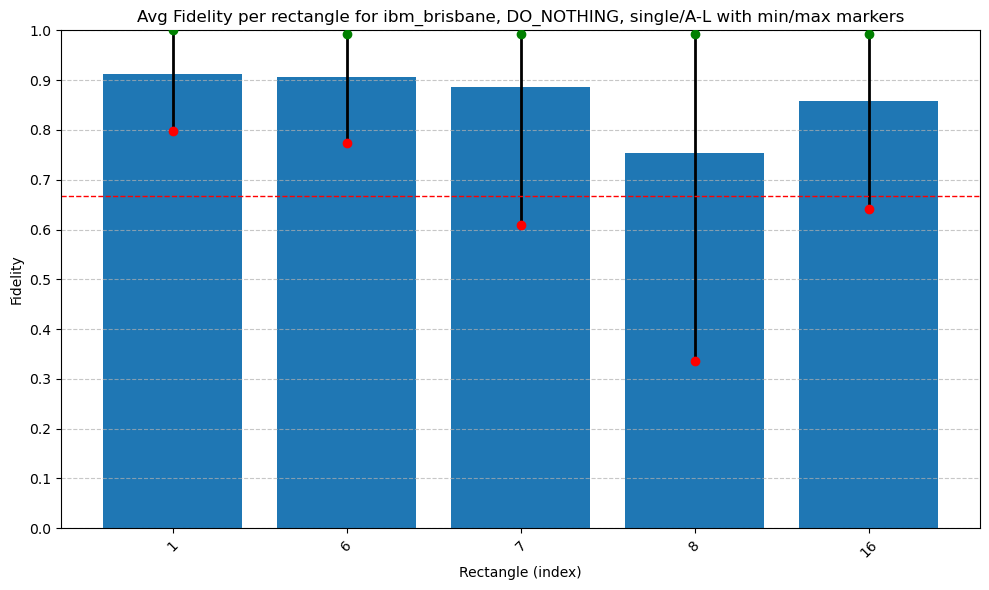}
        \caption{Old Brisbane results from 29 May 2025}
        \label{fig:old_brisbane_new_v_old_do_nothing_AL}
    \end{subfigure}
    \hfill
    \begin{subfigure}[t]{0.48\linewidth}
        \centering
        \includegraphics[width=\linewidth]{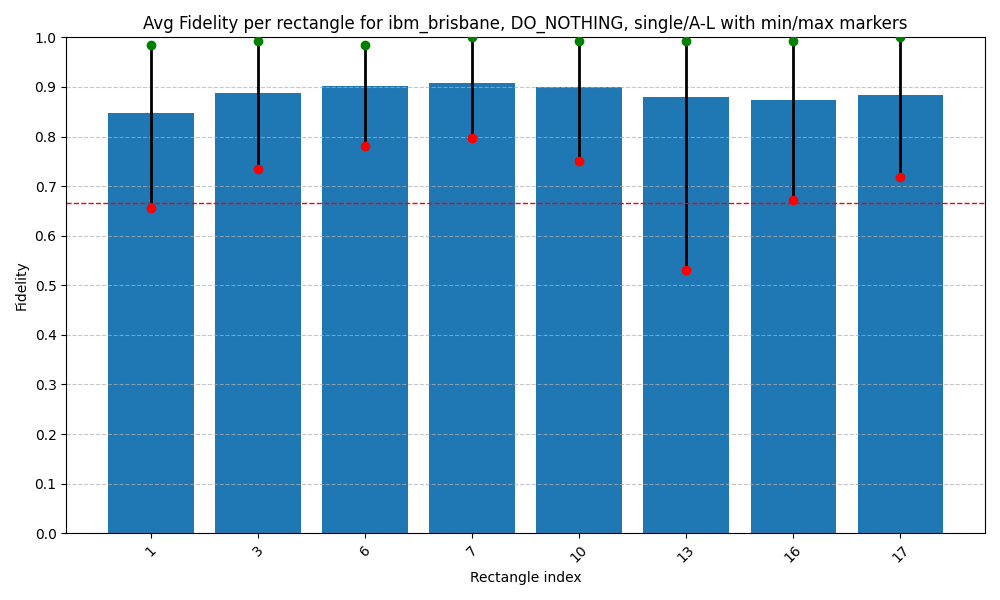}
        \caption{Modified Brisbane results from 28 Sep 2025}
        \label{fig:new_brisbane_new_v_old_do_nothing_AL}
    \end{subfigure}
    
    \caption{Results of Brisbane A-L do nothing protocol on rectangles that passed the M-L stage in figure  \ref{fig:brisbane_old_v_new_do_nothing_ML}. In this stage each rectangle has 144 inner paths that we apply the do-nothing protocol to. This is every minimal inner path between every two qubits}
    \label{fig:brisbane_old_v_new_do_nothing_A-L}
\end{figure}
In the A-L stage (figure~\ref{fig:brisbane_old_v_new_do_nothing_A-L}) the difference between the old and modified chip is best demonstrated. While in the old Brisbane only two sub-chips managed to perform a successful do-nothing protocol on every inner path, which is $\frac{1}{9}$ from the total of 18 rectangles, the modified version managed to produce six do-nothing-capable sub-chips. In the old Brisbane only rectangles 1 and 6 manage to show quantumness while in the modified version rectangles 3, 6, 7, 10, 16, 17 managed to pass the threshold. Rectangle 1 passed in the old version but failed in the modified version of Brisbane. Its minimal fidelity, marked by the red dot on the right figure, is  0.656 which is very close to the $\frac{2}{3}$ threshold but still below it.

The fact that only rectangles 1 and 6 passed the assessment on old Brisbane gave us the motivation to define the simpler protocol - transmit. Then, as can be seen in figures~\ref{fig:new_brisbane_new_v_old_do_nothing_ML} and \ref{fig:new_brisbane_new_v_old_do_nothing_AL}, IBM improved Brisbane significantly and transmit was not necessary any more in the optimal lookup workflow, and we went back to the workflow presented above. We also decided to remove almost all results of old Brisbane and Sherbrooke from the main body of the paper, those can be viewed in appendix~\ref{sec:comparing_brisbane_sherbrooke}. 

\section{IBM's Eagle - Brisbane}\label{sec:Brisbane_results_section}
\subsection{Introduction}
\begin{figure}
    \centering
    \includegraphics[width=0.7\linewidth]{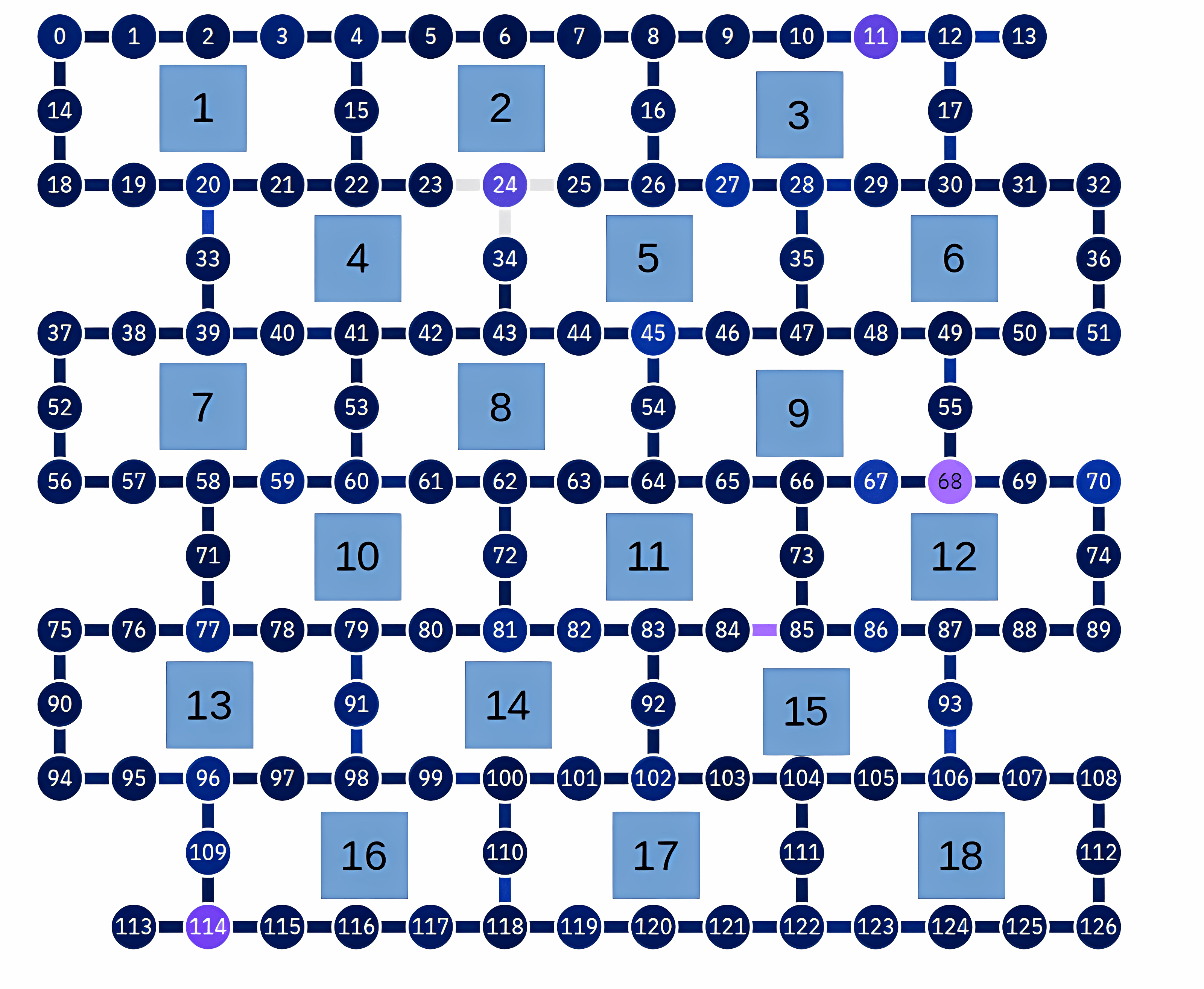}
    \caption{Eagle-r3 series qubits map and rectangle indexes}
    \label{fig:eagle_map}
\end{figure}

The following sections present results of optimal lookup workflow for IBM's Eagle-r3 quantum computer, Brisbane. This chip is composed of 127 qubits, arranged in rectangles of 12 qubits each (as seen in fig \ref{fig:eagle_map}). We tested Brisbane across multiple occasions, and as discussed earlier there was a great improvement in its capabilities. Here we present the results for modified Brisbane (i.e. the newer version of the chip), The results of old Brisbane with a comparison to Sherbrooke (another Eagle-r3 chip) can be view in section \ref{sec:comparing_brisbane_sherbrooke}.

\subsection{Results for Modified Brisbane for Singles and Pairs Sub-Chips}\label{sec:mod_brisbane_results_single_and_pairs}
In this section, we present the optimal lookup workflow on the modified Brisbane. The rectangles selection process is according to the old workflow  (shown in figure \ref{fig:optimal_workflow}), i.e. We run do nothing c2c first and only on the rectangles that passed this test we run a ``Full Assessment" for each of the other protocols. The results of transmit are presented as well but are not part of the selection process.
In the following section we demonstrate a ``Full Assessment" procedure to illustrate this process, which takes place with all protocols. For the subsequent protocol assessments, we present only two charts: A-L for singles and A-L for pairs. The remaining c2c and M-L charts are provided in Appendix~\ref{sec:first_assessment_stages}.

\subsubsection{Transmit}\label{sec:mod_brisbane_transmit}
As said, in this section we demonstrate the full assessment process. 

\begin{figure}[H]
    \centering

    \begin{subfigure}[t]{0.48\linewidth}
        \centering
        \includegraphics[width=\linewidth]{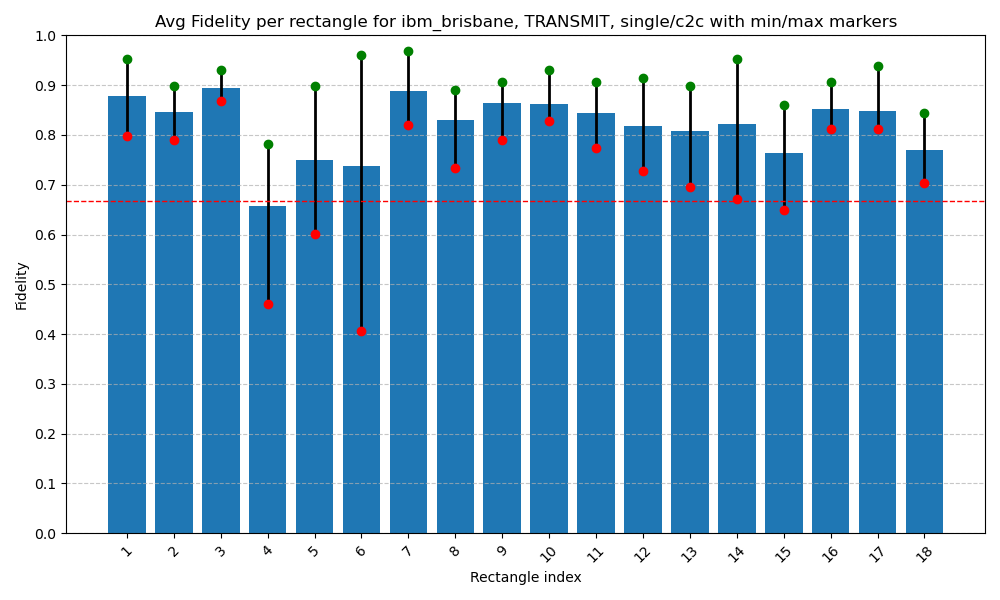}
    \caption{22 Aug 2025: Transmit, single rectangle, corner to  on all 18 rectangles of Brisbane}
    \label{fig:new-brisbane-Transmit-single-c2c}    
    \end{subfigure}
    \hfill
    \begin{subfigure}[t]{0.48\linewidth}
        \centering
        \includegraphics[width=\linewidth]{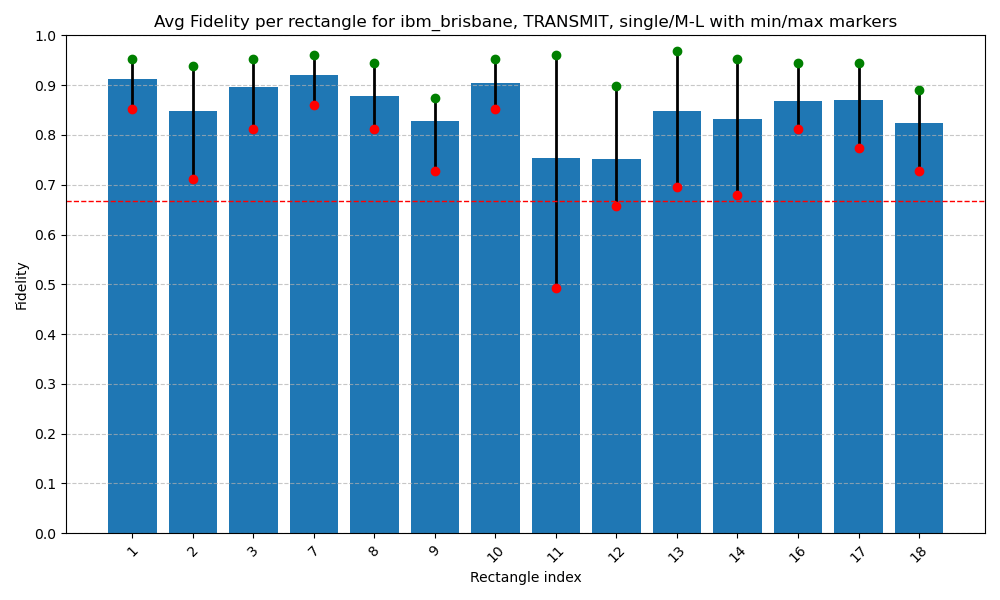}
        \caption{22 Aug 2025: Transmit single rectangle, all rectangles except 4, 5, 6, 15, that failed transmit c2c on 22 Aug 2025 (figure~\ref{fig:new-brisbane-Transmit-single-c2c}). Running max lengths experiment}
        \label{fig:new-brisbane-Transmit-single-M-L}  
    \end{subfigure}
    \caption{The results of c2c and M-L stages on Brisbane}
    \label{new-brisbane-Transmit-single-first-stages}
\end{figure}
On figure~\ref{new-brisbane-Transmit-single-first-stages} in the left sub-figure we see the c2c experiment conducted on all 18 rectangles. The threshold for this protocol is $\frac{2}{3}$, we consider a rectangle to pass the assessment stage if the min fidelity achieved over the tested paths is above this threshold. In the c2c stage the rectangles that passed are 1, 2, 3, 7, 8, 9, 10, 11, 12, 13, 14, 16, 17, 18. Only those rectangles will be tested in the next stage. In the right sub-figure~\ref{fig:new-brisbane-Transmit-single-M-L} we see the following stage in the full assessment, M-L, which only rectangles 11 and 12 failed to pass. We note that in this stage each qubit takes part in multiple paths with multiple roles: in some paths its the initialized qubit, in some the swapping intermediate ancilla qubit and in some the measured qubit. After the first two stages we proceed to the last stage of the full assessment - A-L.

\begin{figure}[H]
    \centering
    \includegraphics[width=0.7\linewidth]{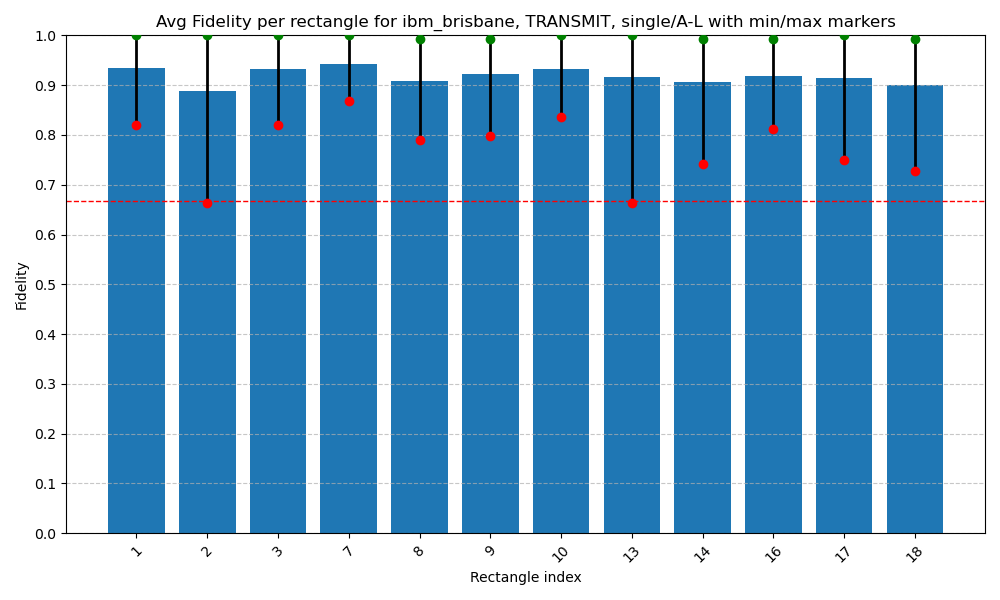}
    \caption{22 Aug 2025: Transmit single rectangle, A-L, selection is done according to M-L experiment in figure~\ref{fig:new-brisbane-Transmit-single-M-L} .}
    \label{fig:new-brisbane-22aug-Transmit-single-A-L}
\end{figure}

The A-L stage is the first heavy load we put on the sub-chips, with 144 different paths in each rectangle. Finally, we obtain the values for the protocol vector for the transmit protocol. We see that only 10 sub-chips managed to exceed the fidelity threshold on every possible inner path. Such a fundamental task, transmitting a state via swap gates, should work on every part of the 127 qubits quantum computer, but as can be seen in the results of the A-L stage in figure~\ref{fig:new-brisbane-22aug-Transmit-single-A-L}, some rectangles did not demonstrate quantumness in this task.
\begin{figure}[H]
    \centering
    \includegraphics[width=0.7\linewidth]{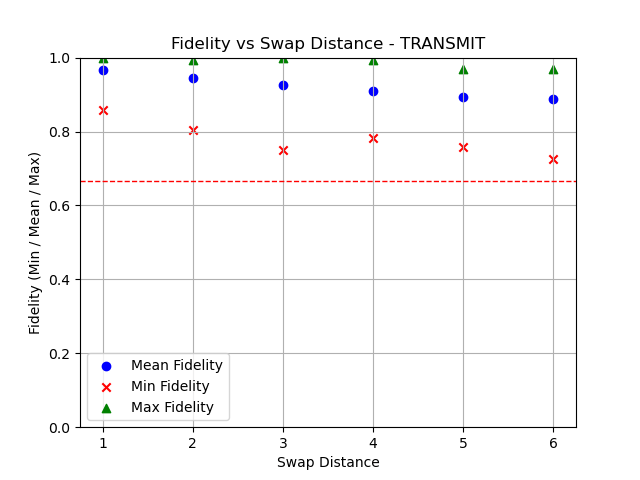}
    \caption{22 Aug 2025: Showing fidelity as function of the swap distance (the distance between Alice and Bob in qubits). Note that this chart contains only results of rectangles that passed transmit in A-L (fig \ref{fig:new-brisbane-22aug-Transmit-single-A-L})}
    \label{fig:brisbane_Aug22_transmit_single_swap_dist}
\end{figure}
In the fidelity as a function of the swap distance chart (figure~\ref{fig:brisbane_Aug22_transmit_single_swap_dist}) we aggregate only on rectangles that passed A-L in the A-L stage (figure \ref{fig:new-brisbane-22aug-Transmit-single-A-L}). The swap distance is defined to be the number of swap gates applied to move the work qubits state from Alice's camp to Bob's. Given a circuit of $n$ qubits the swap distances are calculated as follows:
\begin{table}[H]
    \centering
    \begin{tabular}{c|c}\toprule
        \textbf{Protocol} & \textbf{Swap Distance} \\\midrule
        do-nothing, transmit&  $n-1$ \\
         teleportation, bell-state transfer                 &   $n-3$ \\
         super-dense coding                           & $n-2$ \\
         entanglement swapping                 & $n-5$ \\ \bottomrule
    \end{tabular}
    \caption{The calculation formulas for swap distance on a circuit with $n$ qubits}
    \label{tab:swap_dist_formulas}
\end{table}
In figure \ref{fig:brisbane_Aug22_transmit_single_swap_dist} it is visible that while the 10 rectangles that passed A-L did it successfully for a single rectangle sub-chip, if we extend a linear regression line on the min markers it will soon go below the threshold. The fidelity as function of swap distance chart give the viewer an impression on how far can a protocol be extended and still pass the threshold.
Note that in appendix section \ref{sec:swap_dist_charts} we present the swap distance charts for all the rest of the protocols.
\begin{figure}[H]
    \centering
    \includegraphics[width=\linewidth]{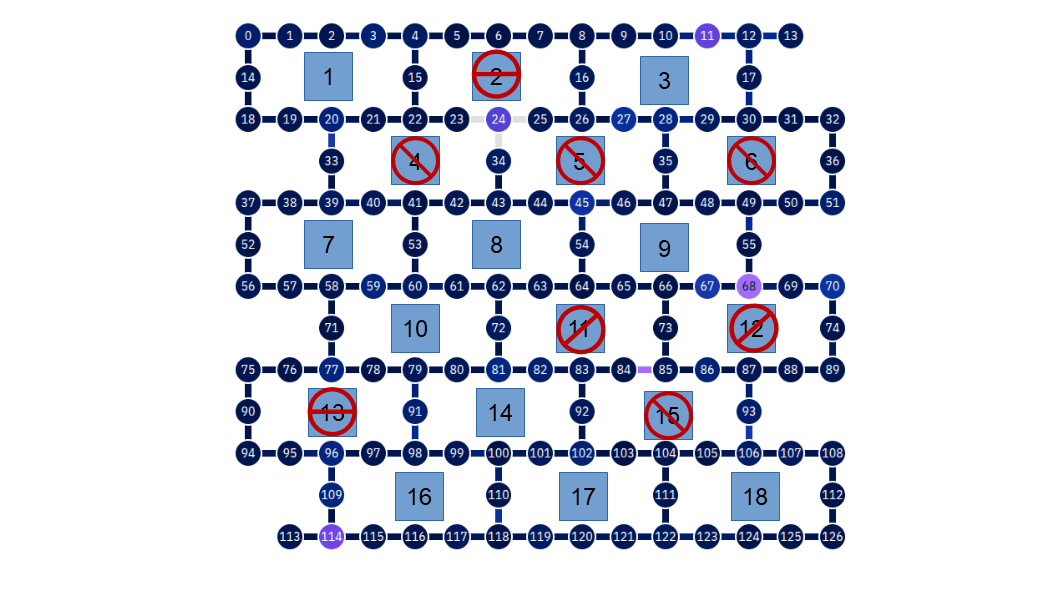}
    \caption{Illustration of the rectangles which passed/failed transmit full assessment for single rectangles. Failing c2c is marked by \rotatebox[origin = c]{-45}{$\ominus$}, M-L by \rotatebox[origin = c]{45}{$\ominus$} and  A-L by $\ominus$.}
    \label{fig:new-brisbane-transmit-full-assessment}
\end{figure}
 After we performed the full assessment of transmit on single rectangles we moved to check a bigger version of a sub-chip, rectangles pairs. The selection method is simple, we test rectangles $R_a$ and $R_b$ as a pair $\{R_a, R_b\}$ if the two following conditions apply:
 \begin{enumerate}
     \item $R_a$ and $R_b$ passed the A-L experiment
     \item $R_a$ and $R_b$ are neighbors on the qubit map of the tested chip
 \end{enumerate}
 For example we see that in figure \ref{fig:new-brisbane-22aug-Transmit-single-A-L} rectangles 7 and 8 passed the threshold, and we can see in the chip map (figure \ref{fig:eagle_map}) that they are indeed neighboring rectangles, so we test them as a pair. 
\begin{figure}[H]
    \centering
    \includegraphics[width=0.7\linewidth]{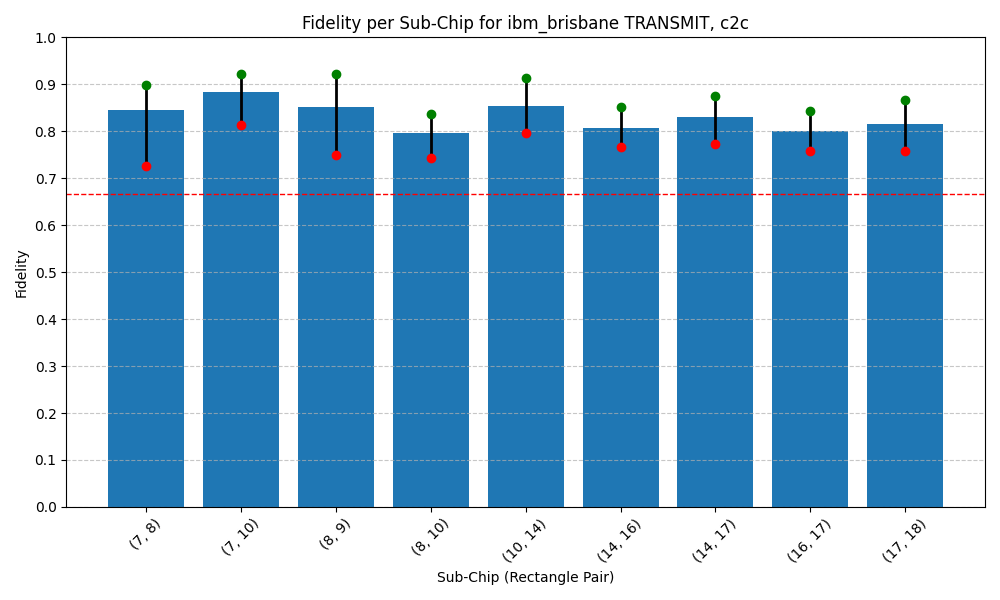} 
    \caption{22 Aug 2025: transmit rectangles pairs, corner to corner, neighboring rectangles that passed transmit A-L on 22 Aug 2025 (figure~\ref{fig:new-brisbane-22aug-Transmit-single-A-L})}
    \label{fig:new-brisbane-Transmit-pairs-c2c}
\end{figure}
Performing the c2c experiment on pairs of rectangles (figure~\ref{fig:new-brisbane-Transmit-pairs-c2c}) we see that none of the pairs failed this experiment. Note that as discussed earlier in section \ref{sec:optimal_lookup_workflow} the number of paths tested in this experiment is dependent on the shape of the sub-chip. Some pairs like \{7, 8\} are side-by-side rectangles and others like \{7,10\} are connected in a diagonal manner. 
\begin{figure}[H]
    \centering
    \includegraphics[width=0.7\linewidth]{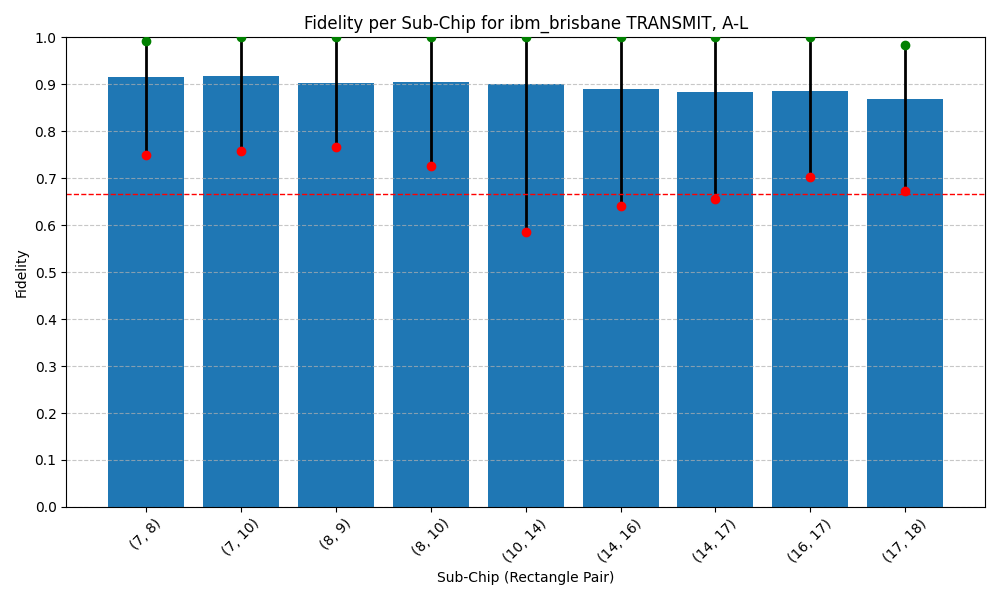} 
    \caption{22 Aug 2025: Transmit all lengths for rectangles pairs that passed transmit c2c on 22 Aug 2025 (figure~\ref{fig:new-brisbane-Transmit-pairs-c2c})}
    \label{fig:new-brisbane-Transmit-pairs-A-L}
\end{figure}
Because in pairs experiments we do not run the M-L experiment (as explained in detail on section~\ref{sec:optimal_lookup_workflow}), next is the A-L stage with pairs. This test checks every minimal inner path between 2 qubits in the pair. We consider this stage as a heavy duty test of the chip capabilities. On this experiment (figure~\ref{fig:new-brisbane-Transmit-pairs-A-L}) only 3 sub-chips failed to hold the threshold, those results can be used to proceed into a triple rectangles sub-chip experiment in the future. 

This is the end of ``Full Assessment" for transmit protocol on singles and pairs, the results are visualized in the protocol vector in section \ref{sec:brisbane_protocol_vector}. From now on for each protocol we only present the A-L charts of singles and pairs, one can view the rest of the assessment charts (c2c and M-L) in appendix section \ref{sec:first_assessment_stages}.

\subsubsection{Do-nothing}

\begin{figure}[H]
    \centering

    \begin{subfigure}[t]{0.48\linewidth}
        \centering
        \includegraphics[width=\linewidth]{figures/new_brisbane_4_comparing_2_old/single/A-L/DO_NOTHING/per_rect_plot_DO_NOTHING_single_A-L.png}
        \caption{28 Sep 2025: Do nothing protocol, all lengths, on all rectangles that passed max length on 28 Sep 2025 (figure~\ref{fig:brisbane_Sep28_do_nothing_single_M-L}.)}
    \label{fig:brisbane_Sep28_do_nothing_single_A-L}
    \end{subfigure}
    \hfill
    \begin{subfigure}[t]{0.48\linewidth}
        \centering
        \includegraphics[width=\linewidth]{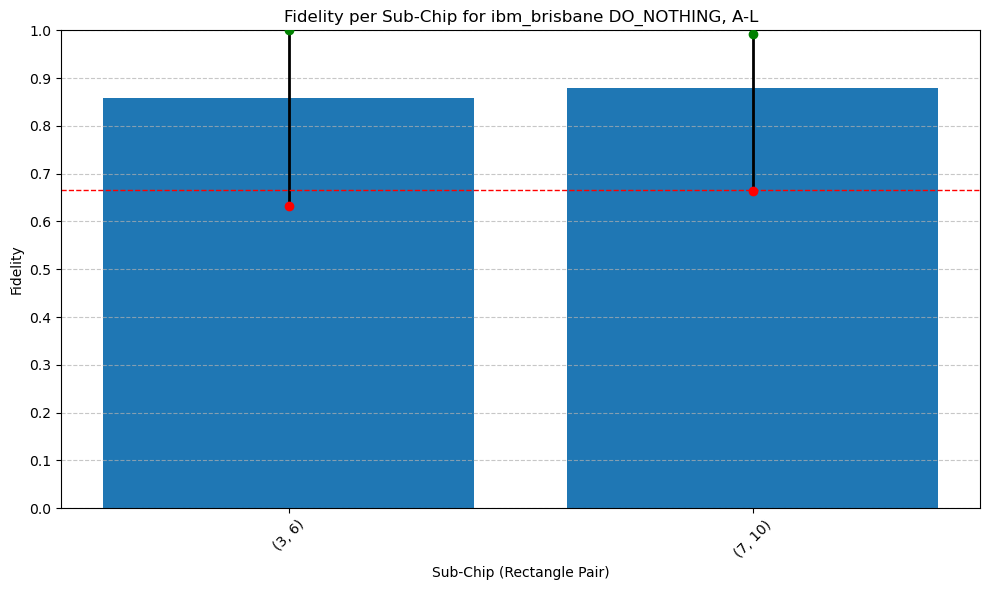}
        \caption{29 Sep 2025: Do nothing protocol on pairs of rectangles, all length experiment, on all pairs that passed c2c on 29 Sep 2025 (figure~\ref{fig:brisbane_Sep29_do_nothing_pair_c2c})}
        \label{fig:brisbane_Sep29_do_nothing_pair_A-L}
    \end{subfigure}

    \begin{subfigure}{0.5\linewidth}
        \centering
    \includegraphics[width=\linewidth]{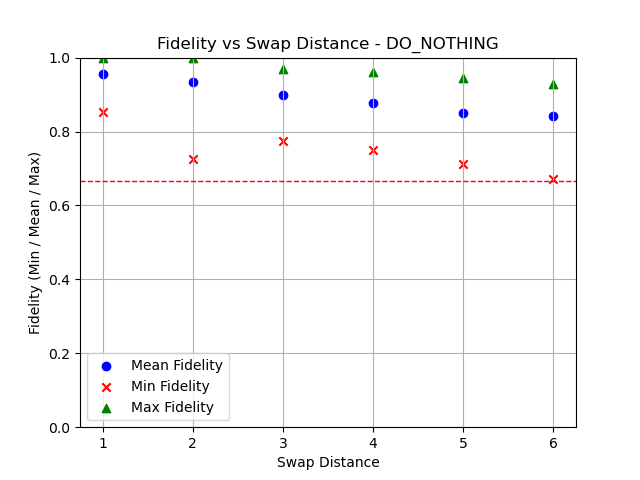}
    \caption{28 Sep 2025: Showing fidelity as a function of the swap distance. Note that this chart contains only results of rectangles that passed do nothing in A-L (fig \ref{fig:brisbane_Sep28_do_nothing_single_A-L})}
    \label{fig:brisbane_Sep28_do_nothing_single_swap_dist}
    \end{subfigure}

    \caption{Results for do nothing on Brisbane for singles and pairs sub-chips} 
\end{figure}
Looking at figure \ref{fig:brisbane_Sep28_do_nothing_single_A-L} we see that modified Brisbane has 6 rectangles that can be considered do-nothing-capable. While this result looks somewhat disappointing, old Brisbane produced only 2 rectangles in the same test, so the improvement is noticeable. 
Figure~\ref{fig:brisbane_Sep29_do_nothing_pair_A-L} is the results for A-L on pairs. We see that pair \{3,6\} failed and \{7,10\} result is $0.664 \pm 0.0001$, that is below the $\frac{2}{3}$ threshold. \textbf{Thus we conclude that none of the possible pairs on modified Brisbane managed to pass the do nothing assessment}.

\subsubsection{Teleportation}

\begin{figure}[H]
    \centering
    \includegraphics[width=0.7\linewidth]{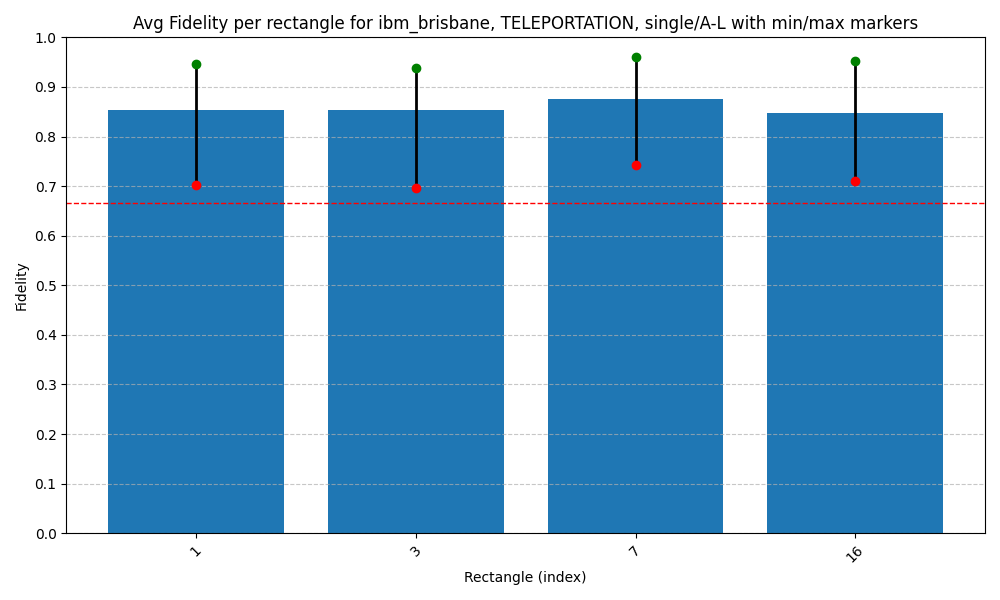} 
    \caption{8 Oct 2025: teleportation protocol, all lengths on all the rectangles that passed the max lengths test of teleportation on 8 Oct 2025 (figure~\ref{fig:brisbane_Oct8_teleportation_single_M-L})}
    \label{fig:brisbane_Oct8_teleportation_single_A-L}
\end{figure}
The teleportation full assessment yielded 4 teleportation-capable rectangles in modified Brisbane. Regarding the teleportation results of modified Brisbane versus old Brisbane and Sherbrooke, we saw that the old Eagle r3 chips performed better then the modified version. This can be seen in the appendix figure \ref{fig:brisbane_vs_sherbrooke_teleportaion}, in the results of old Brisbane in addition to the rectangles that passed in the modified version, rectangle 10 passed as well.
There is no possible pair to check as those rectangles aren't neighbors on the Eagle-r3 chip.

\subsubsection{Bell-state transfer}

\begin{figure}[H]
    \centering
    \includegraphics[width=0.7\linewidth]{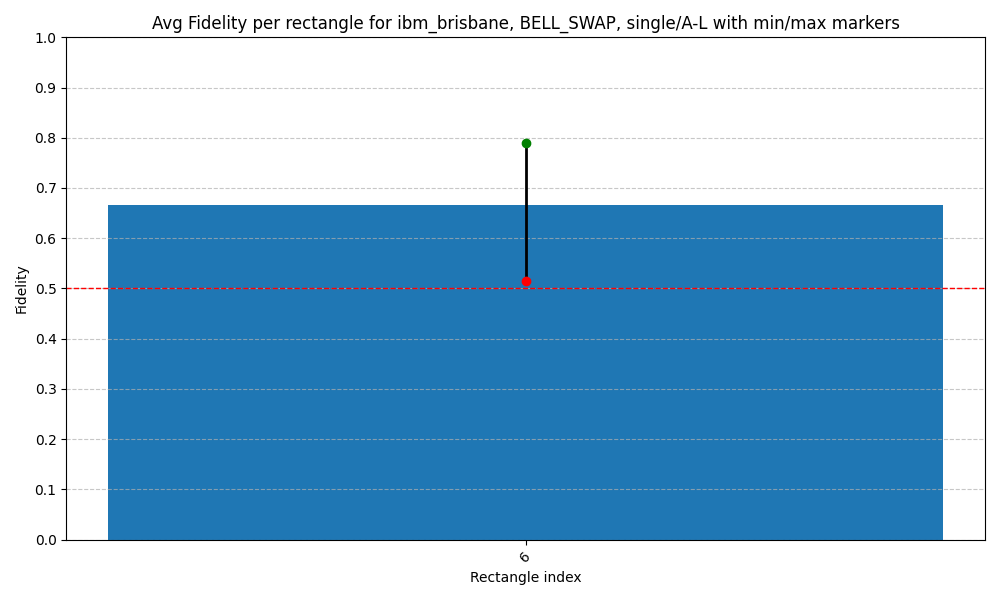}
    \caption{28 Sep 2025: bell-state transfer protocol, all lengths on all the rectangles that passed the max length test of bell-state transfer on 28 Sep 2025 (figure~\ref{fig:brisbane_Sep28_Bell_swap_single_M-L})}
    \label{fig:brisbane_Sep28_Bell_swap_single_A-L}
\end{figure}
In the Bell-state transfer (figure~\ref{fig:brisbane_Sep28_Bell_swap_single_A-L}) we see an improvement in Brisbane, as in the modified version rectangle 6 managed to pass the full assessment unlike the old version where no rectangle passed this assessment.
Regretfully, no pair experiment can be done if only 1 rectangle passed the single rectangle assessment.

\subsubsection{Entanglement swapping}
\begin{figure}[H]
    \centering
    \includegraphics[width=0.7\linewidth]{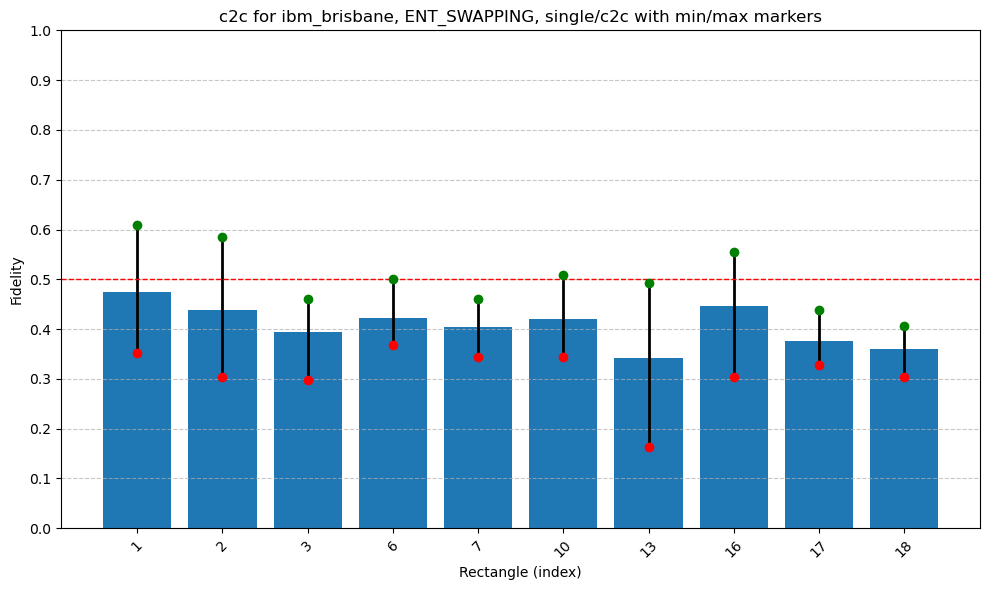} 
    \caption{28 Sep 2025: entanglement swapping protocol, corner to corner on all the rectangles that passed the corner to corner test of do nothing on 28 Sep 2025 (figure~\ref{fig:brisbane_Sep28_do_nothing_single_c2c}).}
    \label{fig:brisbane_28Sep_ent_swap_single_c2c}
\end{figure}

\textbf{Important:} this assessment contains an experimental mistake: according to the workflow that was presented, in addition to the rectangles that appear in figure~\ref{fig:brisbane_28Sep_ent_swap_single_c2c}, rectangle 14 also passed do nothing c2c, it is supposed to be tested on the entanglement swapping c2c stage as well. An experimental error occurred and it wasn't tested, unfortunately IBM took Brisbane down before we managed to fix this error.
Note that the presented chart for entanglement swapping (figure~\ref{fig:brisbane_28Sep_ent_swap_single_c2c}) is not A-L stage but is the c2c stage. This is because no rectangles managed to pass the simplest stage of this full assessment. The chart is presented to illustrate the poor results we got for modified Brisbane in the entanglement swapping task.

\subsubsection{Super-dense coding}\ \\
\textbf{Important:} as explained in the entanglement swapping section above, the results here are missing rectangle 14 due to a similar experimental error as the one mentioned in the previous subsection. 
\begin{figure}[H]
    \centering
    \includegraphics[width=0.7\linewidth]{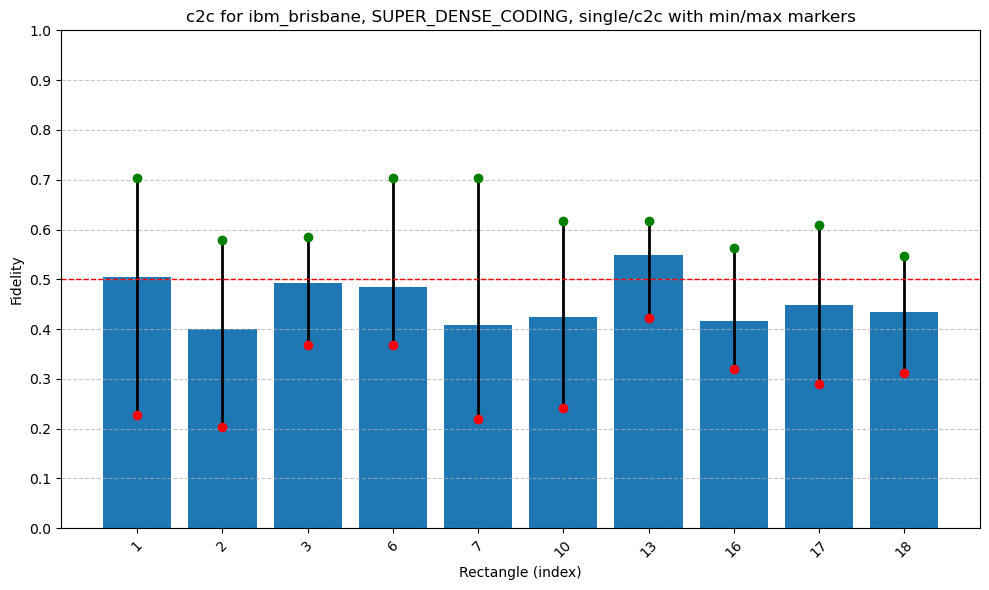}
    \caption{28 Sep 2025: super-dense coding protocol, corner to corner on all the rectangles that passed the corner to corner test of do nothing.}
    \label{fig:brisbane_28Sep_superdense_single_c2c}
\end{figure}
We see a similar result to the entanglement swapping experiment, none of the rectangles in modified Brisbane managed to pass the threshold in the c2c experiment, so we didn't proceed to the next assessment stages.

\subsection{Protocol vector}\label{sec:brisbane_protocol_vector}
In this section, we present an illustration of the protocol vector. This visualization of the protocol vector allow the viewer to have an intuitive understanding of the strengths and weaknesses of the chip, helping prospective IBM users find a suitable sub-chip for their computation.

\begin{figure}[H]
    \centering
    \begin{subfigure}{\linewidth}
        \centering
        \includegraphics[width=0.9\linewidth]{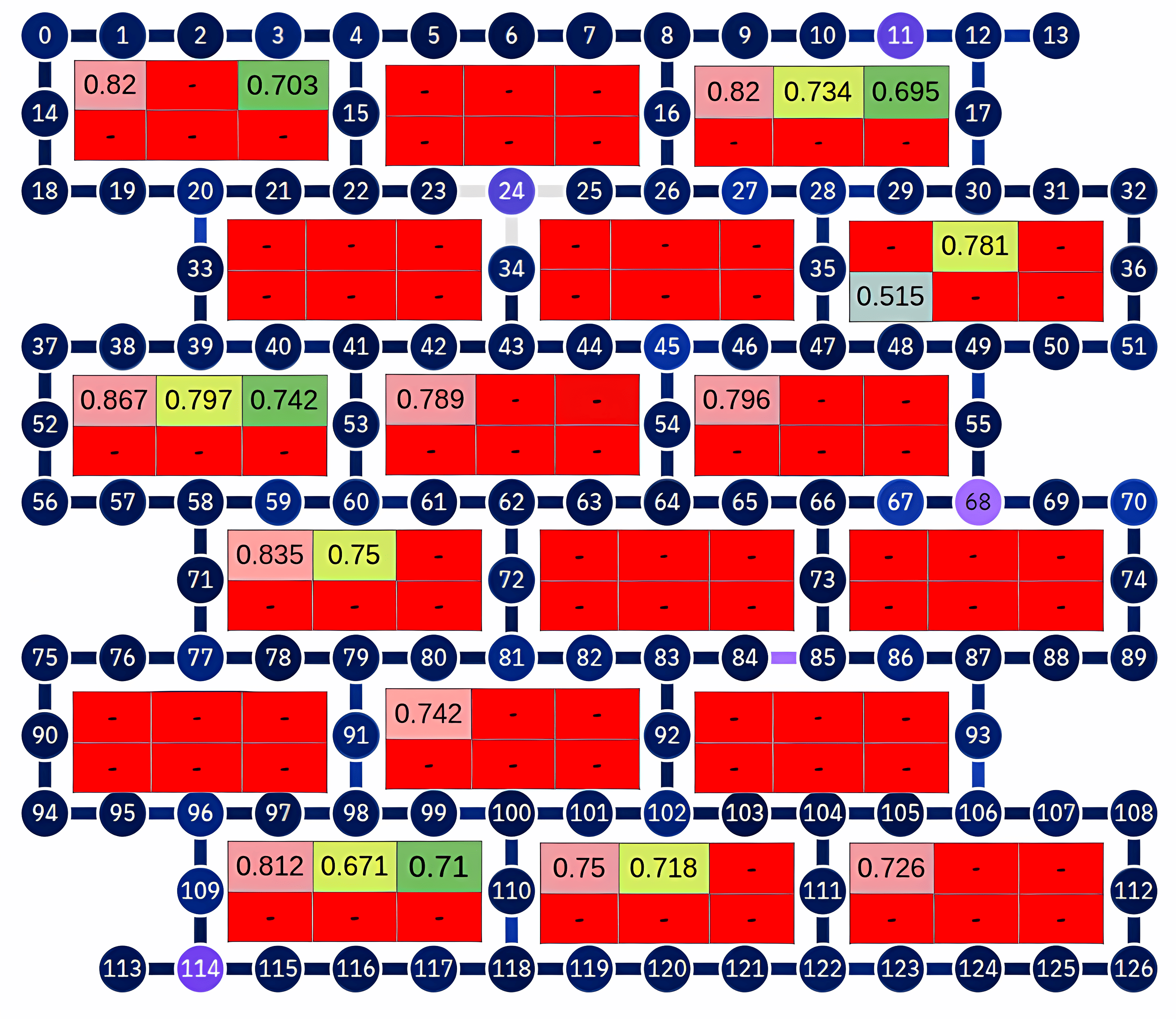}
    \end{subfigure}

    \begin{subfigure}{\linewidth}
        \centering
        \includegraphics[width=0.75\linewidth]{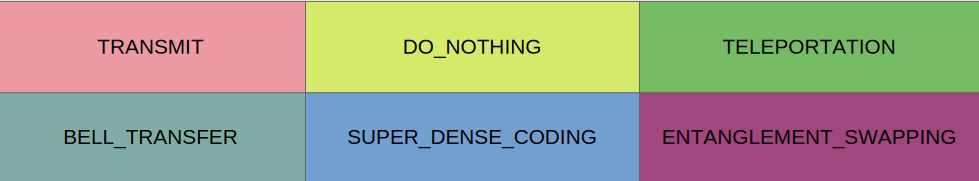}  
    \end{subfigure}

    \caption{Final protocol vector for Brisbane displayed on top of the qubits map. Rectangles index can be seen in fig \ref{fig:eagle_map}. The numbers presented are the minimal fidelity achieved over all of the paths within the rectangle}
    \label{fig:protocl_vector_new_brisbane}
\end{figure}

For modified Brisbane we see a limited performance profile of the quantumness capabilities for each protocol. Most of the rectangles didn't pass our assessments successfully, it's apparent that the only protocol that managed to produce somewhat usable performance is transmit in about half of the rectangles. While the transmit results give an optimistic image of the chip quality, we can't produce a single sub-chip larger than one rectangle that can pass the fidelity threshold for more then three protocols. The three possible pairs of do nothing failed the pairs assessment and the rest of the protocols didn't even have any pair that can be tested. A summary table on the number of successful pairs for each protocol on Brisbane is presented in table~\ref{tab:pairs_count_brisbane}. In the transmit protocol there are a couple of overlapping pairs that passed A-L as pairs, hence the assessment can be further extended into checking the overlapping pairs as triplets.

\begin{table}
    \centering
    \begin{tabular}{c|c} \toprule
        \textbf{Protocol} & \textbf{Count of Successful Pairs} \\ \midrule
         Transmit &  6\\
         Do-nothing & 0\\
         Teleportation & 0\\
         Bell-state transfer & 0\\
         Super-dense coding  & 0\\
         Entanglement swapping & 0 \\ \bottomrule
    \end{tabular}
    \caption{Count of successful pairs sub-chips for each protocol on Brisbane}
    \label{tab:pairs_count_brisbane}
\end{table}

\section{IBM's Heron - Kingston}\label{sec:Kingston_results_section}
\subsection{Introduction}
\begin{figure}[H]
    \centering
    \includegraphics[width=0.7\linewidth]{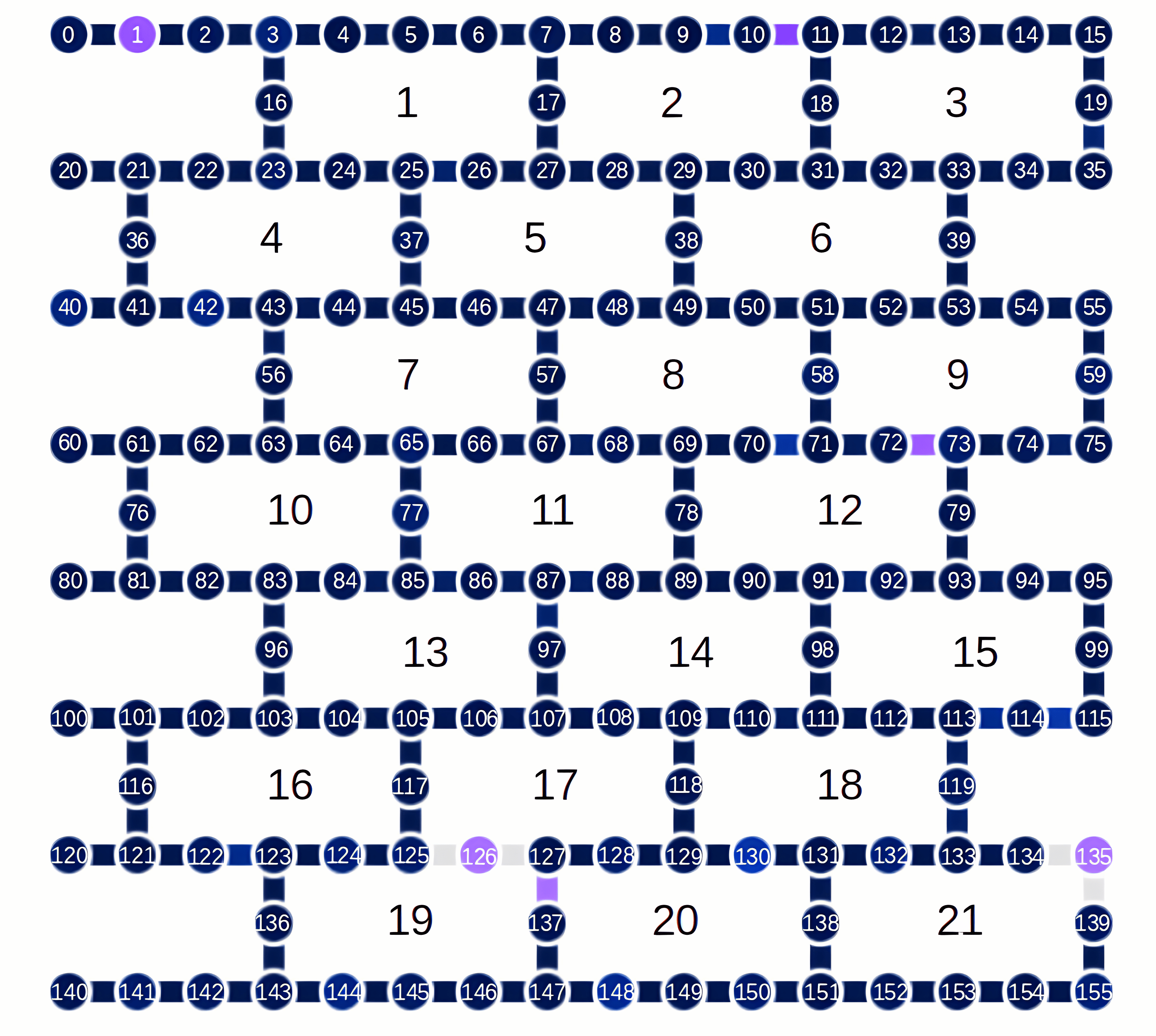}
    \caption{Heron-r2 qubits map and rectangles indexes}
    \label{fig:heron_map}
\end{figure}

Now we present the results of optimal lookup for IBM's Heron-r2 quantum computer. This chip is composed of 156 qubits, arranged in rectangles of 12 qubits each (as seen in fig \ref{fig:heron_map}). We assigned a number for each rectangle, the results presented below show fidelity as function of the rectangle index.

\subsection{Results for Singles and Pairs Sub-Chips}\label{sec:kinsgton_results}

\subsubsection{Transmit}\label{sec:kinsgton_results_transmit}

\begin{figure}[H]
    \centering

    \begin{subfigure}[t]{0.48\linewidth}
        \centering
    \includegraphics[width=\linewidth]{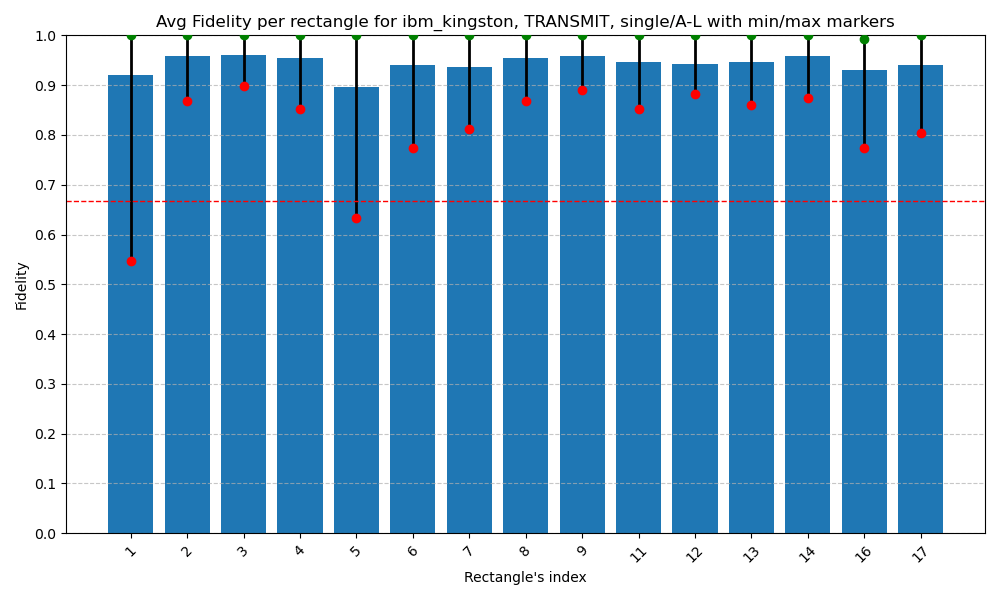}
    \caption{28 Jun 2025: Transmit protocol, all-lengths, on all rectangles but 10,15,18,19,20,21. 144 paths per rectangle. Participating rectangles who passed transmit M-L on 11 Jun 2025 (figure~\ref{fig:kingston_transmit_M-L}).}
    \label{fig:kingston_transmit_A-L}
    \end{subfigure}
    \hfill
    \begin{subfigure}[t]{0.48\linewidth}
        \centering
    \includegraphics[width=\linewidth]{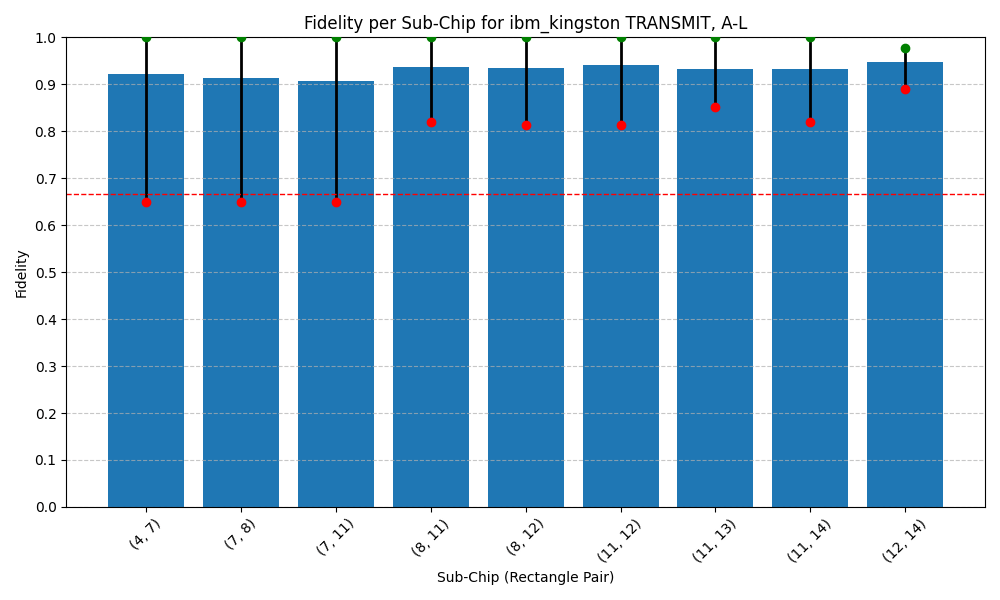}
    \caption{23 Oct 2025: Transmit protocol, all lengths on neighboring pairs of the rectangles that passed transmit all lengths from 12 Oct 2025 (figure~\ref{fig:kingston_transmit_c2c_pairs}) }
    \label{fig:kingston_transmit_A-L_pairs}
    \end{subfigure}

    \caption{Results for transmit on Kingston, results for single and pairs of rectangles of the A-L stage. The corresponding charts of the c2c and M-L stages can be found in appendix section~\ref{sec:first_assessment_stages}} 
\end{figure}
As can be seen, we note a significant difference in the performance of Brisbane and Kingston. The Heron chip aligns more closely with the theoretical expectation that a fundamental operation, such as the Transmit protocol, should be reliably executable across the processor. Indeed in the single rectangles assessment we have 13 rectangles that passed the A-L test successfully. \\
\textbf{Data Integrity Note: } During the execution of the transmit A-L on pairs experiment shown in figure~\ref{fig:kingston_transmit_A-L_pairs}, an undetermined system error caused the cancellation of numerous paths. Consequentially, figure~\ref{fig:kingston_transmit_A-L_pairs} excludes data for five specific pairs: \{13, 14\}, \{13, 16\}, \{13, 17\}, \{14, 17\} and \{16, 17\}. Furthermore, the data for the visualized pairs is partial; specifically, several ``inner paths" - defined as paths between two qubits that are completely contained within the designated sub-chip - are missing from the analysis. Overall, Figure~\ref{fig:kingston_transmit_A-L_pairs} depicts the aggregation of the recoverable results, representing approximately one-third of the complete experimental dataset.
\begin{figure}[H]
    \centering
    \includegraphics[width=0.7\linewidth]{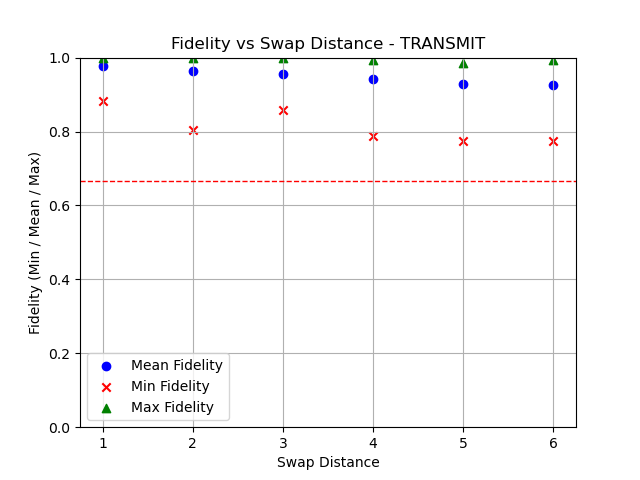}
    \caption{28 Jun 2025: Showing fidelity as function of the swap distance (the distance between Alice and Bob in qubits). Note that this chart contains only results of rectangles that passed transmit in A-L (fig \ref{fig:kingston_transmit_A-L})}
    \label{fig:kingston_transmit_single_A-L_old_workflow_swap_dist}
\end{figure}
In the swap distance chart we see that Kingston remains somewhat stable when the swap distance is increasing. 
\begin{figure}[H]
    \centering
    \includegraphics[width=\linewidth]{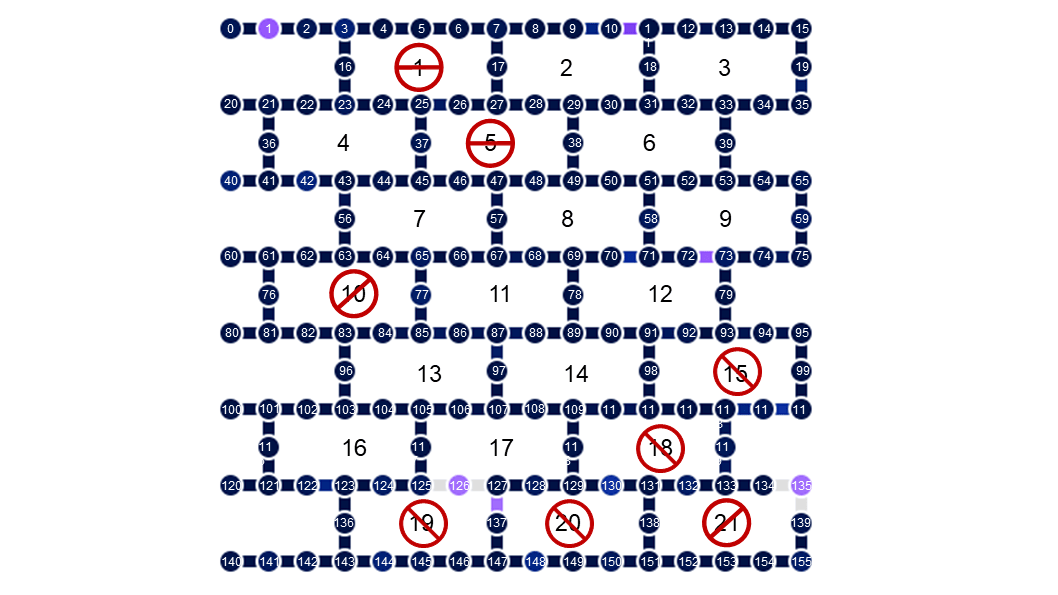}
    \caption{Illustration of the rectangles that passed/failed transmit full assessment for single rectangles. Failing c2c is marked by \rotatebox[origin = c]{-45}{$\ominus$}, M-L by \rotatebox[origin = c]{45}{$\ominus$} and  A-L by $\ominus$.}
    \label{fig:kingston-transmit-full-assessment}
\end{figure}

\subsubsection{Do nothing}

\begin{figure}[H]
    \centering

    \begin{subfigure}[t]{0.48\linewidth}
        \centering
        \includegraphics[width=\linewidth]{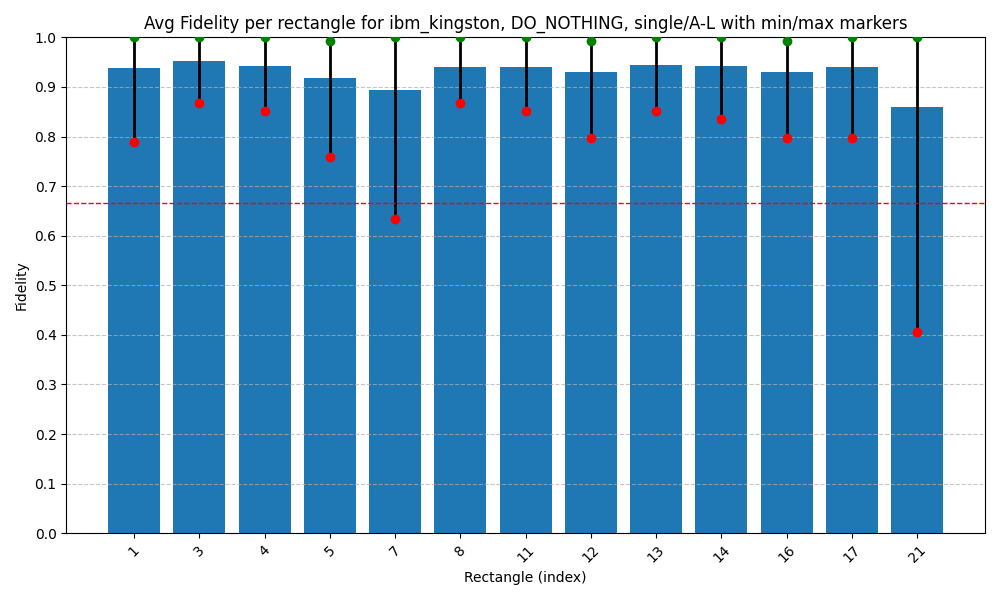} 
        \caption{9 Oct 2025: Do nothing protocol, all lengths on rectangles that passed do-nothing maximal length (figure~\ref{fig:kingston_do_nothing_single_M-L_old_workflow})}
        \label{fig:kingston_do_nothing_single_A-L_old_workflow}
        
    \end{subfigure}
    \hfill
    \begin{subfigure}[t]{0.48\linewidth}
        \centering
        \includegraphics[width=\linewidth]{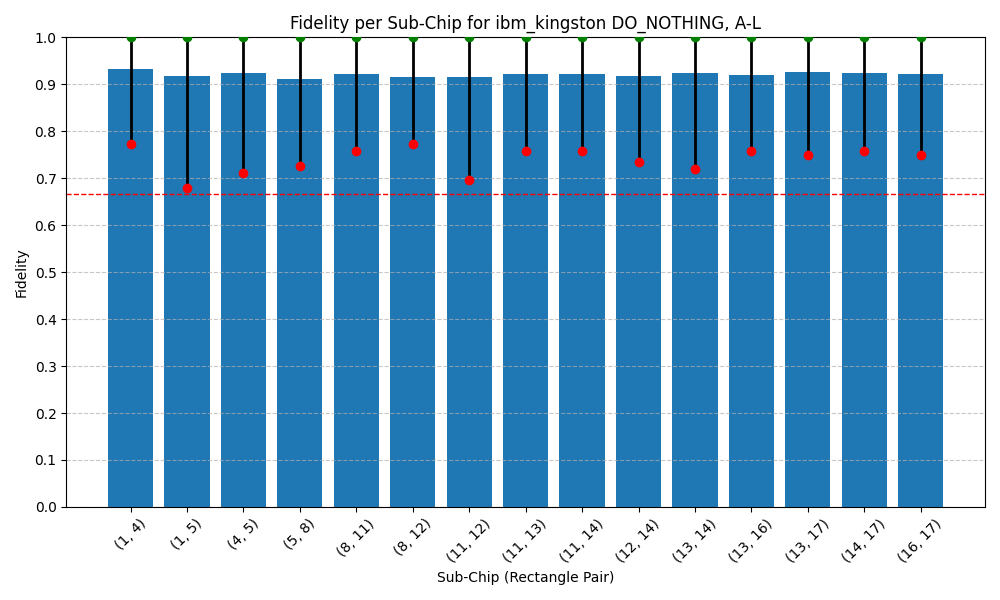}
        \caption{23 Oct 2025: do-nothing protocol, all lengths on pairs that passed do-nothing c2c (figure~\ref{fig:kingston_do_nothing_pair_c2c_old_workflow}).}
        \label{fig:kingston_do_nothing_pair_A-L_old_workflow}
    \end{subfigure}

    \caption{Do-nothing protocol on Kingston, results for single and pairs of rectangles of the A-L stage. The corresponding charts of the c2c and M-L stages can be found in appendix section~\ref{sec:first_assessment_stages}} 
\end{figure}
As can be seen, the chip produces good results with many rectangles that pass the threshold for a successful do-nothing. The most significant performance improvements are evident in the right panel, where all possible pairs exceeded the quantumness threshold, maintaining this level of performance even after the sub-chip size was doubled. This is a good demonstration of scalability from the Kingston chip.
\begin{figure}[H]
    \centering
    \includegraphics[width=0.7\linewidth]{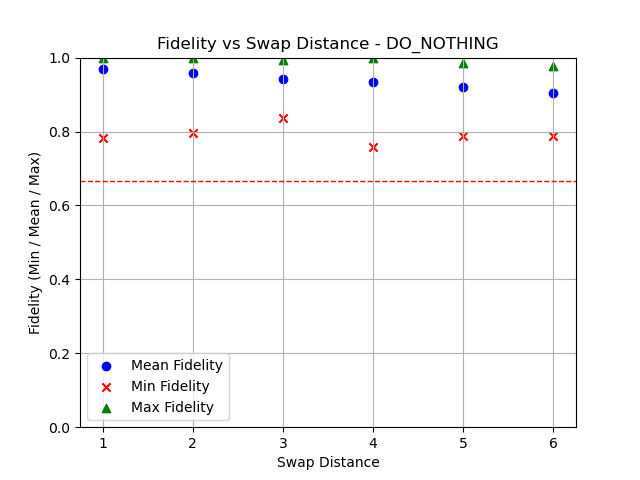}
    \caption{9 Oct 2025: Showing fidelity as function of the swap distance (the distance between Alice and Bob in qubits). Note that this chart contains only results of rectangles that passed do-nothing in A-L (fig \ref{fig:kingston_do_nothing_single_A-L_old_workflow})}
    \label{fig:kingston_do_nothing_single_A-L_old_workflow_swap_dist}
\end{figure}
Looking at the swap distance chart there is an interesting effect where in the larger distances we see a plateau in the minimum fidelity at those distances. This may explain the great results we got on the pairs assessment.

\subsubsection{Teleportation}

\begin{figure}[H]
    \centering

    \begin{subfigure}[t]{0.48\linewidth}
        \centering
        \includegraphics[width=\linewidth]{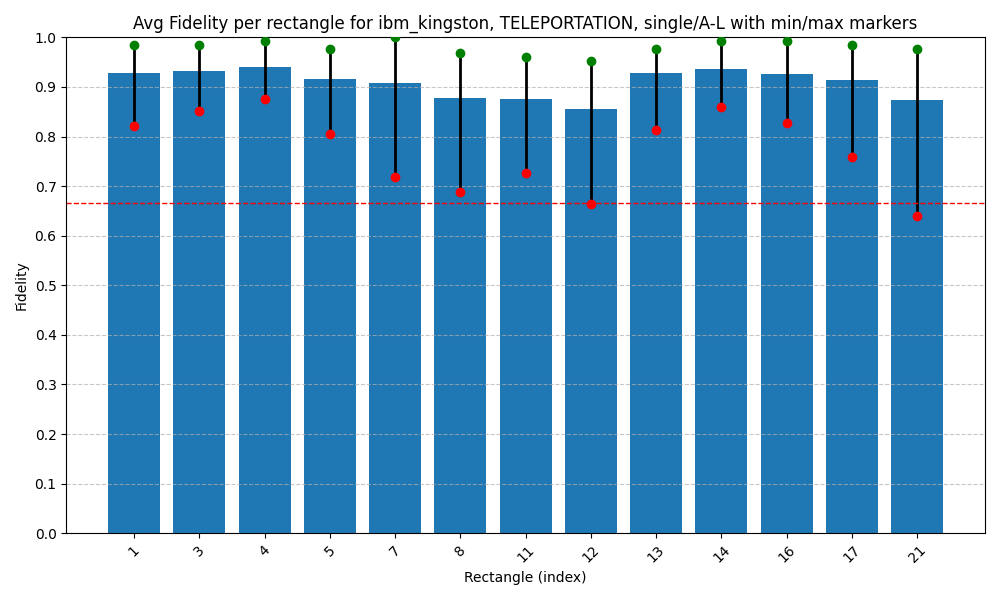}
        \caption{9 Oct 2025: Teleportation protocol, all lengths on rectangles that passed teleportation maximal length (figure~\ref{fig:kingston_teleportation_single_M-L_old_workflow})}
        \label{fig:kingston_teleportation_single_A-L_old_workflow}
    \end{subfigure}
    \hfill
    \begin{subfigure}[t]{0.48\linewidth}
        \centering
        \includegraphics[width=\linewidth]{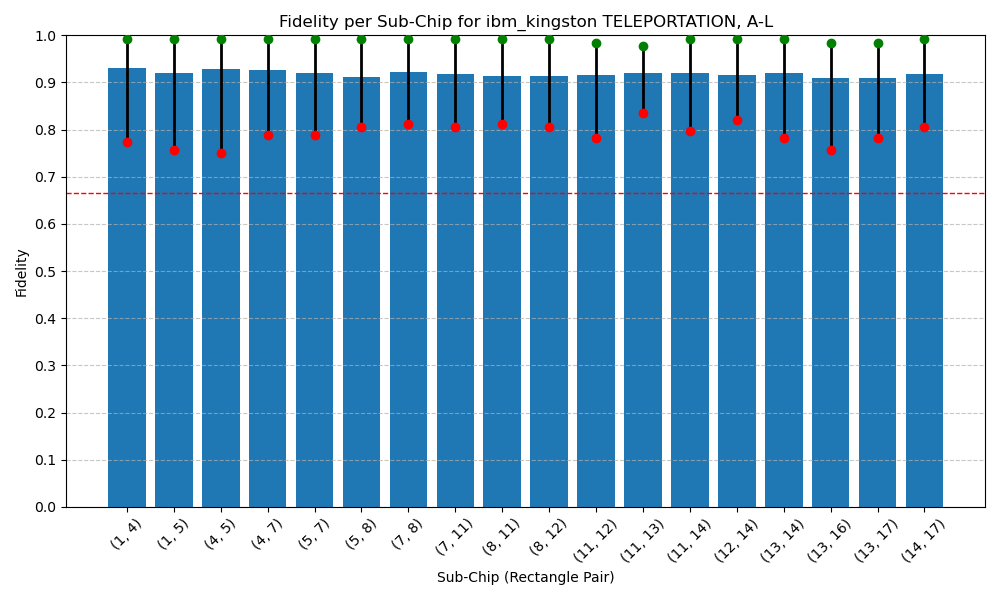}
        \caption{19 Nov 2025: Teleportation protocol, all lengths on rectangle pairs that passed teleportation c2c (figure~\ref{fig:kingston_teleportation_pair_c2c_old_workflow}). In the original experiment some paths were canceled by IBM for an unknown reason to us. We re-ran the failed paths again on 4'th of December and merged the results}
        \label{fig:kingston_teleportation_pair_A-L_old_workflow}
    \end{subfigure}

    \caption{Results for teleportation on Kingston, results for single and pairs of rectangles of the A-L stage. The corresponding charts of the c2c and M-L stages can be found in appendix section~\ref{sec:first_assessment_stages}} 
    \label{fig:kingston_teleportation_A-L_single_and_pairs}
\end{figure}
In the teleportation singles and pairs assessments we see that Kingston remains stable under a more complex protocol. As can be seen in figure~\ref{fig:kingston_teleportation_A-L_single_and_pairs} there are 12 teleportation-capable rectangles and 19 teleportation-capable rectangle pairs, which managed to successfully teleport a state between every two qubits in the sub-chip (in both ways). 

\subsubsection{Bell-state transfer}

\begin{figure}[H]
    \centering

    \begin{subfigure}[t]{0.48\linewidth}
        \centering
        \includegraphics[width=\linewidth]{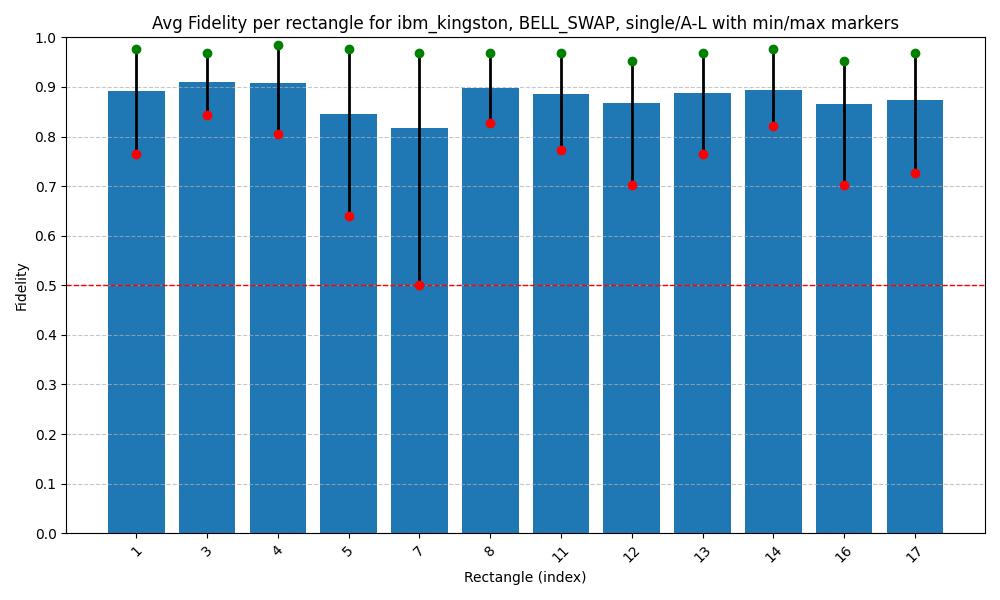}
        \caption{9 Oct 2025: bell-state transfer protocol all lengths on rectangles that passed bell-state transfer maximal lengths (figure~\ref{fig:kingston_bell_swap_single_M-L_old_workflow})}
        \label{fig:kingston_bell_swap_single_A-L_old_workflow}
    \end{subfigure}
    \hfill
    \begin{subfigure}[t]{0.48\linewidth}
        \centering
        \includegraphics[width=\linewidth]{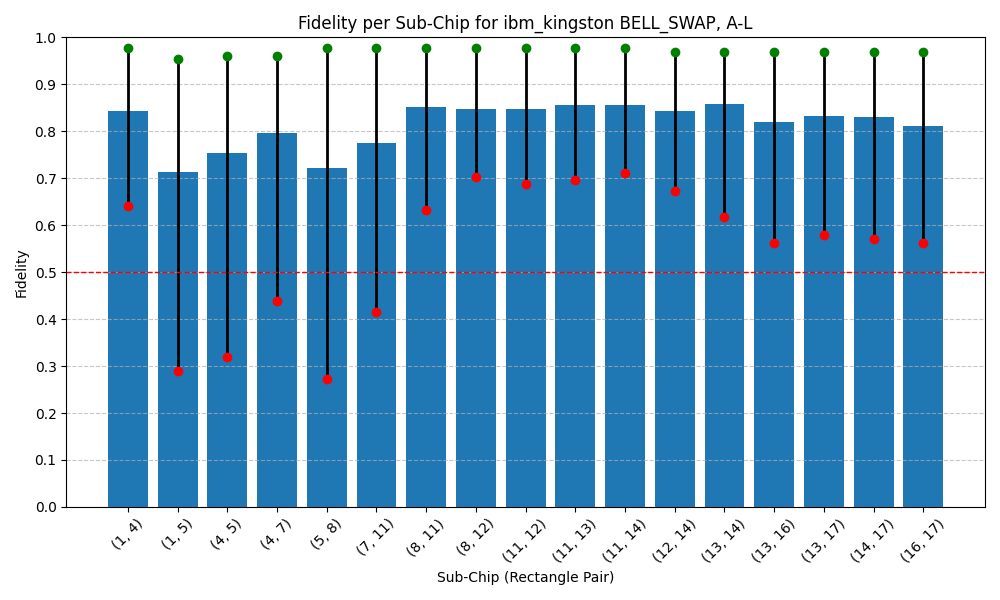}
        \caption{28 Oct 2025: bell-state transfer protocol, A-L on pairs that passed bell-state transfer c2c on pairs (figure~\ref{fig:kingston_bell_swap_pair_c2c_old_workflow})}
    \label{fig:kingston_bell_swap_pair_A-L_old_workflow}
    \end{subfigure}

    \caption{Results for bell-state transfer on Kingston, results for single and pairs of rectangles of the A-L stage. The corresponding charts of the c2c and M-L stages can be found in appendix section~\ref{sec:first_assessment_stages}} 
\end{figure}
The bell-state transfer protocol who failed badly on Brisbane have much better results on the Heron chip, with 12 successful sub-chips of single rectangles (the fidelity of rectangle 7 is 0.5 so is considered above the threshold). 
In the pairs assessment we see an interesting phenomenon where the variance of the pairs that passed is larger than in other experiments. Like in teleportation we see that many rectangle pairs managed to transmit a bell pair from every two neighboring qubits in the circuit to every other two neighboring pair. These results further illustrate the substantial performance differences between the Heron and Eagle architectures. 

\subsubsection{Entanglement swapping}

\begin{figure}[H]
    \centering

    \begin{subfigure}[t]{0.48\linewidth}
        \centering
        \includegraphics[width=\linewidth]{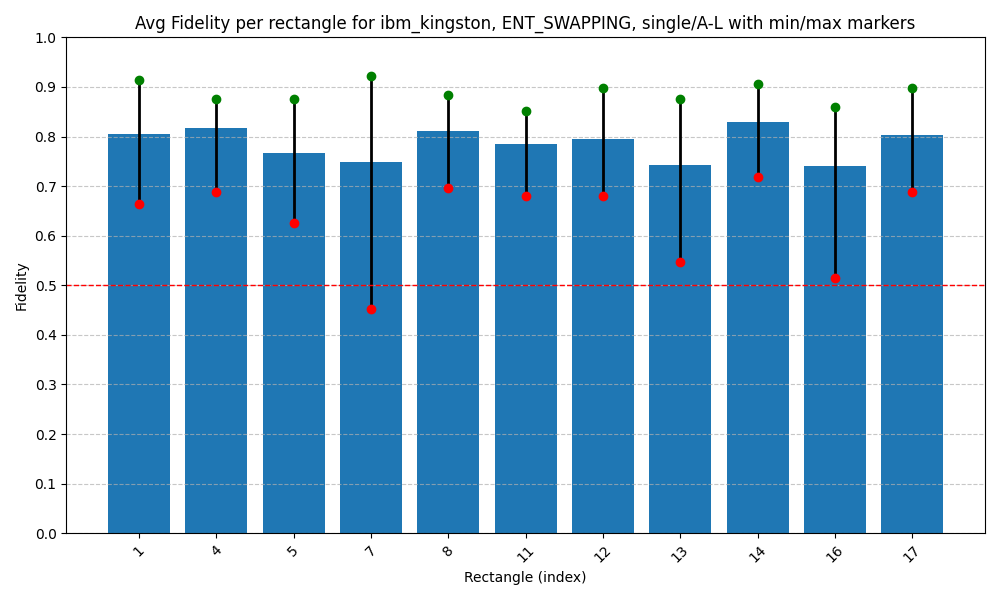}
        \caption{9 Oct 2025: entanglement swapping  protocol all lengths on rectangles that passed entanglement swapping maximal lengths (figure~\ref{fig:kingston_ent_swap_single_M-L_old_workflow})}
        \label{fig:kingston_ent_swap_single_A-L_old_workflow}
    \end{subfigure}
    \hfill
    \begin{subfigure}[t]{0.48\linewidth}
        \centering
        \includegraphics[width=\linewidth]{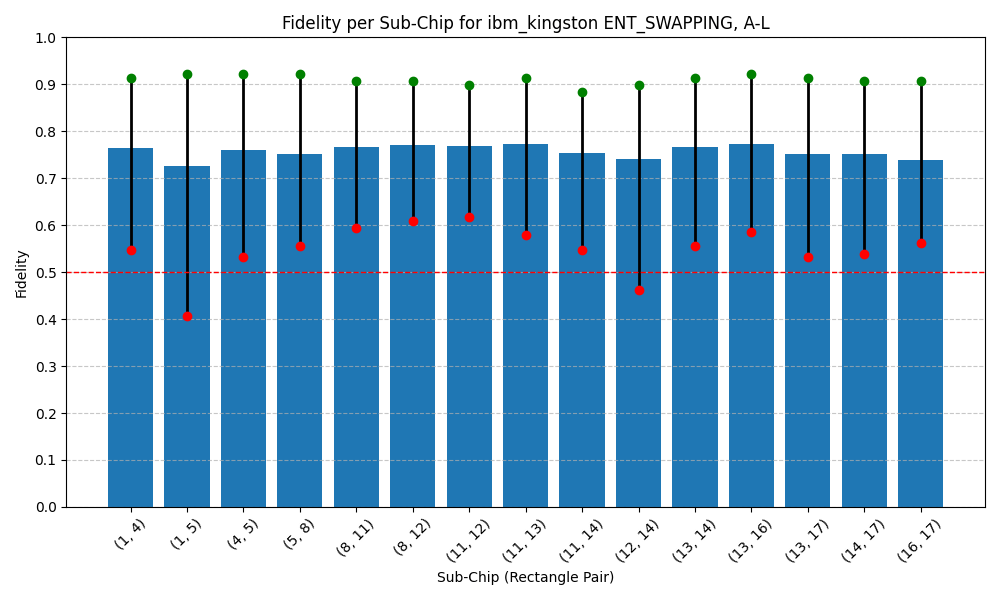}
        \caption{30 Oct 2025: entanglement swapping protocol, A-L on neighboring pairs that passed entanglement swapping c2c on rectangle pairs (figure~\ref{fig:kingston_ent_swap_pair_c2c_old_workflow})}
        \label{fig:kingston_ent_swap_pair_A-L_old_workflow}
    \end{subfigure}

    \caption{Results for entanglement swapping on Kingston, results for single and pairs of rectangles of the A-L stage. The corresponding charts of the c2c and M-L stages can be found in appendix section~\ref{sec:first_assessment_stages}} 
\end{figure}
Contrary to Brisbane results in which no rectangle passed the c2c stage, we see 10 rectangles that passed on the Heron chip, this is almost half of the chip rectangles. Another illustration of the great improvement over the Eagle chip. 
In the pairs assessment, we see that only 2 of the possible pairs we can check failed the assessment. 

\subsubsection{Super-dense coding}
\begin{figure}[H]
    \centering

    \begin{subfigure}[t]{0.48\linewidth}
        \centering
        \includegraphics[width=\linewidth]{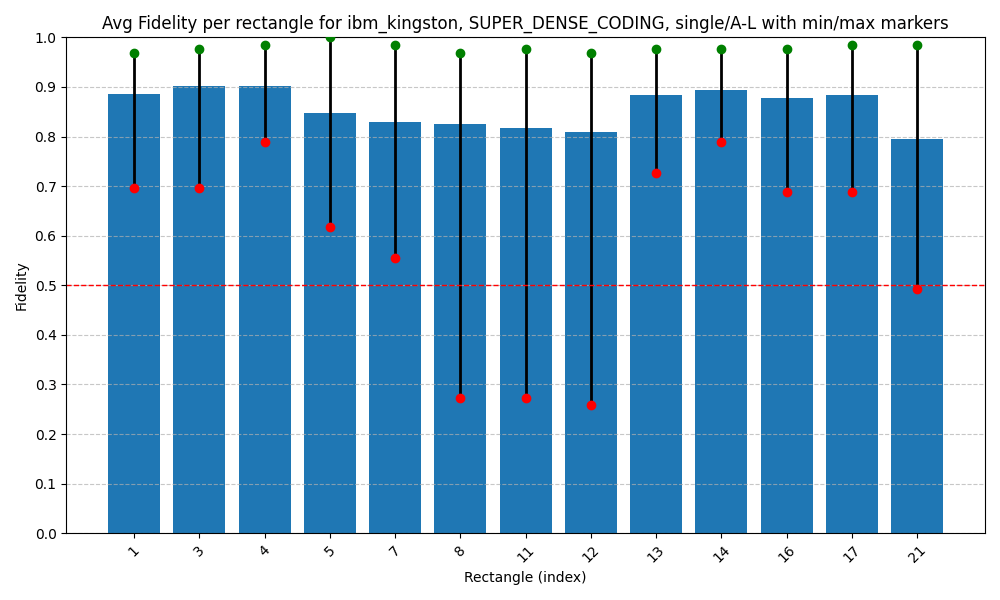}
        \caption{9 Oct 2025: super-dense coding  protocol all lengths on rectangles that passed super-dense coding maximal lengths (figure~\ref{fig:kingston_superdense_single_M-L_old_workflow})}
        \label{fig:kingston_superdense_single_A-L_old_workflow}
    \end{subfigure}
    \hfill
    \begin{subfigure}[t]{0.48\linewidth}
        \centering
        \includegraphics[width=\linewidth]{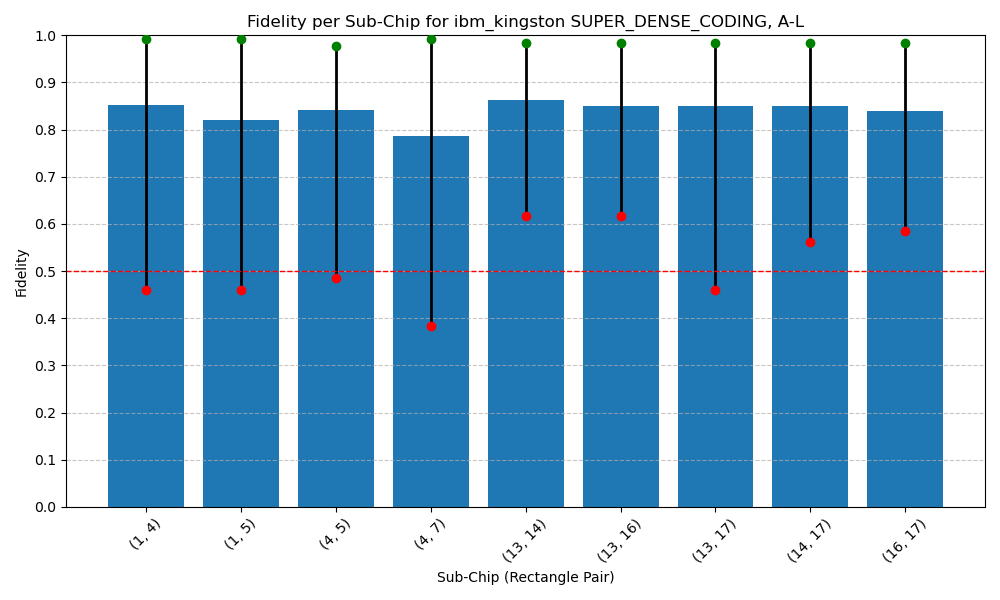}
        \caption{11 Nov 2025: Super-dense coding protocol, A-L on neighboring pairs that passed super-dense coding c2c on rectangle pairs (figure~\ref{fig:kingston_superdense_pair_c2c_old_workflow})}
        \label{fig:kingston_superdense_pair_A-L_old_workflow}
    \end{subfigure}

    \caption{Results for super-dense coding on Kingston, results for single and pairs of rectangles of the A-L stage. The corresponding charts of the c2c and M-L stages can be found in appendix section~\ref{sec:first_assessment_stages}} 
\end{figure}
The last but not least protocol we present is super-dense coding, on the Heron chip one can see that 9 rectangles managed to do this task successfully on every inner path of the sub-chip. We do note that while in the previous protocols the chip performed with ease here we see less optimistic results both in the single and pairs results. In the singles chart rectangles 8, 11 and 12 fidelity values were unexpectedly low, those 3 specific rectangles are neighbors and share a qubit. This can explain the similarity in the minimum fidelity, but need to be further tested. In the pairs chart only 4 pairs out of the 10 possible pairs passed the full assessment. 

\subsection{Protocol vector}
\begin{figure}[H]
    \centering
    \begin{subfigure}{\linewidth}
        \centering
    \includegraphics[width=0.8\linewidth]{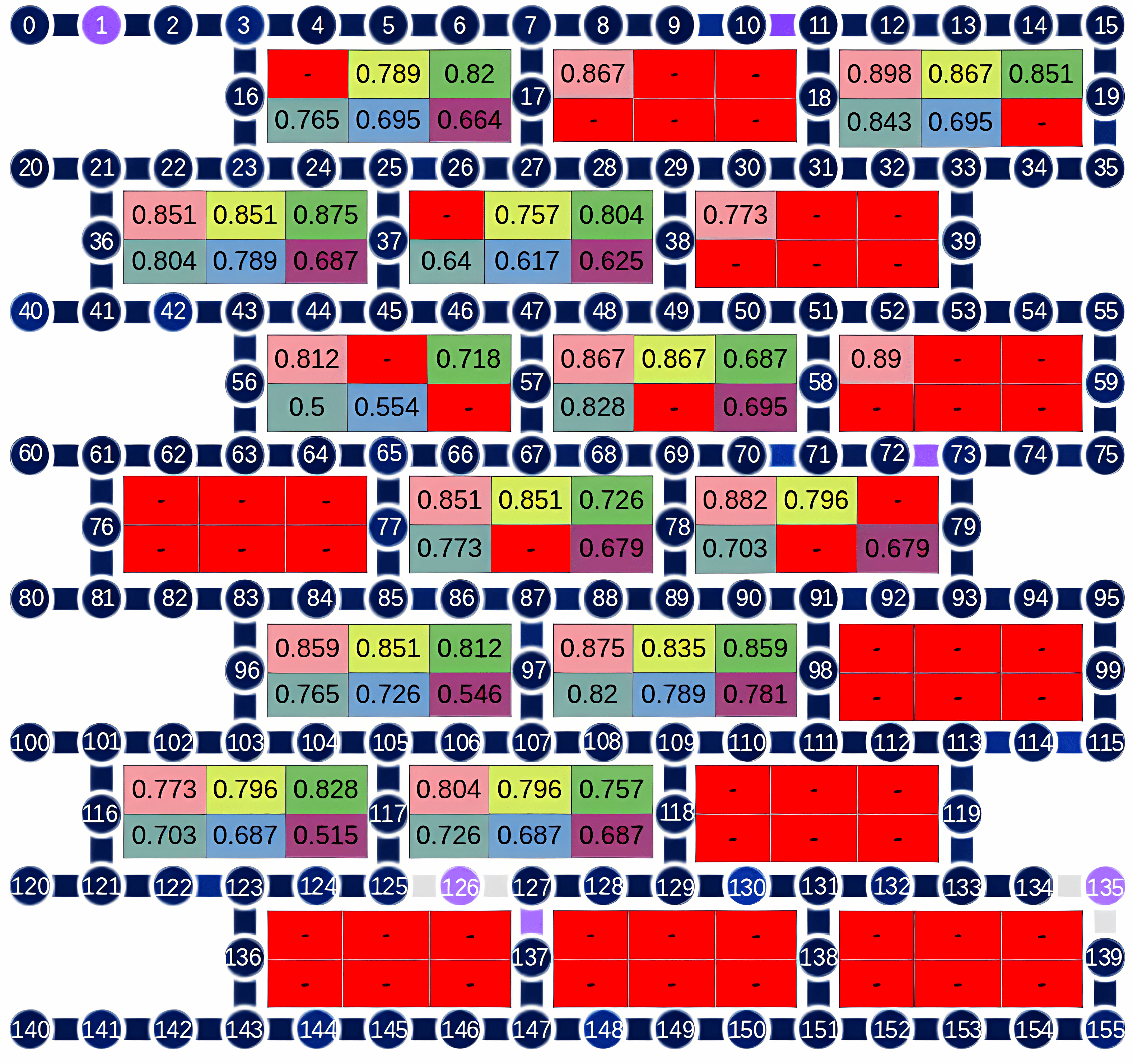}
    
    \end{subfigure}

    \begin{subfigure}{\linewidth}
        \centering
        \includegraphics[width=0.6\linewidth]{figures/protocol_vectors/protocol_vector_legend.png}  
    \end{subfigure}

    \caption{Protocol vector for Kingston chip}
    \label{fig:protocl_vector_kingston_old_workflow}
\end{figure}

The Kingston optimal lookup workflow output an interesting protocol vector, we can intuitively compare this protocol vector to the one of the Eagle chip in figure \ref{fig:protocl_vector_new_brisbane} and see the great difference. On a deeper look into the results that produced this vector we highlight the following matters:
\begin{enumerate}
    \item Some rectangles are all red, this is not because no protocol managed to pass them but because they didn't managed to pass c2c of do-nothing thus according to the workflow method we don't proceed to other protocols on those rectangles. Note that transmit is presented independent to the workflow, so where the transmit protocol is red it means that the sub-chip failed the transmit full assessment.
    \item It is possible to choose a big sub-set of rectangles on this map that theoretically should perform much better then the whole chip for some tasks, this is the main goal of the ``Optimal Lookup Workflow". The sub-set of rectangles can be chosen according to the desired calculation nature and the data presented in the protocol vector above. 
    \item In rectangles 1, 5 and 7 we got unexpected results in which some protocols passed although do-nothing or transmit failed. This is a counter intuitive especially regarding transmit because in other protocols the function of transmitting a state from one qubit to another is used. The results of transmit protocol were taken months apart from the other protocols, this can potentially explain this unexpected outcome.
\end{enumerate}

Kingston managed to show significant improvement on the pairs experiment over Brisbane. As presented in table~\ref{tab:pairs_count_kingston}, every protocol have some pairs that passed our assessment, reaching as high as 18 successful pairs in the teleportation protocol.

\begin{table}
    \centering
    \begin{tabular}{c|c} \toprule
        \textbf{Protocol} & \textbf{Count of Successful Pairs} \\ \midrule
         Transmit &  6*\\
         Do-nothing & 15\\
         Teleportation & 18\\
         Bell-state transfer & 11\\
         Super-dense coding  & 4\\
         Entanglement swapping & 13 \\ \bottomrule
    \end{tabular}
    \caption{Count of successful pairs sub-chips for each protocol on Kingston. *The transmit results, as explained in section~\ref{sec:kinsgton_results_transmit}, are only partial due to system errors that canceled most of this experiment}
    \label{tab:pairs_count_kingston}
\end{table}

\section{Summary - Comparing Eagle to Heron}\label{sec:comparison_section}
While working on the benchmarking of the chips, both Eagle and Heron got significant improvements that made the comparison less trivial. Here, we present comparisons across the whole chip and across every protocol. This should provide to the reader a clear understanding of the strengths and weaknesses of each chip in comparison to the other.

\subsection{Optimal Sub-chip Comparison}
\begin{figure}[H]
    \centering

    \begin{subfigure}[t]{0.48\linewidth}
        \centering
        \includegraphics[width=\linewidth]{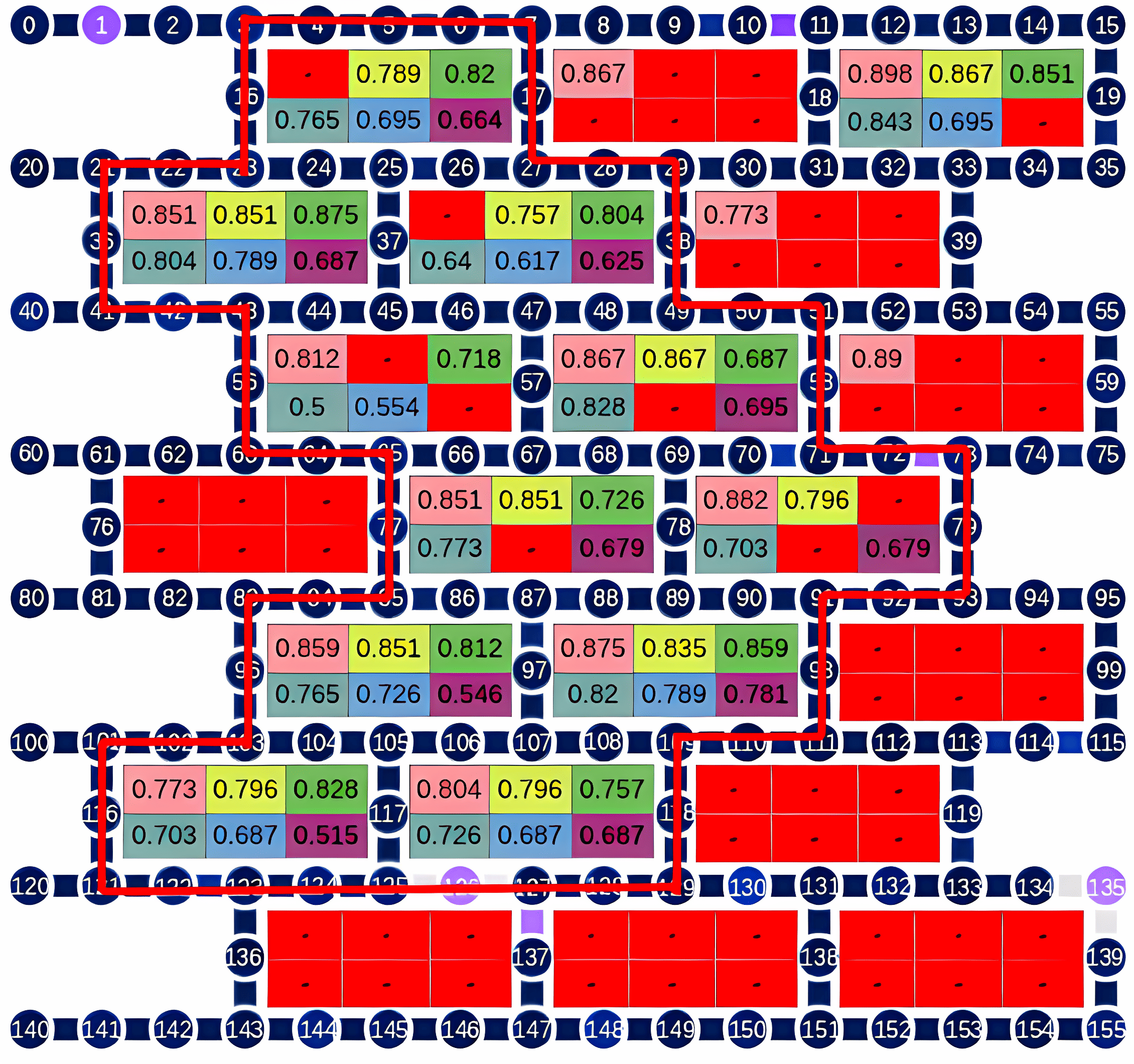}
        \caption{Heron r2 - Kingston example of sub-chip choice}
    \end{subfigure}
    \hfill
    \begin{subfigure}[t]{0.48\linewidth}
        \centering
        \includegraphics[width=\linewidth]{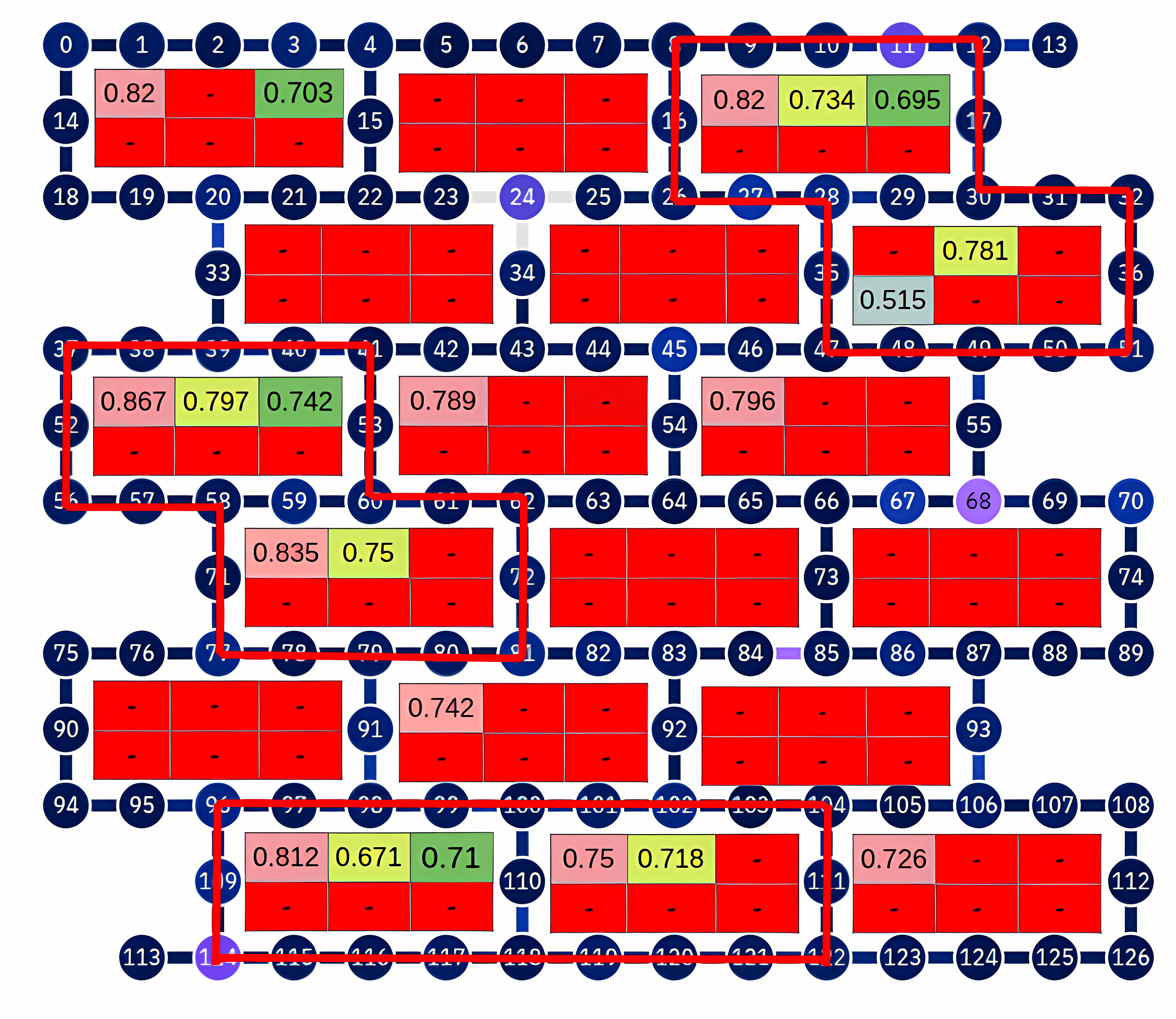}
        \caption{Eagle r3 - Brisbane example of sub-chip choice. In this case it is not clear where is the optimal sub-chip or if it even exists, this example is given for sake of comparison and illustration of a choice}
    \end{subfigure}
    
    \caption{An example of a choice of a sub-set of rectangles according to the protocol vectors of each quantum computer}
    \label{fig:delimiting_optimal_sub_chips}
\end{figure}

As said, the main objective of the optimal lookup workflow is to locate the best ``parts" in the chip that will provide high fidelity in calculations. The protocol vectors allow us to identify a high-performance sub-set of rectangles in the chip, as seen in figure \ref{fig:delimiting_optimal_sub_chips}. As the optimal sub-chip of Kingston contains 11 rectangles that showed good performance in the assessments of various protocols, Brisbane has only 9 rectangles that showed quantumness solely in the basic protocols - transmit and do-nothing, which is of course insufficient for multi-qubit protocols. 

In this comparison we see a huge difference between the 2 chips. While Kingston shows a sub-chip which is practical in terms of size and capabilities, Brisbane fails to have any of those in the optimal sub-chip.

\subsection{Successful Sub-Chips Comparison}

\begin{table}[H]
    \centering
    \begin{tabular}{c|cc|cc} 
        \toprule
        \multicolumn{1}{c|}{\multirow{2}{*}{\textbf{Protocol}}} & \multicolumn{2}{c|}{\textbf{Kingston}} & \multicolumn{2}{c}{\textbf{Brisbane}} \\ 
         & \textbf{Count} & \textbf{Avg} & \textbf{Count} & \textbf{Avg}\\ 
        \midrule    
        Transmit & 13 & 0.847 & 10 & 0.8 \\
        Do-nothing & 11 & 0.824 & 6 & 0.715 \\
        Teleportation & 11 & 0.795 & 4 & 0.713 \\
        Bell-state transfer & 12 & 0.74 & 1 & 0.515 \\
        Super-dense coding & 9 & 0.694 & 0 & 0 \\
        Entanglement swapping & 10 & 0.65 & 0 & 0 \\
        \bottomrule
    \end{tabular}
    \caption{Comparison between Kingston and Brisbane, presenting for each protocol the count of successful single rectangles and \textbf{their} average minimum fidelity}
    \label{tab:pass_singles_count_avg}
\end{table}

Table~\ref{tab:pass_singles_count_avg} presents a comparative performance analysis of the Kingston and Brisbane systems across the six protocols. For every quantum computer the table shows the count of single rectangles that passed the full assessment procedure and the average of their achieved minimum fidelity. In general, the Kingston device demonstrates superior operation fidelity for all the protocols over Brisbane. Both in the average minimum fidelity and successful rectangles count, we note a significant improvement especially in the complex protocols such as super-dense coding and entanglement swapping, where all of Brisbane's rectangles failed these protocols, about half of Kingston rectangles passed them successfully. 

\subsection{By Protocol Comparison}
To facilitate a holistic comparison, we established a scoring metric that assigns a single numerical value to each protocol. This approach enables us to characterize the chip's aggregated performance rather than focusing on isolated sub-regions. This specific formulation was adopted to address budget limitations and to mitigate consistency issues arising from data collection across different dates. In scenarios where such resource constraints are absent, we recommend a comprehensive performance measurement of every rectangle rather than relying on this aggregated scoring method.

The scoring formula incorporates the following parameters
\begin{enumerate}
    \item $N_0$ - The total number of rectangles in the chip, specifically, 18 for Eagle and 21 for Heron
    \item $x_i$ - The minimum fidelity obtained by an inner path in rectangle $i$ in the given protocol. For rectangles that failed to meet the threshold of the A-L stage we assign $x_i=0$
\end{enumerate}

The global score is defined as:
\begin{equation}\label{eq:chip_score_formula}
    \text{Score} = \frac{1}{N_0} \sum_{i=1}^{N_0} x_i
\end{equation}

This metric can be equivalently interpreted as the mean fidelity of valid rectangles scaled by the yield ratio:
\begin{equation}
    \text{Score} = \frac{N}{N_0} \cdot \text{mean}(\{ x_i \mid x_i > 0\})
\end{equation}
where $N$ represents the count of rectangles with non-zero $x_i$.

\begin{table}[H]
    \centering
    \begin{tabular}{|c|c|c|}\hline
        \textbf{} & \textbf{Eagle-r3 - Brisbane} & \textbf{Heron-r2 - Kingston} \\\hline
         \textbf{transmit} &  \colcell{0.442} & \colcell{0.524} \\\hline
         \textbf{do-nothing} &  \colcell{0.247} & \colcell{0.431} \\ \hline
         \textbf{teleportation} &  \colcell{0.158} & \colcell{0.416} \\ \hline
         \textbf{bell-state transfer} &  \colcell{0.028} & \colcell{0.422} \\\hline
         \textbf{entanglement swapping} &  \colcell{0} & \colcell{0.309} \\\hline
         \textbf{super-dense coding} &  \colcell{0} & \colcell{0.297} \\ \hline
    \end{tabular}
    \caption{Table showing score results of each chip with each protocol. Formula \ref{eq:chip_score_formula} is used to calculate these values}
    \label{tab:by_protocol_comparison}
\end{table}
Table~\ref{tab:by_protocol_comparison} shows the great performance difference between the two chips. While Kingston demonstrates broad protocol viability, Brisbane fails to meet the quantumness threshold across the majority of the tested protocols. Note that the score values \textbf{does not} represent fidelity any more so whether the score is above or below the threshold is meaningless. 

\section{Limitations and Malfunctions}\label{sec:limitations_and_malfunctiuons}
Our experimental work faced several operational and resource-related challenges that influenced the data acquisition strategy and the scope of this work.

\subsection{Hardware Availability}
The primary constraint regarding data continuity was the availability of the IBM Brisbane and Kingston systems. Both devices underwent extended maintenance periods during our research window, these periods resulted in significant shifts in performance characteristics. This discontinuity rendered previous datasets obsolete, forcing the re-acquisition of data and, in the case of Brisbane, the distinct categorization of results into 'Old' and 'Modified' datasets to account for the performance shift.

\subsection{Temporal Consistency}
Both systems exhibited notable temporal instability. Repeated benchmarks executed with a one-week interval  occasionally yielded divergent performance metrics, a phenomenon most pronounced in Brisbane but also observed in Kingston. This variance poses a challenge for benchmarking via protocols, as our methodology categorizes sub-chips based on a binary criterion of quantumness advantage (passing or failing a strict fidelity threshold). Consequently, the classification of a specific sub-chip quantum performance is sensitive to the specific calibration window in which the data was acquired, as the underlying performance fluctuations can shift a sub-chip across the pass/fail threshold between sessions.

\subsection{Cancellation of Experiments}
Large-scale characterization jobs (such as A-L experiments) frequently experienced involuntary termination, resulting from either system errors or unprompted cancellation by the provider. In most instances, the system returned partial datasets rather than a complete failure. Due to budgetary constraints, re-executing full experiments was not feasible. Consequently, we adopted a data aggregation strategy where the failed portion of a job was re-executed (sometimes days apart) and merged with the original job's dataset to form a complete dataset. This necessity introduced an unavoidable dependency on the temporal consistency issues noted above, as different regions of the same chip map may represent performance states from different dates.

\section{Conclusions}\label{sec:conclusions}
In this paper we demonstrated a method to assess a quantum chip more practically and intuitively. Demonstrating the method on both Heron and Eagle gave clear and conclusive results, Heron-r2 chip Kingston far out-performed the Eagle-r3 Brisbane chip. The difference was apparent in both size and quality of the optimal sub-chip found. The overall performance comparison of each chip on each protocol also showed the superiority of Kingston over Brisbane. We note that the benchmarking via protocols method shows a \textbf{significant improvement} in the capabilities of IBM's quantum hardware.
It is important to mention that because IBM renews and improves their hardware regularly, the results presented here are correct and precise to the dates written in every chart caption. As hardware progresses, the proposed method in this paper will surely still help in the quality assurance (QA) of each capability that the quantum computer has. The benchmarking via protocols method allows one to have a more relevant and grounded view on the true nature of the chip, as, in general, gate-based benchmarking methods cannot do. 

We would like to add a word about blinding. The issue of ``blinding to the task'' is sometimes relevant in quantum computing. It is also potentially important to guarantee independent and reliable benchmarking methods.

This research could not have been done without IBM's generous policy of allowing independent and objective assessment on their quantum computers and sub-chips. In the name of science, we encourage other companies to emulate the independent qubit availability and fair pricing that allow researchers to perform such assessments.

\section{Acknowledgment}
The authors thank the Quantum Computing Consortium of the Israel Innovation Authority for financial
support. This research project was partially supported by the Helen Diller Quantum Center at the Technion.
TM and YW thank Chen Mechel and Rotem Liss for their contributions to
the very early stage of experimenting teleportation and entanglement swapping 

\printbibliography

\appendix
\section[Appendix - Comparing old Brisbane and Sherbrooke]{Appendix - Comparing two Eagle chips, old Brisbane and Sherbrooke}\label{sec:comparing_brisbane_sherbrooke}
In this section we compare two different Eagle-r3 chips. The comparison is between Sherbrooke and old Brisbane (old Brisbane refers to Brisbane results before the beginning of August 2025, as discussed in section \ref{sec:motivation_to_transmit}). To illustrate the comparison without going into much detail, we present the A-L experiment done on each chip on all protocols except transmit (which was not defined in the time of the measurements). The selection process method for these experiments is the old ``Optimal Lookup Workflow", in which we initially run do-nothing c2c on the chip and eliminate all the failed rectangles from the chip. Then we proceed to run a full assessment for the rest of the protocols. 
This comparison show that the two chips have similar results for most of the protocols, with a small advantage to Sherbrooke over old Brisbane.

\begin{figure}[H]
    \centering

    \begin{subfigure}[t]{0.48\linewidth}
        \centering
        \includegraphics[width=1\linewidth]{figures/brisbane_optimal_lookup/single_rect/full_vec/DO_NOTHING/per_rect_plot_DO_NOTHING_single_A-L.png}
        \caption{30 May 2025: old Brisbane, do-nothing, A-L on single rectangles}
        \label{fig:brisbane_DO_NOTHING_A-L}
    \end{subfigure}
    \hfill
    \begin{subfigure}[t]{0.48\linewidth}
        \centering
        \includegraphics[width=1\linewidth]{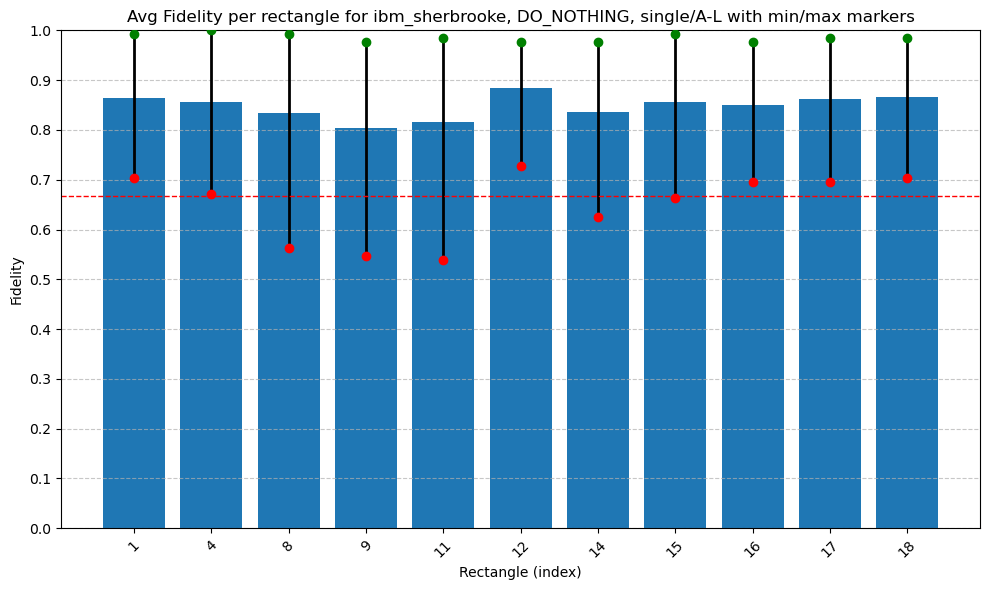}
        \caption{31 May 2025: Sherbrooke, do-nothing, A-L on single rectangles}
        \label{fig:sherbrooke-do-Nothing-single-A-L}
    \end{subfigure}

    \caption{Comparing the results of A-L for do-nothing on Old Brisbane and Sherbrooke. Assessing only the rectangles that passed the earlier stages of the assessment}
\end{figure}
Viewing both charts side by side it is apparent that many more rectangles managed to get further in the assessment on Sherbrooke, as many more rectangles were tested in the A-L experiment. Looking at the results of the do-nothing full assessment  old Brisbane got 2 successful rectangles while Sherbrooke have 6. This protocol is the simplest one done in this comparison, still old Brisbane success rate is $\frac{1}{9}$ and Sherbrooke's is $\frac{1}{3}$.
\begin{figure}[H]
    \centering

    \begin{subfigure}[t]{0.48\linewidth}
        \centering
        \includegraphics[width=\linewidth]{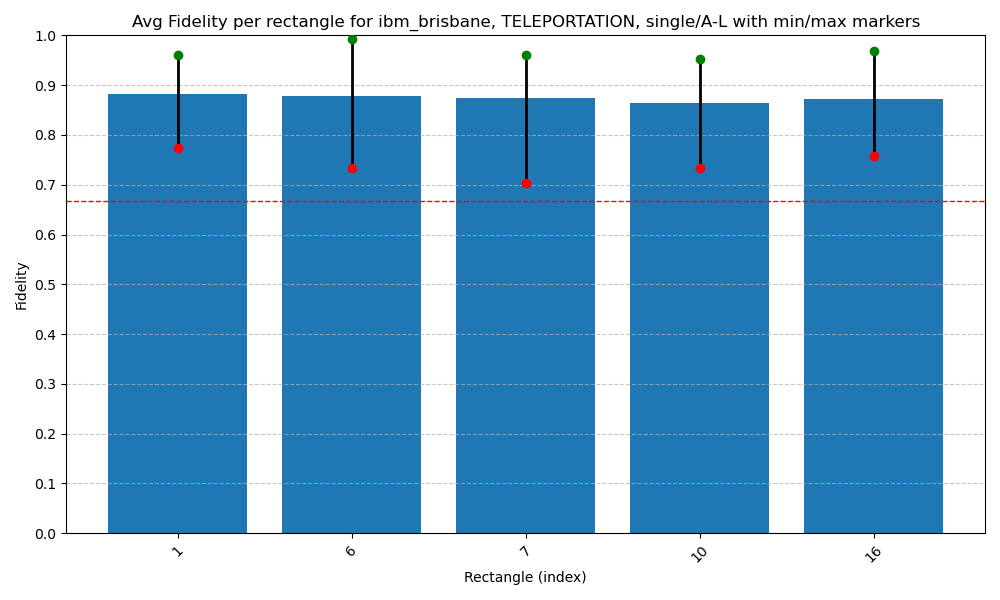}
        \caption{30 May 2025: Brisbane teleportation, A-L on single rectangles}
        \label{fig:brisbane_teleportation_A-L}
    \end{subfigure}
    \hfill
    \begin{subfigure}[t]{0.48\linewidth}
        \centering
        \includegraphics[width=1\linewidth]{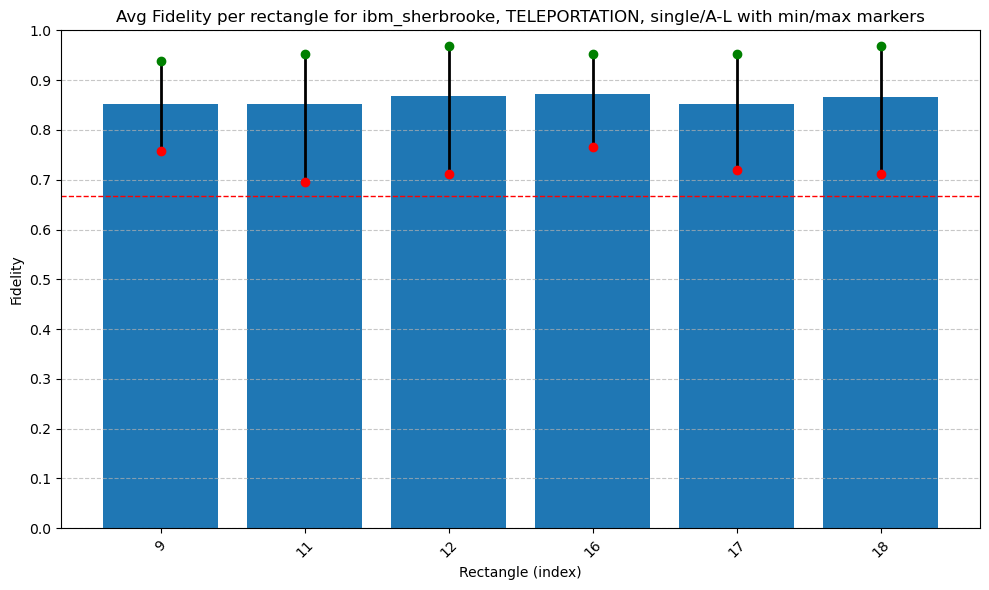}
        \caption{31 May 2025: Sherbrooke teleportation, A-L on single rectangles}
        \label{fig:sherbrooke_teleportation_A-L}
    \end{subfigure}

    \caption{Comparing the results of A-L for teleportation on Old Brisbane and Sherbrooke. Assessing only the rectangles that passed the earlier stages of the assessment}
    \label{fig:brisbane_vs_sherbrooke_teleportaion}
\end{figure}
Surprisingly, in figure~\ref{fig:brisbane_vs_sherbrooke_teleportaion} the teleportation experiment on old Brisbane got better results then do-nothing. 
\begin{figure}[H]
    \centering

    \begin{subfigure}[t]{0.48\linewidth}
        \centering
        \includegraphics[width=1\linewidth]{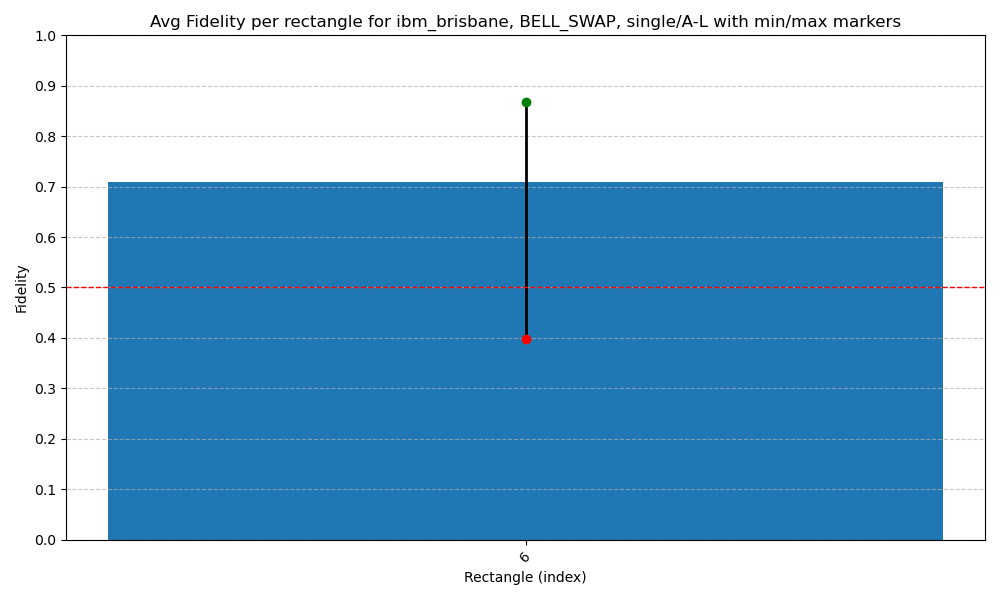}
        \caption{30 May 2025: Brisbane, bell-state transfer, A-L on single rectangles}
        \label{fig:brisbane_bell_state_A-L}
    \end{subfigure}
    \hfill
    \begin{subfigure}[t]{0.48\linewidth}
        \centering
        \includegraphics[width=1\linewidth]{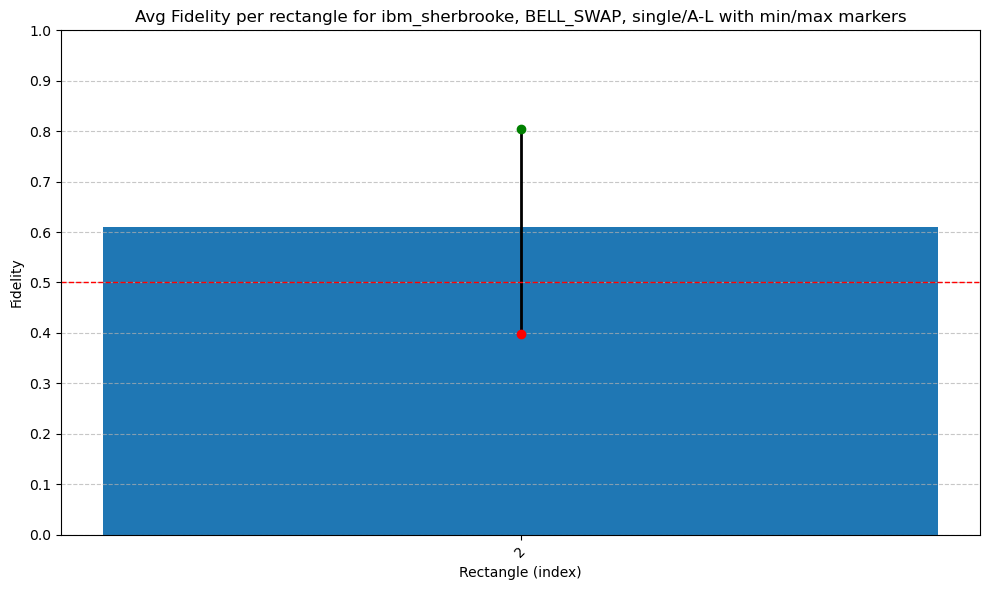}
        \caption{31 May 2025: Sherbrooke, bell-state transfer, A-L on single rectangles}
        \label{fig:sherbrooke_bell_state_A-L}
    \end{subfigure}

    \caption{Comparing the results of A-L for bell-state transfer on Old Brisbane and Sherbrooke. Assessing only the rectangles that passed the earlier stages of the assessment}
\end{figure}
The results for this protocol and the two that follow show poor results from the two chips. None of the rectangles in both old Brisbane and Sherbrooke managed to pass bell-state transfer, entanglement swapping or super-dense coding. While in the last two protocols none of the rectangles managed to pass c2c experiment. This demonstrates the poor performance of the two chips which are older versions of IBM's hardware.
\begin{figure}[H]
    \centering

    \begin{subfigure}[t]{0.48\linewidth}
        \centering
        \includegraphics[width=1\linewidth]{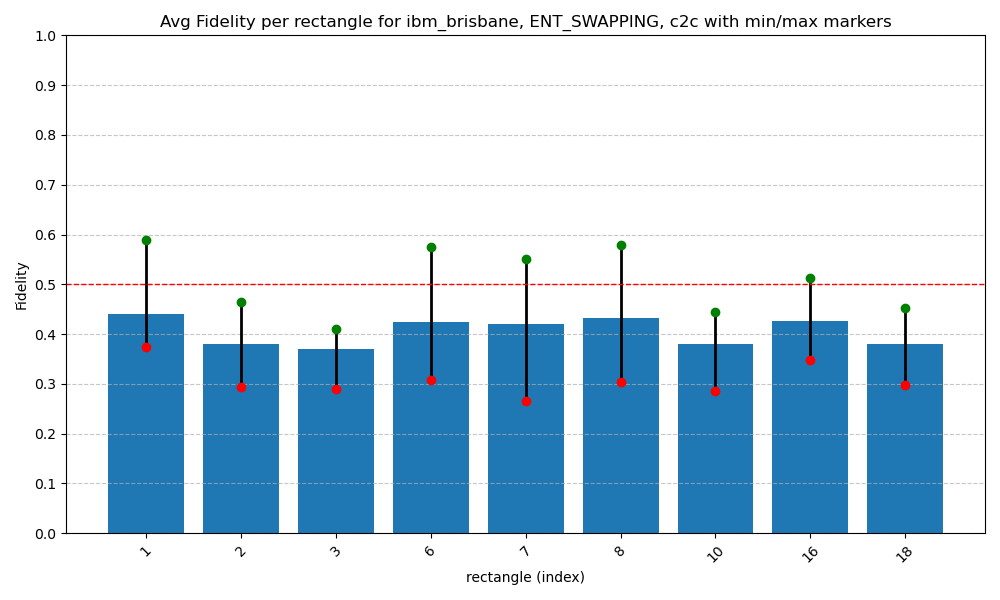}
        \caption{29 May 2025: Brisbane entanglement swapping, c2c on single rectangles}
        \label{fig:brisbane_ent_swapping_c2c}
    \end{subfigure}
    \hfill
    \begin{subfigure}[t]{0.48\linewidth}
        \centering
        \includegraphics[width=1\linewidth]{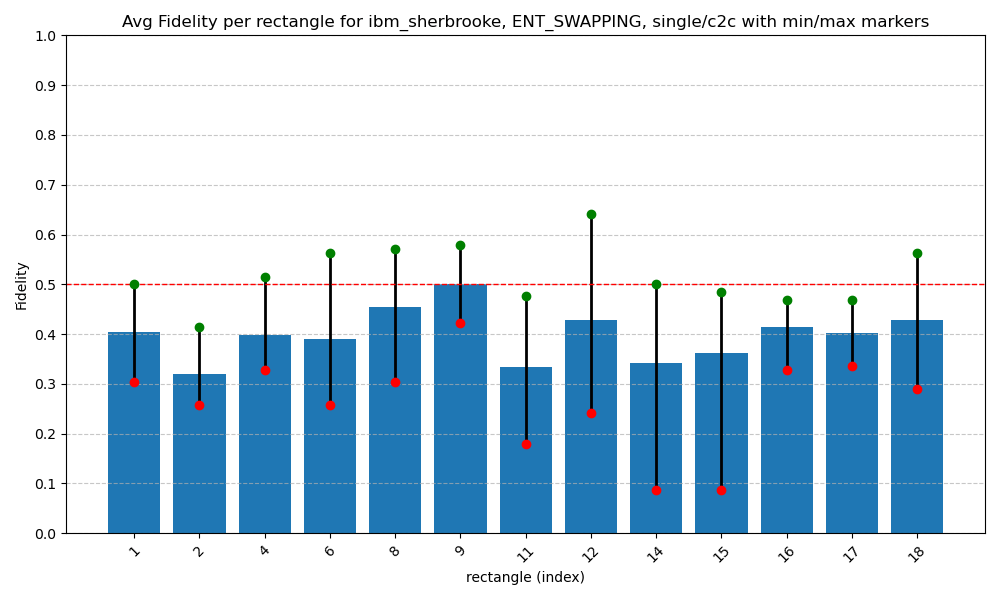}
        \caption{31 May 2025: Sherbrooke entanglement swapping, c2c on single rectangles}
        \label{fig:sherbrooke_ent_swapping_c2c}
    \end{subfigure}

    \caption{Comparing the results of A-L for entanglement swapping on Old Brisbane and Sherbrooke. Assessing only the rectangles that passed the earlier stages of the assessment}
\end{figure}

\begin{figure}[H]
    \centering

    \begin{subfigure}[t]{0.48\linewidth}
        \centering
        \includegraphics[width=1\linewidth]{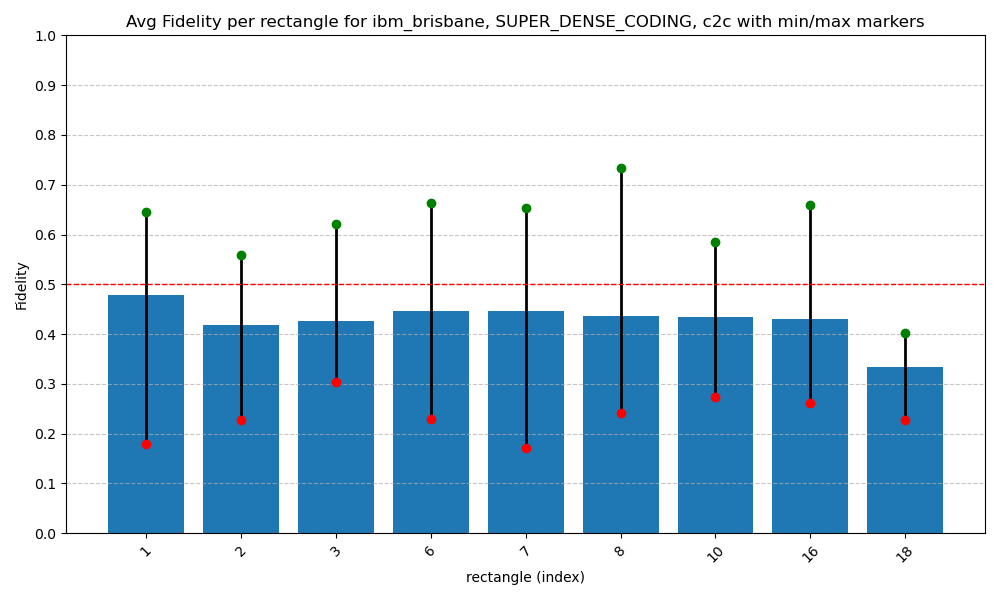}
        \caption{29 May 2025: Brisbane super-dense coding, c2c on single rectangles}
        \label{fig:brisbane_superdense_coding_c2c}
    \end{subfigure}
    \hfill
    \begin{subfigure}[t]{0.48\linewidth}
        \centering
        \includegraphics[width=1\linewidth]{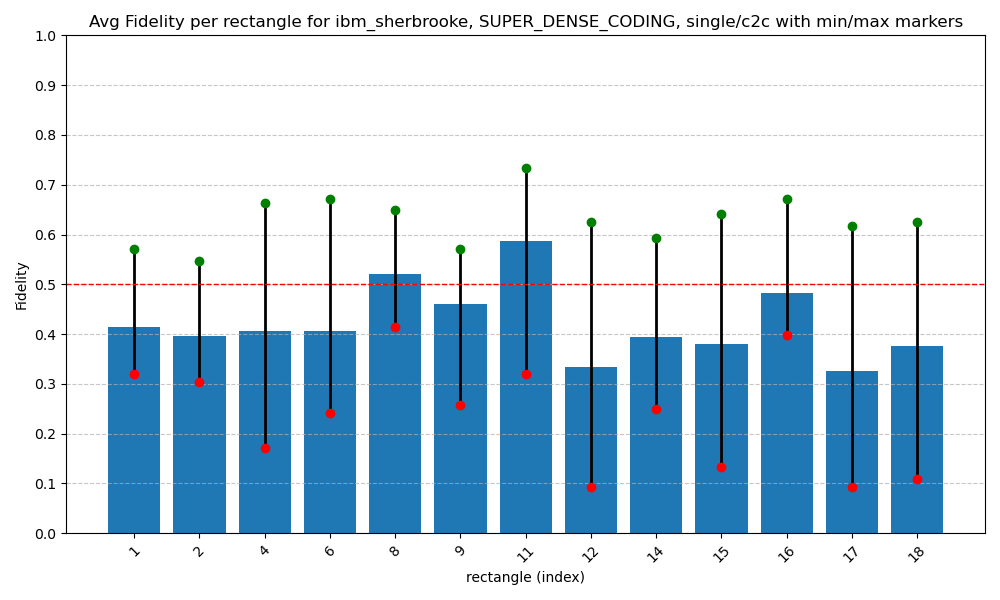}
        \caption{31 May 2025: Sherbrooke super-dense coding, c2c on single rectangles}
        \label{fig:sherbrooke_superdense_coding_c2c}
    \end{subfigure}
        
    \caption{Comparing the results of A-L for entanglement swapping on Old Brisbane and Sherbrooke. Assessing only the rectangles that passed the earlier stages of the assessment}
\end{figure}

\section[Appendix - Consistency of Modified Brisbane]{Appendix - Consistency Check on Modified Brisbane}
In this section we show results that were taken a month apart in the end of Aug and the end of Sep only for do-nothing. This experiment was conducted to help us assess the consistency of the hardware across a time interval of one month. The results show consistency with six rectangles that passed each assessment, with an overlap of four rectangles that passed both assessments.

\begin{figure}[H]
    \centering

    \begin{subfigure}[t]{0.48\linewidth}
        \centering
        \includegraphics[width=\linewidth]{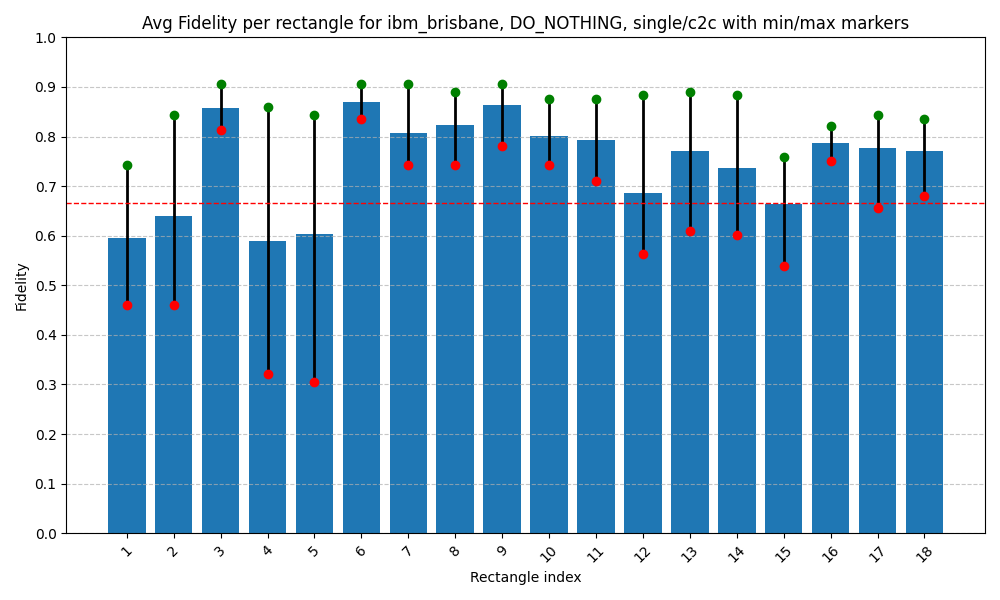}
        \caption{28 Aug 2025: do-nothing experiment on all of Brisbane's rectangles}
        \label{fig:con_check_28Aug_c2c}
    \end{subfigure}
    \hfill
    \begin{subfigure}[t]{0.48\linewidth}
        \centering
        \includegraphics[width=\linewidth]{figures/new_brisbane_4_comparing_2_old/single/c2c/DO_NOTHING/per_rect_plot_DO_NOTHING_single_c2c.png}
        \caption{28 Sep 2025: do-nothing experiment on all of Brisbane's rectangles}
        \label{fig:con_check_28Sep_c2c}
    \end{subfigure}

    \caption{c2c experiments done on Brisbane 1 month apart}
\end{figure}
The full assessment of do-nothing starts with c2c on all 18 rectangles. We see that in the August assessment nine rectangles passed while in the September assessment there were eleven successful rectangles, the two successful rectangles sets have an overlap of six rectangles.
\begin{figure}[H]
    \centering

    \begin{subfigure}[t]{0.48\linewidth}
        \centering
        \includegraphics[width=\linewidth]{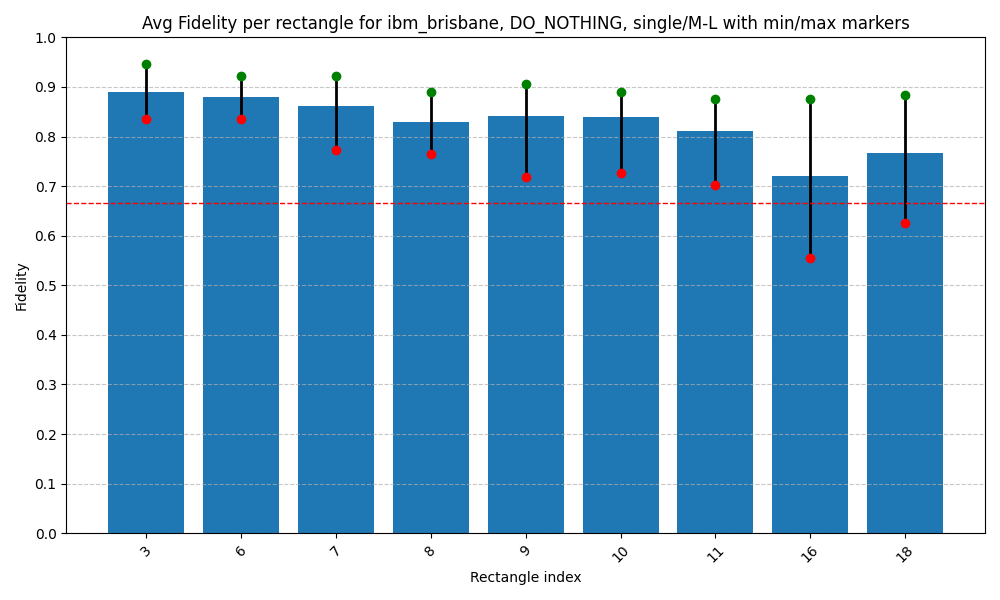}
        \caption{28 Aug 2025: do-nothing experiment on all rectangles that passed in figure \ref{fig:con_check_28Aug_c2c}}
        \label{fig:con_check_28Aug_ML}
    \end{subfigure}
    \hfill
    \begin{subfigure}[t]{0.48\linewidth}
        \centering
        \includegraphics[width=\linewidth]{figures/new_brisbane_4_comparing_2_old/single/M-L/DO_NOTHING/per_rect_plot_DO_NOTHING_single_M-L.png}
        \caption{28 Sep 2025: do-nothing experiment on all rectangles that passed in figure \ref{fig:con_check_28Sep_c2c}}
        \label{fig:con_check_28Sep_ML}
    \end{subfigure}

    \caption{M-L experiments done on Brisbane 1 month apart}
\end{figure}
In the M-L test the two result sets are almost equivalent, with seven successful rectangles for August results and nine for September. Rectangles 3, 6, 7 and 10 passed both experiments, about half of the total successful rectangles.
\begin{figure}[H]
    \centering

    \begin{subfigure}[t]{0.48\linewidth}
        \centering
        \includegraphics[width=\linewidth]{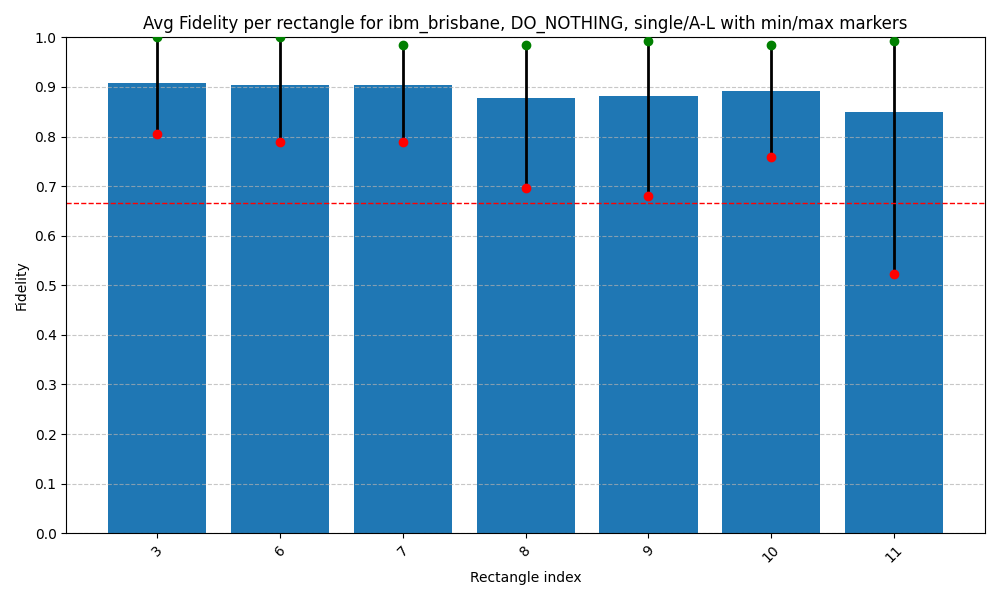}
        \caption{28 Aug 2025: do-nothing experiment on all rectangles that passed in figure \ref{fig:con_check_28Aug_ML}}
        \label{fig:con_check_28Aug_AL}
    \end{subfigure}
    \hfill
    \begin{subfigure}[t]{0.48\linewidth}
        \centering
        \includegraphics[width=\linewidth]{figures/new_brisbane_4_comparing_2_old/single/A-L/DO_NOTHING/per_rect_plot_DO_NOTHING_single_A-L.png}
        \caption{28 Sep 2025: do-nothing experiment on all rectangles that passed in figure \ref{fig:con_check_28Sep_ML}}
        \label{fig:con_check_28Sep_AL}
    \end{subfigure}

    \caption{A-L experiments done on Brisbane 1 month apart}
\end{figure}
The final results of this comparison are six do-nothing-capable rectangles in the August results and seven in September's. With rectangles 3, 6, 7 and 10 that passed both experiments, we note consistency in the performance of Brisbane across one month period.

\section[Appendix - Results of First Assessment Stages]{Appendix - Results of First Assessment Stages - c2c and M-L}\label{sec:first_assessment_stages}
This section contains all the c2c and M-L charts that are part of the assessments shown in sections \ref{sec:mod_brisbane_results_single_and_pairs} and \ref{sec:kinsgton_results}.
\subsection{Brisbane - Do-nothing - Singles and Pairs}
\begin{figure}[H]
    \centering

    \begin{subfigure}[t]{0.48\linewidth}
        \centering
        \includegraphics[width=\linewidth]{figures/new_brisbane_4_comparing_2_old/single/c2c/DO_NOTHING/per_rect_plot_DO_NOTHING_single_c2c.png}
    \caption{28 Sep 2025: do-nothing protocol, corner to corner, all rectangles.}
    \label{fig:brisbane_Sep28_do_nothing_single_c2c}
    \end{subfigure}
    \hfill
    \begin{subfigure}[t]{0.48\linewidth}
        \centering
        \includegraphics[width=\linewidth]{figures/new_brisbane_4_comparing_2_old/single/M-L/DO_NOTHING/per_rect_plot_DO_NOTHING_single_M-L.png}
    \caption{28 Sep 2025: do-nothing protocol, max length, all rectangles that passed corner to corner on 28 Sep 2025 (figure~\ref{fig:brisbane_Sep28_do_nothing_single_c2c}).}
    \label{fig:brisbane_Sep28_do_nothing_single_M-L}
    \end{subfigure}
    \hfill
    \begin{subfigure}[t]{0.48\linewidth}
        \centering
        \includegraphics[width=\linewidth]{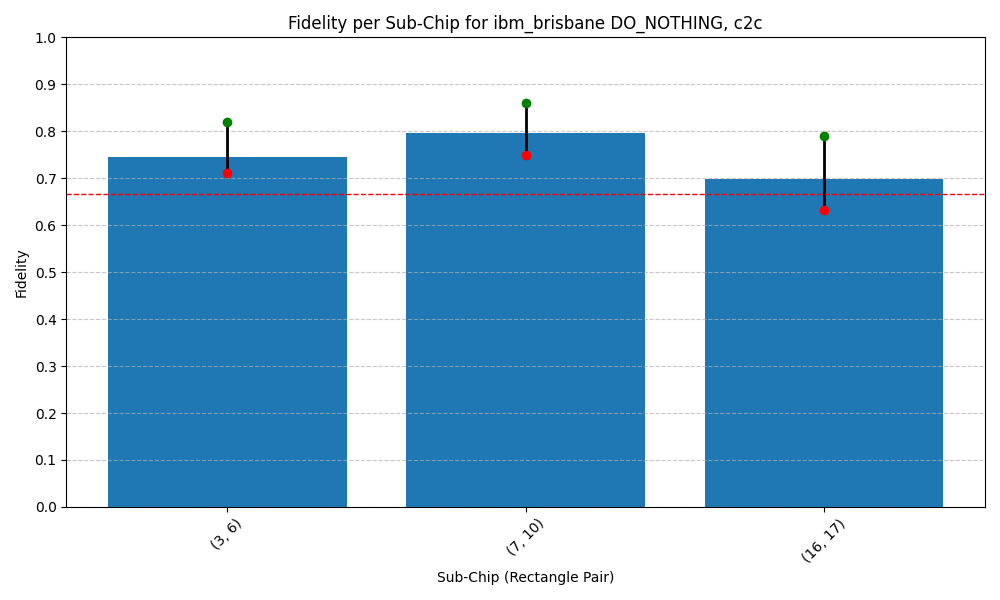}
    \caption{29 Sep 2025: do-nothing protocol on pairs of rectangles, c2c, testing only pairs where each of the single rectangles passed all length experiment of do-nothing on 28 Sep 2025 (figure~\ref{fig:brisbane_Sep28_do_nothing_single_A-L})}
    \label{fig:brisbane_Sep29_do_nothing_pair_c2c}
    \end{subfigure}

    \caption{Results of first assessment stages of do-nothing for single rectangle}
\end{figure}

\subsection{Brisbane - Teleportation - Singles}
\begin{figure}[H]
    \centering

    \begin{subfigure}[t]{0.48\linewidth}
    \centering
        \includegraphics[width=\linewidth]{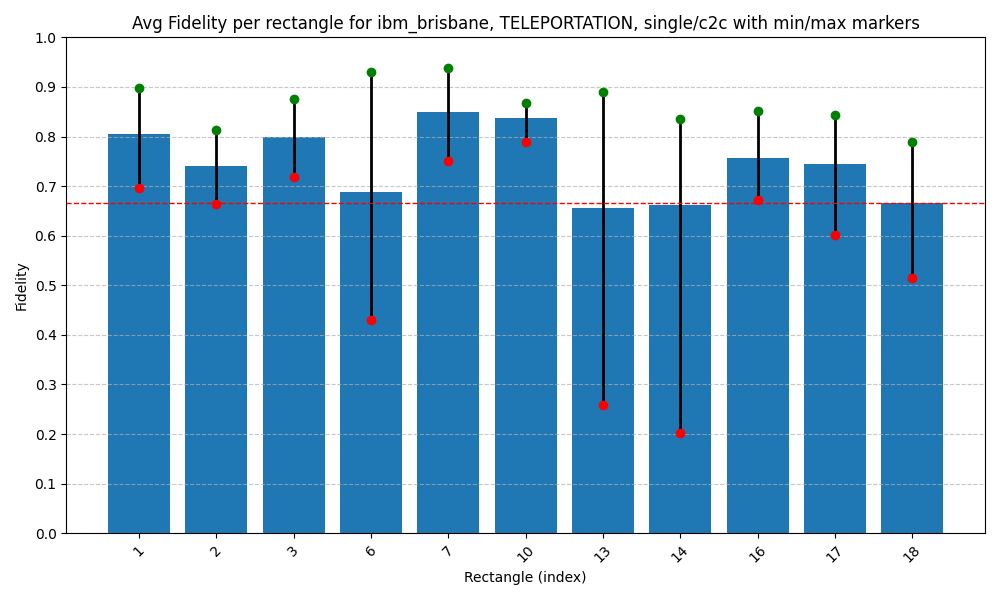} 
    \caption{7 Oct 2025: teleportation protocol, corner to corner on all the rectangles that passed the corner to corner test of do-nothing on 28 Sep 2025 (figure~\ref{fig:brisbane_Sep28_do_nothing_single_c2c})}
    \label{fig:brisbane_Oct7_teleportation_single_c2c}
    \end{subfigure}
    \hfill
    \begin{subfigure}[t]{0.48\linewidth}
    \centering
    \includegraphics[width=\linewidth]{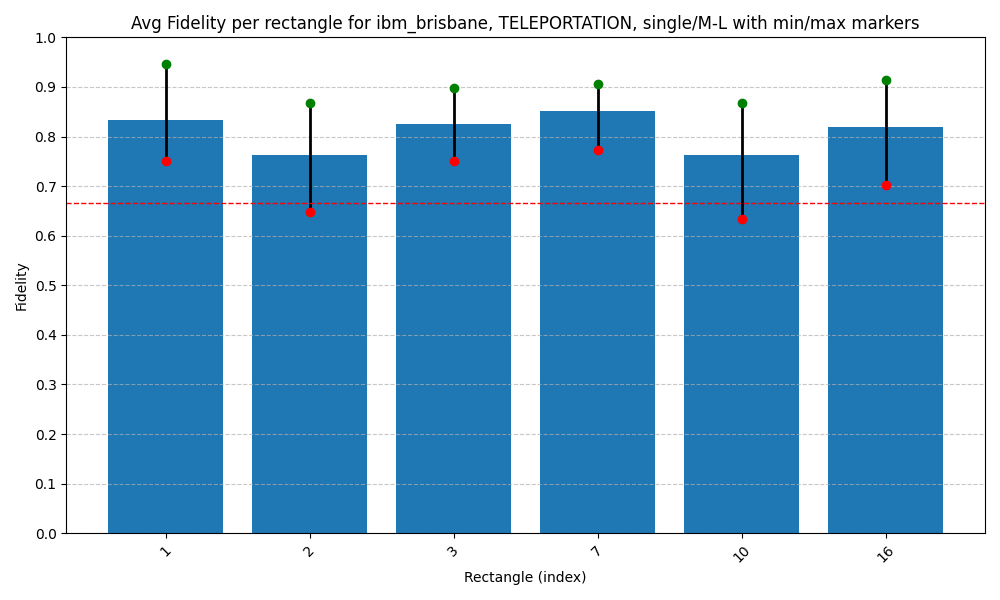} 
    \caption{8 Oct 2025: teleportation protocol, max lengths on all the rectangles that passed the corner to corner test of teleportation on 7 Oct 2025 (figure~\ref{fig:brisbane_Oct7_teleportation_single_c2c})}
    \label{fig:brisbane_Oct8_teleportation_single_M-L}
    \end{subfigure}

    \caption{Results of first assessment stages of teleportation for single rectangle}
\end{figure}

\subsection{Brisbane - bell-state transfer - singles}
\begin{figure}[H]
    \centering

    \begin{subfigure}[t]{0.48\linewidth}
        \centering
        \includegraphics[width=\linewidth]{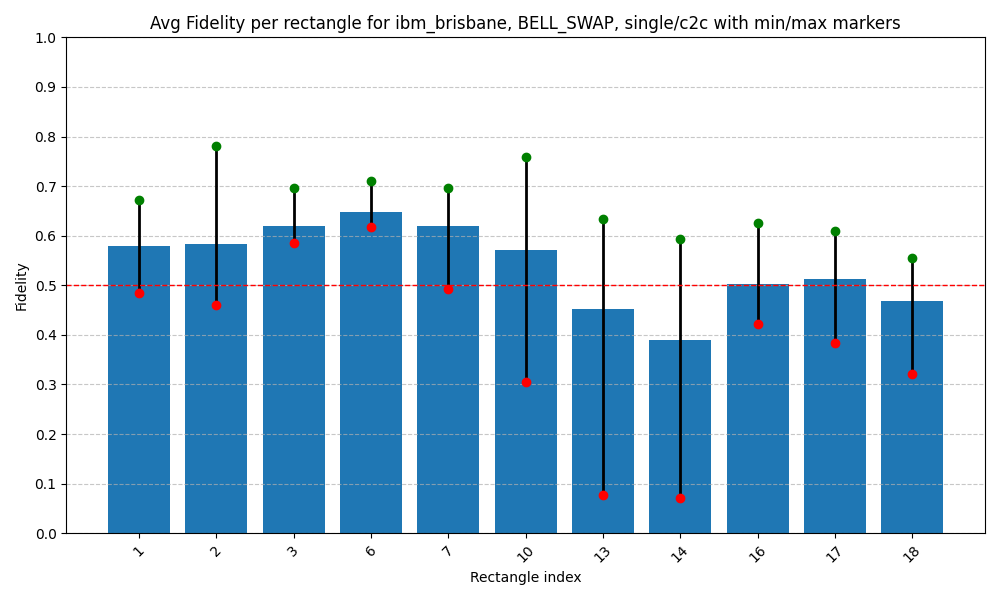}
    \caption{28 Sep 2025: bell-state transfer protocol, corner to corner on all the rectangles that passed the corner to corner test of do-nothing on 28 Sep 2025 (figure~\ref{fig:brisbane_Sep28_do_nothing_single_c2c}). }
    \label{fig:brisbane_Sep28_Bell_swap_single_c2c}
    \end{subfigure}
    \hfill
    \begin{subfigure}[t]{0.48\linewidth}
        \centering
    \includegraphics[width=\linewidth]{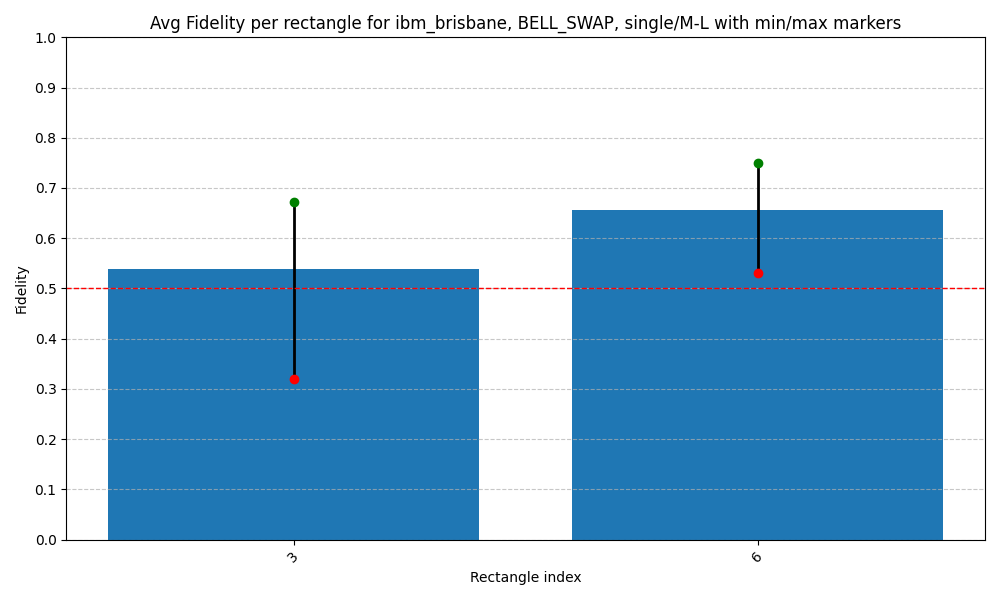}
    \caption{28 Sep 2025: bell-state transfer protocol, max lengths on all the rectangles that passed the corner to corner test of bell-state transfer on 28 Sep 2025 (figure~\ref{fig:brisbane_Sep28_Bell_swap_single_c2c}).}
    \label{fig:brisbane_Sep28_Bell_swap_single_M-L}    
    \end{subfigure}

    \caption{Results of first assessment stages of bell-state transfer on Brisbane for single rectangles}
\end{figure}

\subsection{Kingston - Transmit on Singles}

\begin{figure}[H]
    \centering

    \begin{subfigure}[t]{0.48\linewidth}
        \centering
        \includegraphics[width=\linewidth]{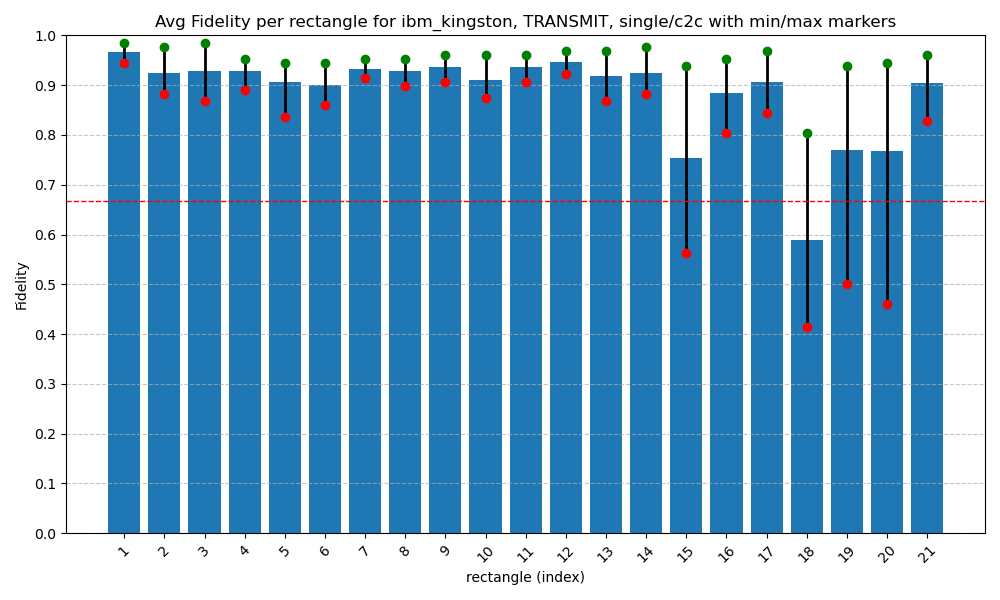}
    \caption{7 June 2025: Transmit protocol, c2c, all rectangles. 8 paths per rectangle. all rectangles passed but 15,18,19,20.}
    \label{fig:kingston_transmit_c2c}
    \end{subfigure}
    \hfill
    \begin{subfigure}[t]{0.48\linewidth}
        \centering
        \includegraphics[width=\linewidth]{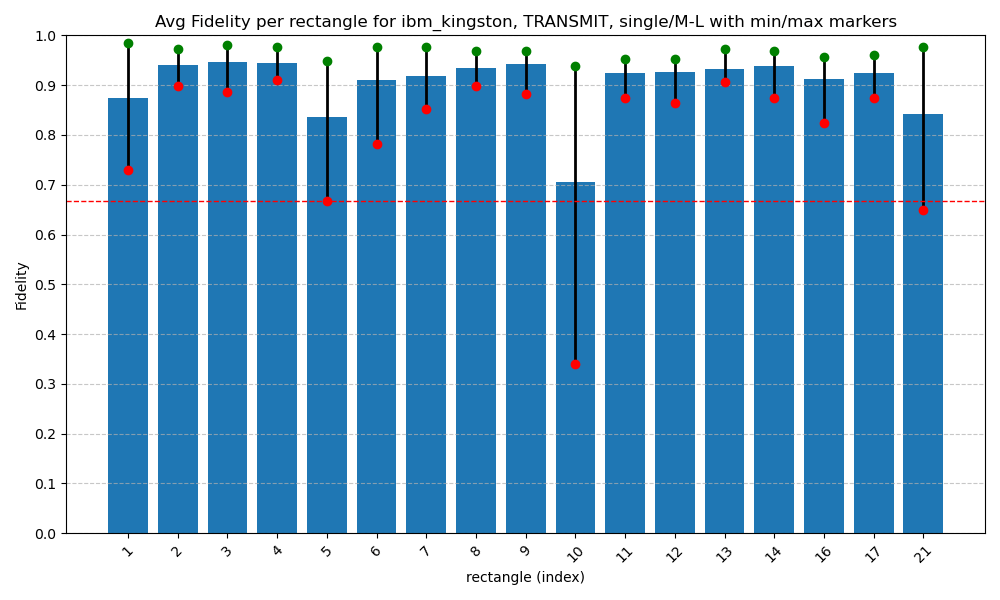}
    \caption{11 Jun 2025: Transmit protocol, max-lengths, 24 paths per rectangle. We see that rectangles 10 and 21 failed in this test. Participating rectangles that passed transmit c2c on 7 June 2025 (figure~\ref{fig:kingston_transmit_c2c})}
    \label{fig:kingston_transmit_M-L}
    \end{subfigure}

    \caption{Results of first assessment stages of transmit on Kingston for single rectangle}
\end{figure}

\subsection{Kingston - Do-nothing on Singles}
\begin{figure}[H]
    \centering

    \begin{subfigure}[t]{0.48\linewidth}
        \centering
        \includegraphics[width=\linewidth]{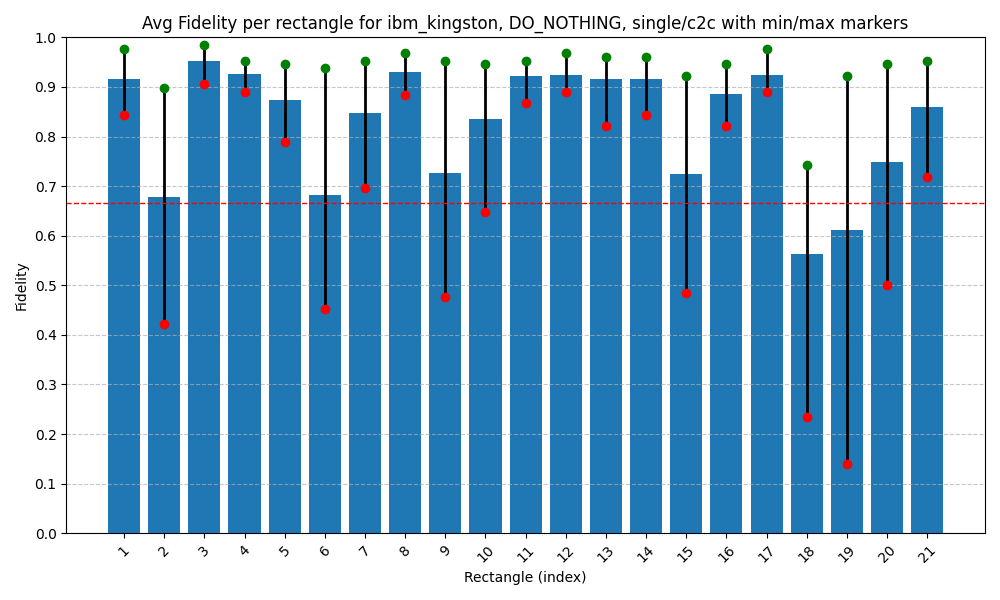}
        \caption{8 Oct 2025: do-nothing protocol, corner to corner, all rectangles, according to old workflow (before transmit)} 
        \label{fig:kingston_do_nothing_single_c2c_old_workflow}    
    \end{subfigure}
    \hfill
    \begin{subfigure}[t]{0.48\linewidth}
        \centering
        \includegraphics[width=\linewidth]{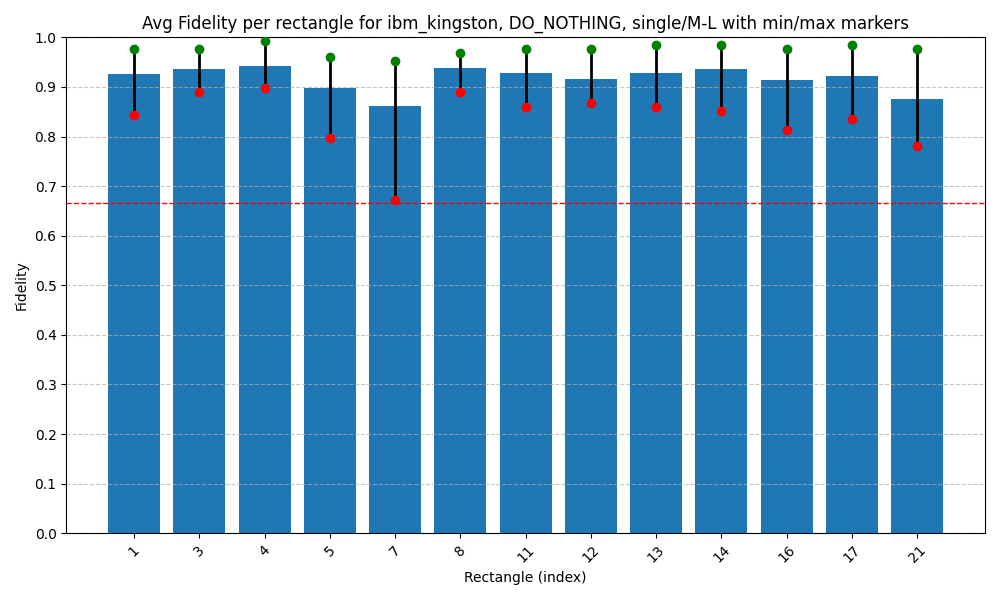}
        \caption{9 Oct 2025: do-nothing protocol, maximal lengths on rectangles that passed do-nothing corner to corner (figure~\ref{fig:kingston_do_nothing_single_c2c_old_workflow})}
        \label{fig:kingston_do_nothing_single_M-L_old_workflow}
    \end{subfigure}
    
    \caption{Results of first assessment stages of do-nothing on Kingston for single rectangle}
\end{figure}

\subsection{Kingston - Teleportation on Singles}
\begin{figure}[H]
    \centering

    \begin{subfigure}[t]{0.48\linewidth}
        \centering
        \includegraphics[width=\linewidth]{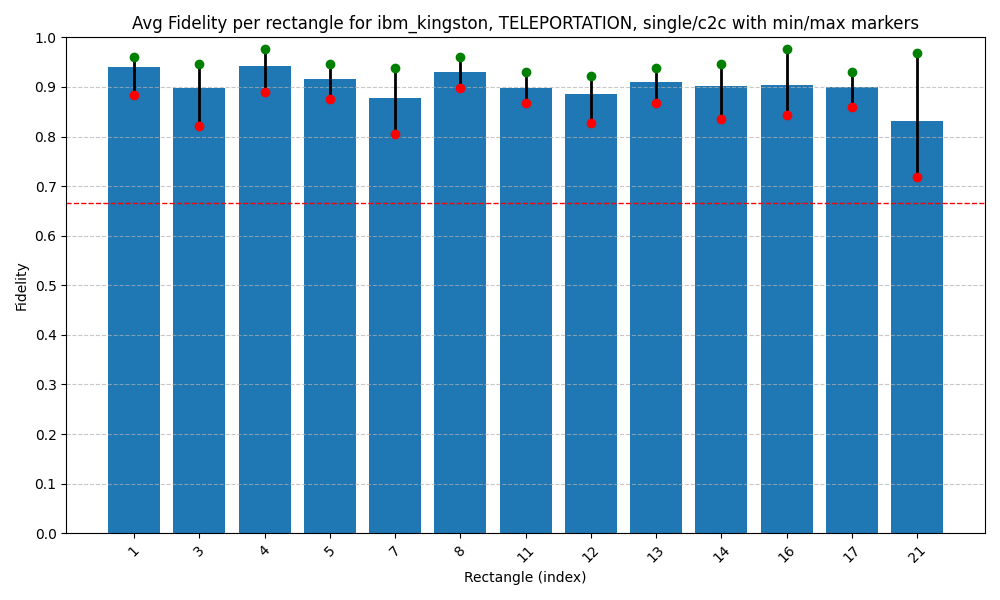}
    \caption{9 Oct 2025: teleportation protocol corner to corner on rectangles that passed do-nothing corner to corner (figure~\ref{fig:kingston_do_nothing_single_c2c_old_workflow}) according to the old workflow (before transmit)}
    \label{fig:kingston_teleportation_single_c2c_old_workflow}
    \end{subfigure}
    \hfill
    \begin{subfigure}[t]{0.48\linewidth}
        \centering
        \includegraphics[width=\linewidth]{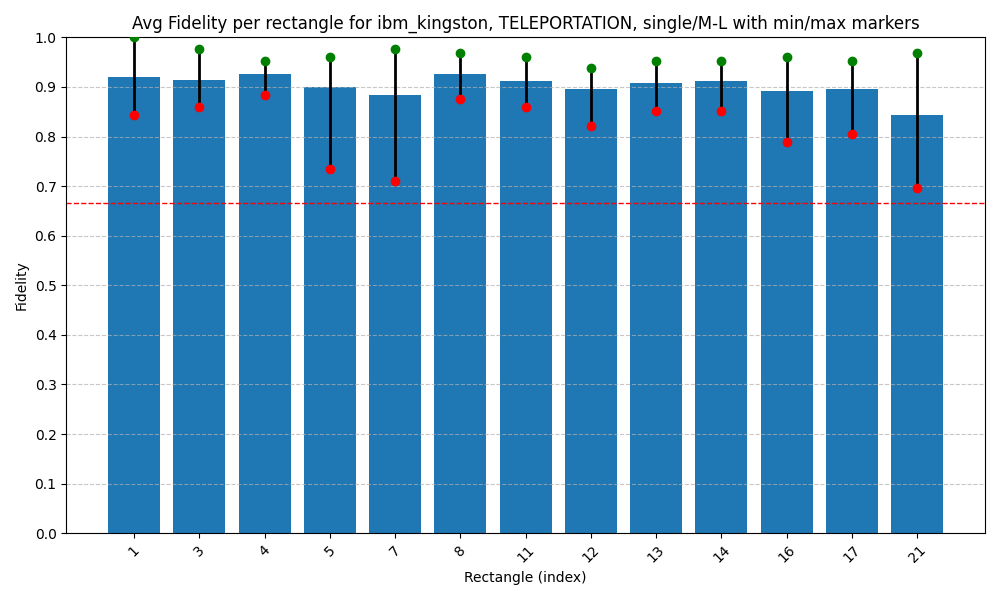}
    \caption{9 Oct 2025: teleportation protocol, maximal lengths on rectangles that passed teleportation corner to corner (figure~\ref{fig:kingston_teleportation_single_c2c_old_workflow})}
        \label{fig:kingston_teleportation_single_M-L_old_workflow}
    \end{subfigure}
    
    \caption{Results of first assessment stages of teleportation on Kingston for single rectangle}
\end{figure}

\subsection{Kingston - Bell-State Transfer on singles}

\begin{figure}[H]
    \centering

    \begin{subfigure}[t]{0.48\linewidth}
        \centering
        \includegraphics[width=\linewidth]{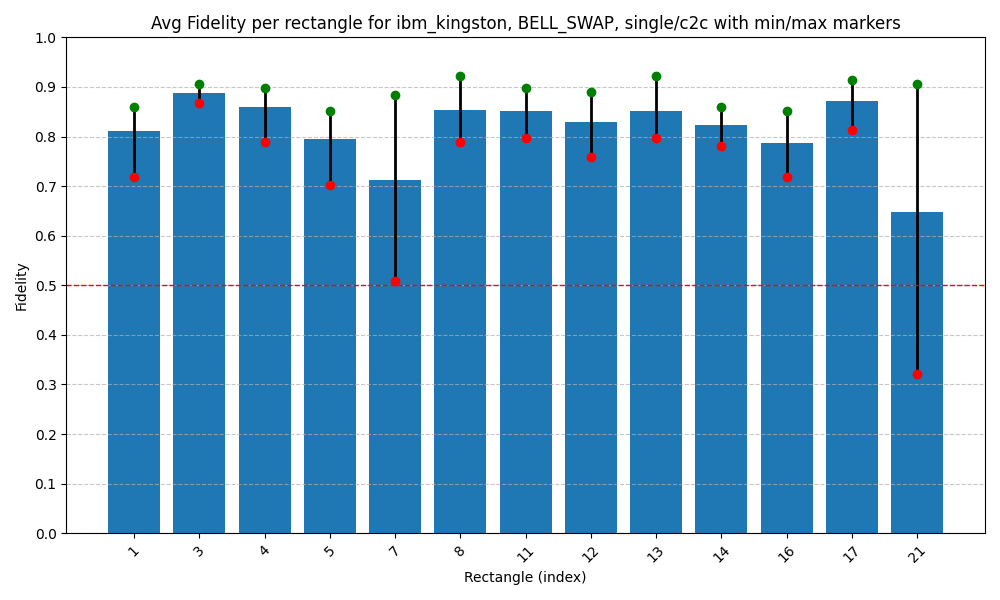}
    \caption{9 Oct 2025: bell-state transfer protocol corner to corner on rectangles that passed do-nothing corner to corner (figure~\ref{fig:kingston_do_nothing_single_c2c_old_workflow})}
    \label{fig:kingston_bell_swap_single_c2c_old_workflow}
    \end{subfigure}
    \hfill
    \begin{subfigure}[t]{0.48\linewidth}
        \centering
        \includegraphics[width=\linewidth]{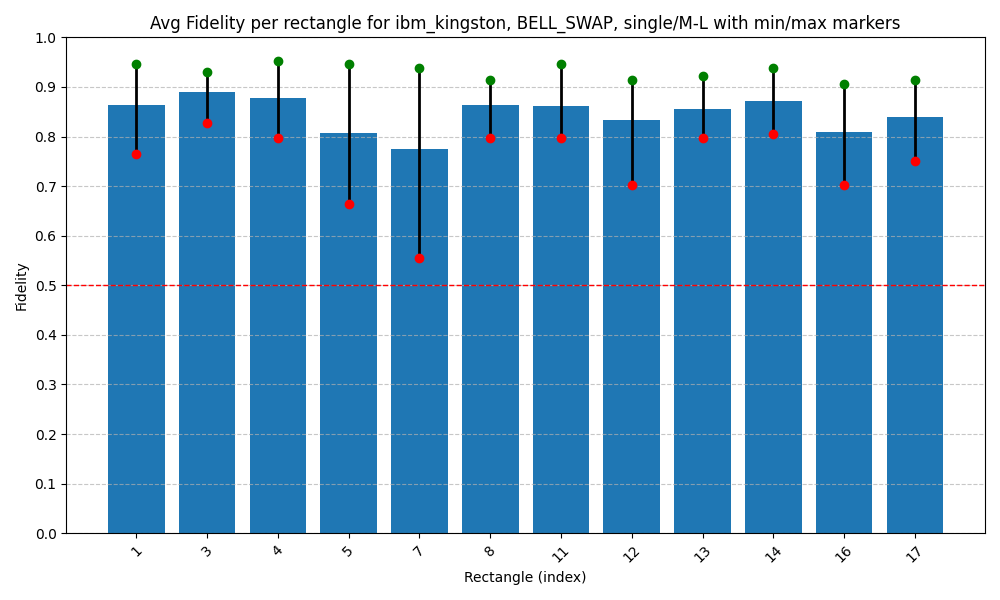}
    \caption{9 Oct 2025: bell-state transfer protocol maximal length on rectangles that passed bell-state transfer corner to corner (figure~\ref{fig:kingston_bell_swap_single_c2c_old_workflow})}
    \label{fig:kingston_bell_swap_single_M-L_old_workflow}
    \end{subfigure}

    \caption{Results of first assessment stages of teleportation on Kingston for single rectangle}
\end{figure}

\subsection{Kingston - Entanglement Swapping on Singles}
\begin{figure}[H]
    \centering

    \begin{subfigure}[t]{0.48\linewidth}
        \centering
        \includegraphics[width=\linewidth]{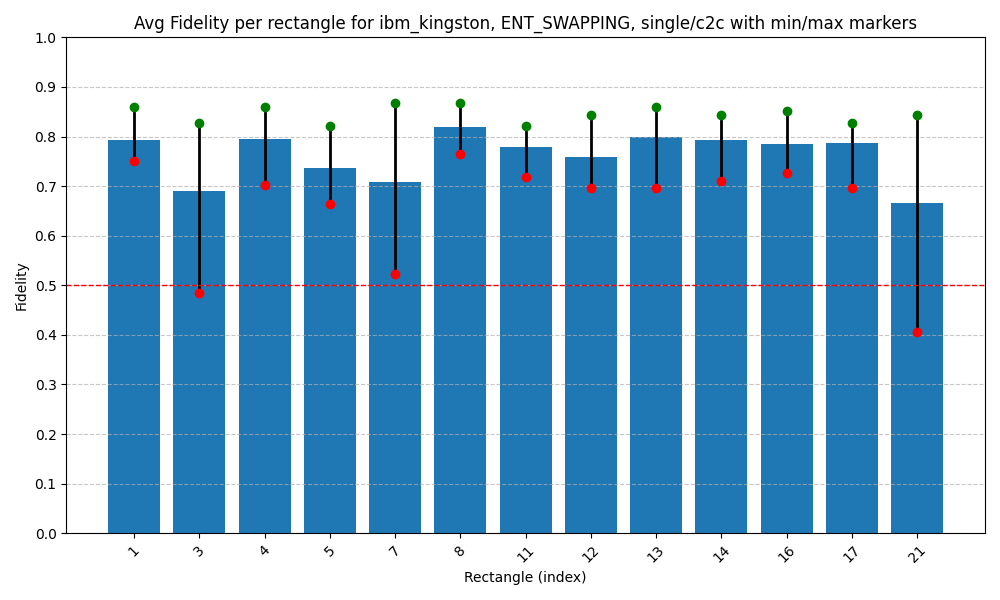}
    \caption{9 Oct 2025: entanglement swapping protocol corner to corner on rectangles that passed do-nothing corner to corner (figure~\ref{fig:kingston_do_nothing_single_c2c_old_workflow})}
    \label{fig:kingston_ent_swap_single_c2c_old_workflow}
    \end{subfigure}
    \hfill
    \begin{subfigure}[t]{0.48\linewidth}
        \centering
        \includegraphics[width=\linewidth]{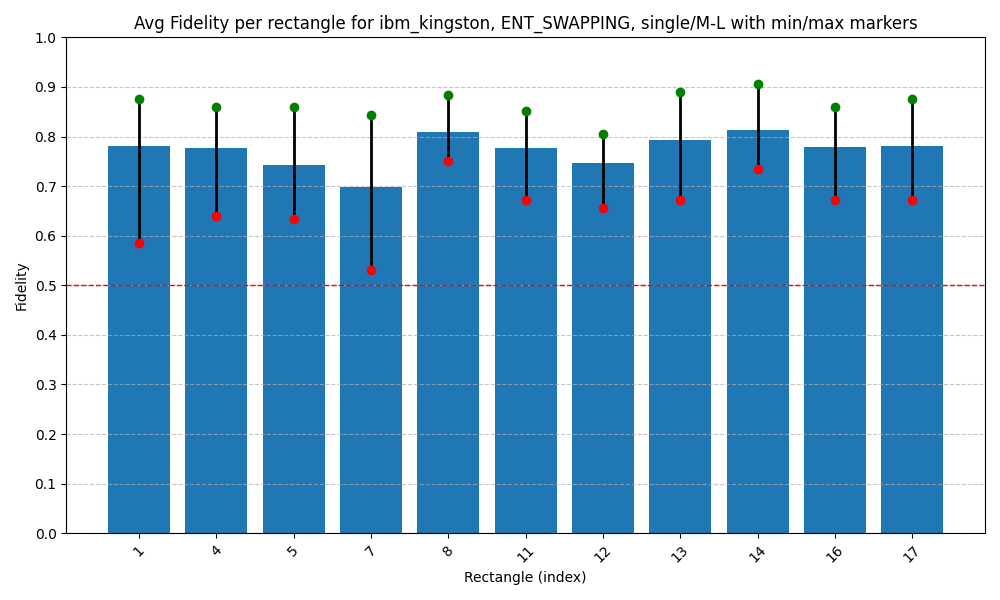}
    \caption{9 Oct 2025: entanglement swapping protocol maximal lengths on rectangles that passed entanglement swapping corner to corner (figure~\ref{fig:kingston_ent_swap_single_c2c_old_workflow})}
    \label{fig:kingston_ent_swap_single_M-L_old_workflow}
    \end{subfigure}    
    
    \caption{Results of first assessment stages of entanglement swapping on Kingston for single rectangle}
\end{figure}

\subsection{Kingston - Super-Dense Coding on Singles}
\begin{figure}[H]
    \centering

    \begin{subfigure}[t]{0.48\linewidth}
        \centering
        \includegraphics[width=\linewidth]{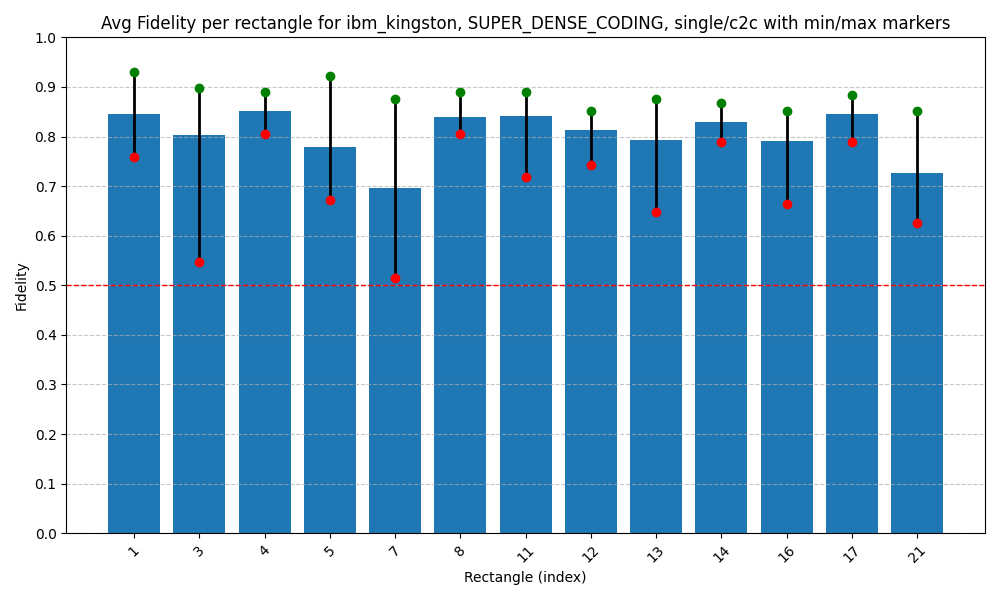}
    \caption{9 Oct 2025: super-dense coding protocol corner to corner on rectangles that passed do-nothing corner to corner (figure~\ref{fig:kingston_do_nothing_single_c2c_old_workflow})}
    \label{fig:kingston_superdense_single_c2c_old_workflow}
    \end{subfigure}
    \hfill
    \begin{subfigure}[t]{0.48\linewidth}
        \centering
        \includegraphics[width=\linewidth]{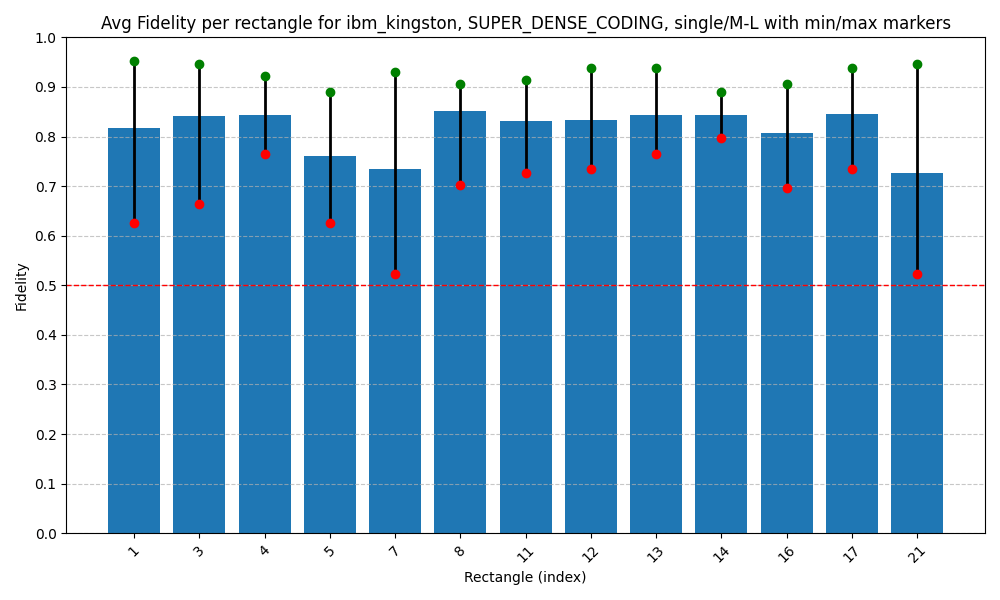}
    \caption{9 Oct 2025:  super-dense coding protocol maximal lengths on rectangles that passed  super-dense coding corner to corner (figure~\ref{fig:kingston_superdense_single_c2c_old_workflow})}
    \label{fig:kingston_superdense_single_M-L_old_workflow}
    \end{subfigure}

    \caption{Results of first assessment stages of teleportation on Kingston for single rectangle}
\end{figure}

\subsection{Kingston - All Protocols on Pairs}

\begin{figure}[H]
    \centering

    \begin{subfigure}[t]{0.48\linewidth}
        \centering
        \includegraphics[width=\linewidth]{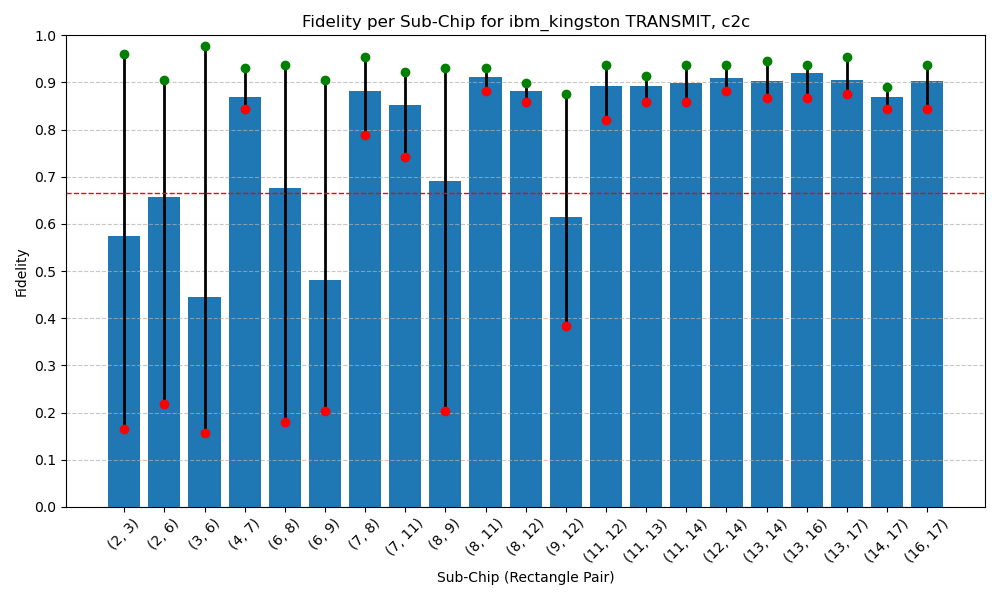}
    \caption{12 Oct 2025: transmit protocol, corner to corner on neighboring pairs that passed transmit all lengths from 28 Jun 2025 (figure~\ref{fig:kingston_transmit_A-L})}
    \label{fig:kingston_transmit_c2c_pairs}
    \end{subfigure}
    \hfill
    \begin{subfigure}[t]{0.48\linewidth}
        \centering
        \includegraphics[width=\linewidth]{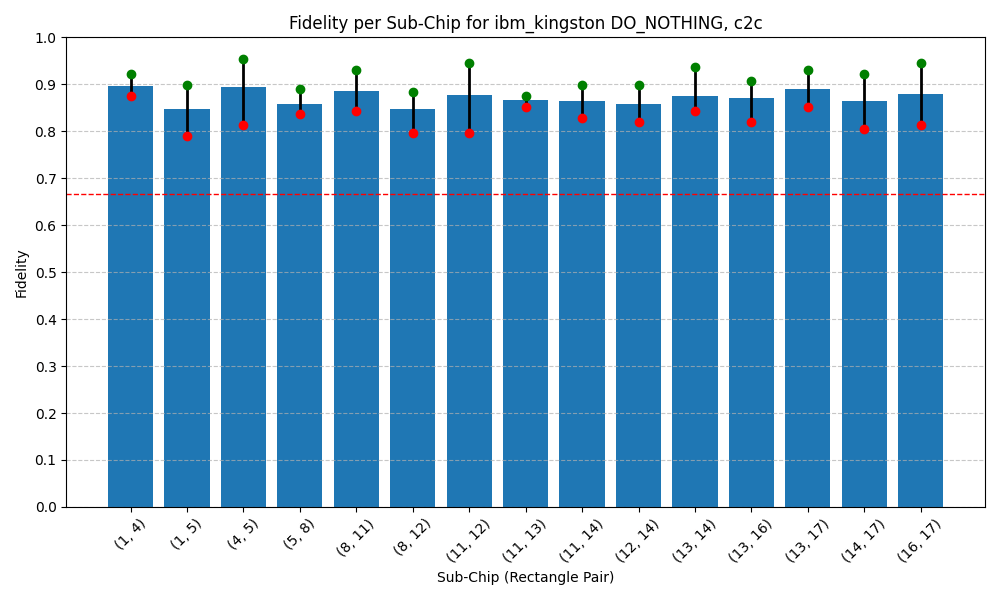}
    \caption{12 Oct 2025: do-nothing protocol, c2c on neighboring pairs of rectangles that passed do-nothing all lengths (figure~\ref{fig:kingston_do_nothing_single_A-L_old_workflow})}
    \label{fig:kingston_do_nothing_pair_c2c_old_workflow}
    \end{subfigure}

    \begin{subfigure}[t]{0.48\linewidth}
        \centering
        \includegraphics[width=\linewidth]{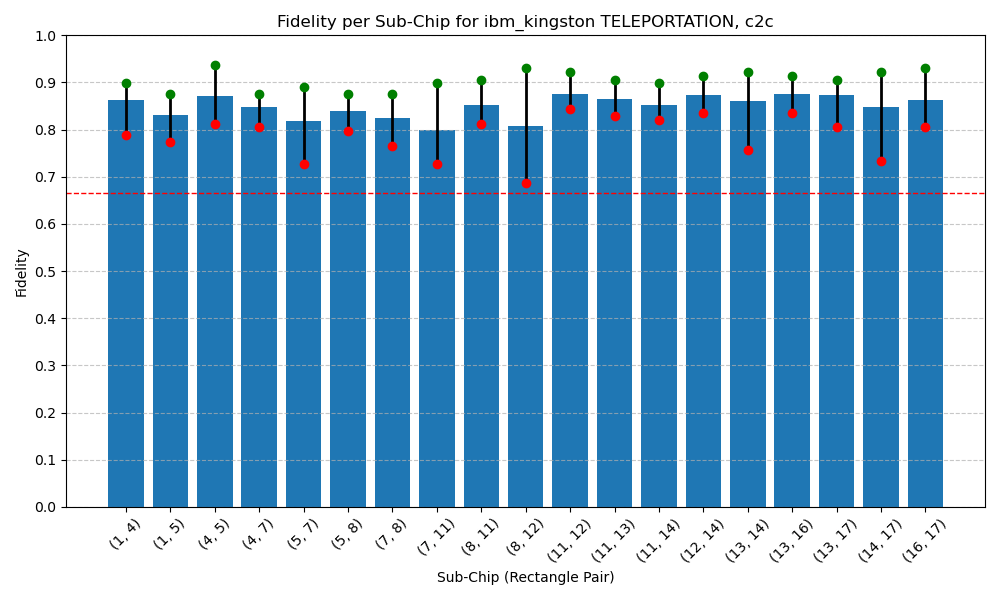}
        \caption{26 Oct 2025: teleportation protocol, c2c on neighboring pairs of rectangles that passed teleportation all lengths (figure~\ref{fig:kingston_teleportation_single_A-L_old_workflow})}
        \label{fig:kingston_teleportation_pair_c2c_old_workflow}
    \end{subfigure}
    \hfill
    \begin{subfigure}[t]{0.48\linewidth}
        \centering
        \includegraphics[width=\linewidth]{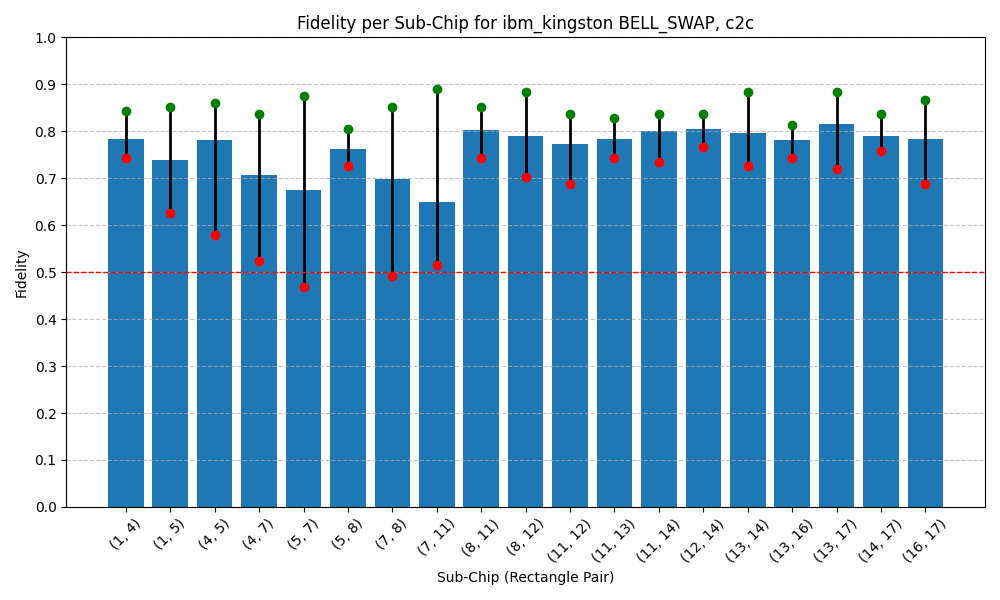}
    \caption{12 Oct 2025: bell-state transfer protocol, c2c on neighboring pairs of rectangles that passed bell-state transfer all lengths (figure~\ref{fig:kingston_bell_swap_single_A-L_old_workflow})}
    \label{fig:kingston_bell_swap_pair_c2c_old_workflow}
    \end{subfigure}

    \begin{subfigure}[t]{0.48\linewidth}
        \centering
        \includegraphics[width=\linewidth]{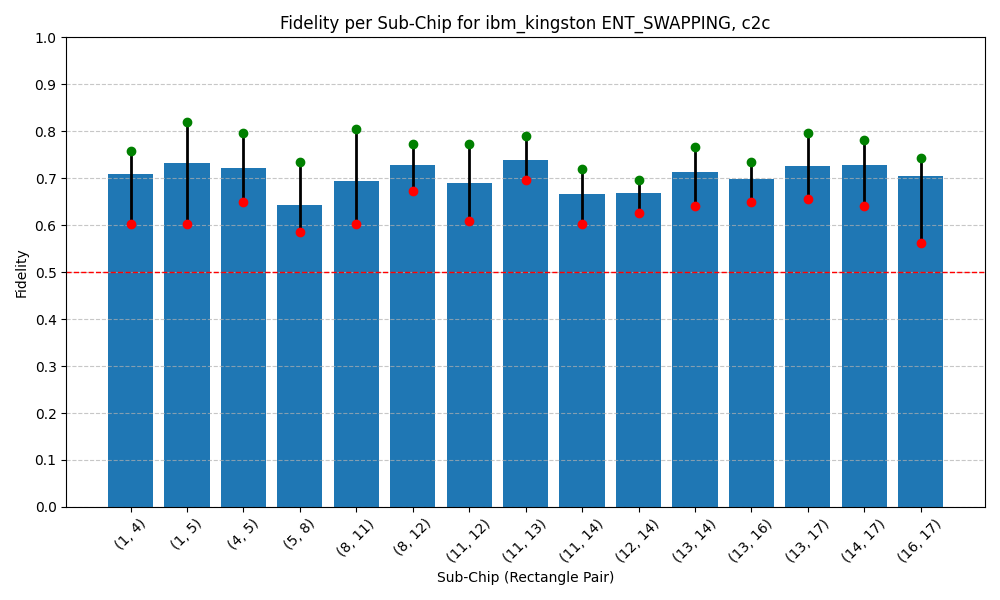}
    \caption{12 Oct 2025: entanglement swapping protocol, c2c on neighboring pairs of rectangles that passed entanglement swapping all lengths (figure~\ref{fig:kingston_ent_swap_single_A-L_old_workflow})}
    \label{fig:kingston_ent_swap_pair_c2c_old_workflow}
    \end{subfigure}
    \hfill
    \begin{subfigure}[t]{0.48\linewidth}
        \centering
        \includegraphics[width=\linewidth]{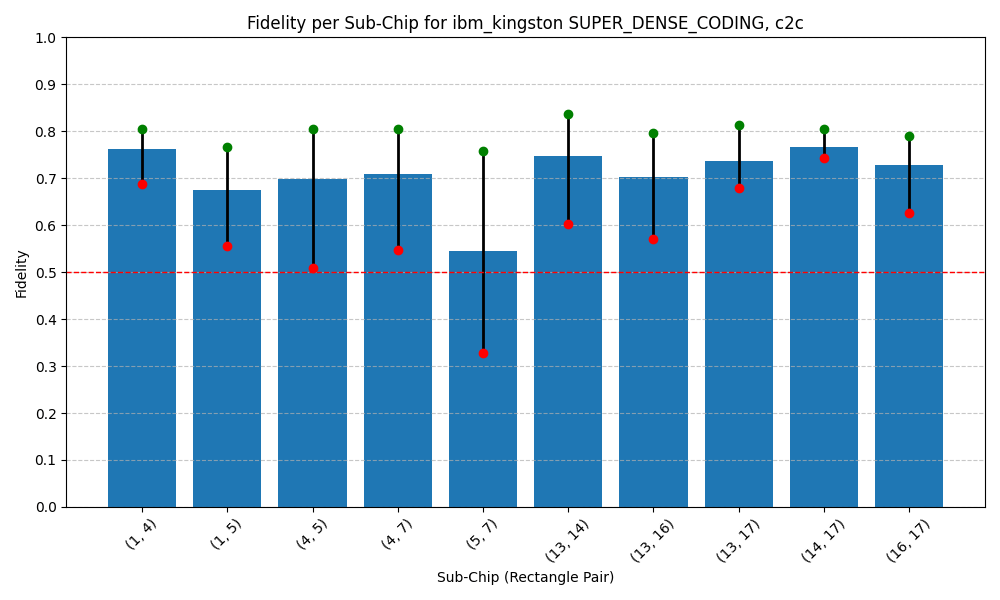}
    \caption{12 Oct 2025: super-dense coding protocol, c2c on neighboring pairs of rectangles that passed super-dense coding all lengths (figure~\ref{fig:kingston_superdense_single_A-L_old_workflow})}
    \label{fig:kingston_superdense_pair_c2c_old_workflow}
    \end{subfigure}
    
    \caption{Charts of all protocols c2c stage on pairs for Kingston}
    \label{fig:kingston_all_paris_c2c}
\end{figure}

\section{Appendix - Swap Distance Charts}\label{sec:swap_dist_charts}
The swap distance charts based on the A-L experiments of the full assessments presented in sections \ref{sec:mod_brisbane_results_single_and_pairs} and \ref{sec:kinsgton_results}.  A detailed explanation about those charts and how to understand them in section \ref{sec:mod_brisbane_transmit}.
\begin{figure}[H]
\centering

    \begin{subfigure}[t]{0.32\linewidth}
        \centering
        \includegraphics[width=1\linewidth]{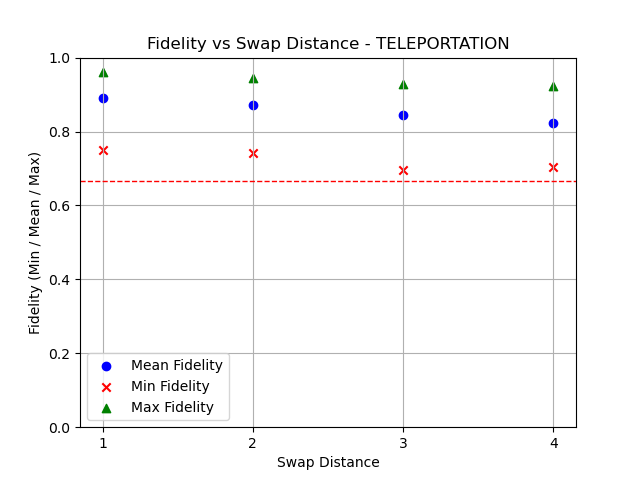}
        \caption{8 Oct 2025: Brisbane, Showing fidelity as function of the swap distance (the distance between Alice and Bob in qubits). Note that this chart contains only results of rectangles that passed teleportation in A-L (fig \ref{fig:brisbane_Oct8_teleportation_single_A-L})}
        \label{fig:brisbane_Oct8_teleportation_single_swap_dist}
    \end{subfigure}
    \hfill
    \begin{subfigure}[t]{0.32\linewidth}
        \centering
        \includegraphics[width=1\linewidth]{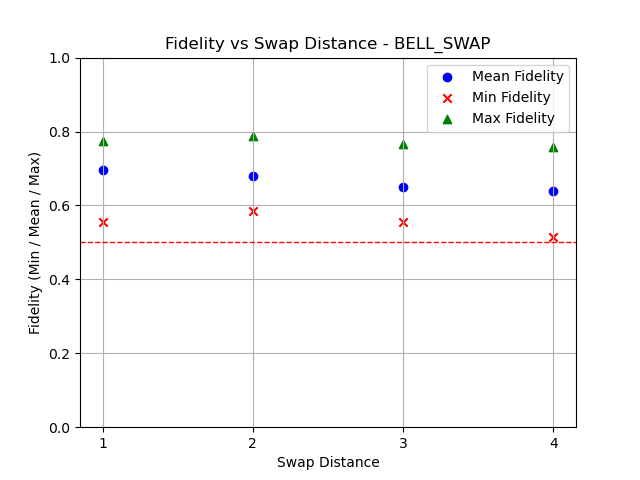}
        \caption{28 Sep 2025: Brisbane, Showing fidelity as function of the swap distance (the distance between Alice and Bob in qubits). Note that this chart contains only results of rectangles that passed bell-state transfer in A-L (fig \ref{fig:brisbane_Sep28_Bell_swap_single_A-L})}
        \label{fig:brisbane_Sep28_bell_swap_single_swap_dist}
    \end{subfigure}
    \hfill
    \begin{subfigure}[t]{0.32\linewidth}
        \centering
        \includegraphics[width=1\linewidth]{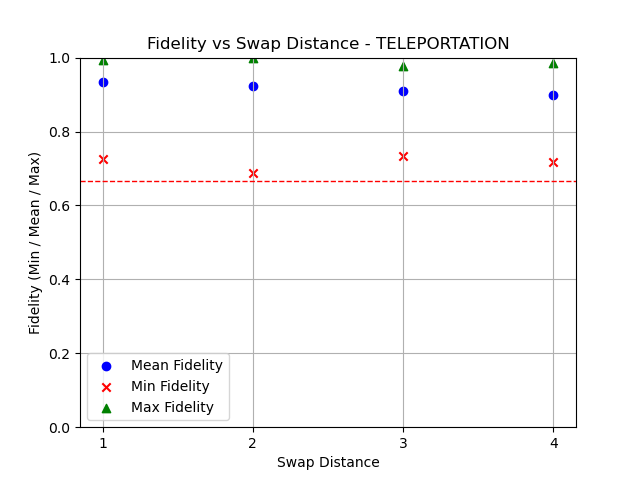}
        \caption{9 Oct 2025: Kingston, Showing fidelity as function of the swap distance (the distance between Alice and Bob in qubits). Note that this chart contains only results of rectangles that passed teleportation in A-L (fig \ref{fig:kingston_teleportation_single_A-L_old_workflow})}
        \label{fig:kingston_teleportation_single_A-L_old_workflow_swap_dist}
    \end{subfigure}

    \begin{subfigure}[t]{0.32\linewidth}
        \centering
        \includegraphics[width=1\linewidth]{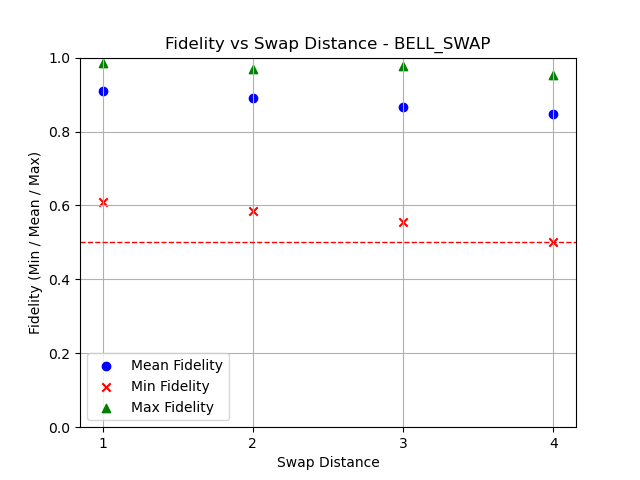}
        \caption{9 Oct 2025: Kingston, Showing fidelity as function of the swap distance (the distance between Alice and Bob in qubits). Note that this chart contains only results of rectangles that passed bell-state transfer in A-L (fig \ref{fig:kingston_bell_swap_single_A-L_old_workflow})}
        \label{fig:kingston_bell_swap_single_A-L_old_workflow_swap_dist}
    \end{subfigure}
    \hfill
    \begin{subfigure}[t]{0.32\linewidth}
        \centering
        \includegraphics[width=1\linewidth]{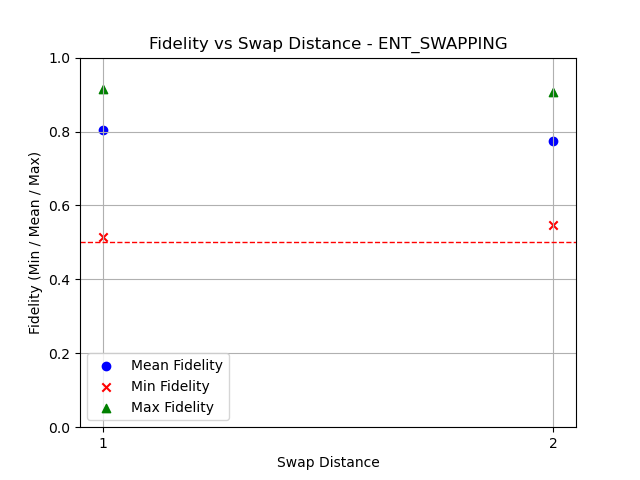}
        \caption{9 Oct 2025: Kingston, Showing fidelity as function of the swap distance (the distance between Alice and Bob in qubits). Note that this chart contains only results of rectangles that passed entanglement swapping in A-L (fig \ref{fig:kingston_ent_swap_single_A-L_old_workflow})}
        \label{fig:kingston_ent_swapping_single_A-L_old_workflow_swap_dist}
    \end{subfigure}
    \hfill
    \begin{subfigure}[t]{0.32\linewidth}
        \centering
        \includegraphics[width=1\linewidth]{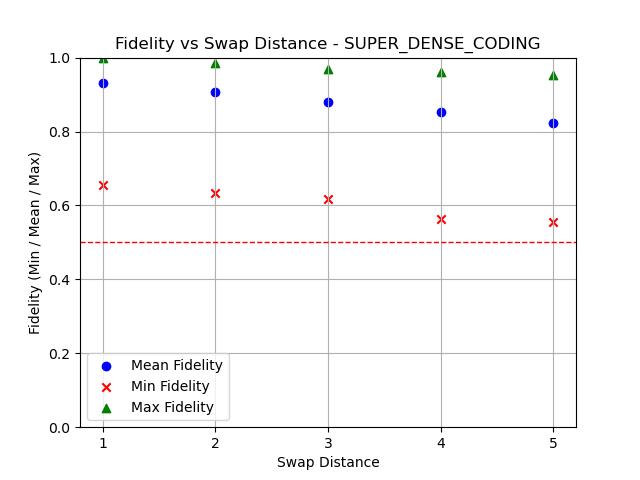}
        \caption{9 Oct 2025: Kingston, Showing fidelity as function of the swap distance (the distance between Alice and Bob in qubits). Note that this chart contains only results of rectangles that passed super-dense coding in A-L (fig \ref{fig:kingston_superdense_single_A-L_old_workflow})}
        \label{fig:kingston_superdense_single_A-L_old_workflow_swap_dist}
    \end{subfigure}

\end{figure}

\end{document}